\documentclass[12pt]{article}
\pdfoutput=1
\usepackage[DIV13]{typearea}
\usepackage{indentfirst}
\usepackage{amsmath}
\usepackage{amsfonts}
\usepackage{mathrsfs}
\usepackage{amssymb}
\usepackage{mathtools}
\usepackage{bbm}
\usepackage{epsfig}
\usepackage{graphicx}
\usepackage{subfigure}
\usepackage{slashed}
\usepackage{multicol}
\usepackage[usenames,dvipsnames]{color}
\usepackage{cite}
\RequirePackage[colorlinks=true,urlcolor=blue,anchorcolor=blue,citecolor=blue,filecolor=blue,
               linkcolor=blue,menucolor=blue,linktocpage=true,pdfproducer=medialab]{hyperref}
\usepackage{cancel}
\usepackage{feynmp}
\usepackage{lscape}
\usepackage{multirow}
\usepackage{array}
\usepackage{pdfpages}
\usepackage{fancyhdr}
\usepackage{pifont}

\usepackage[left=.9in, right=.9in]{geometry}

\newcommand{\email}[1]{\href{mailto:#1}{\tt #1}}

\numberwithin{equation}{section}
\newcommand{\LL}{\mathscr{L}}

\def\cD{{\cal D}}
\def\cF{{\cal F}}

\def\cS{{\cal S}}

\def\cY{{\bf Y}}

\def\be{\begin{equation}}
\def\ee{\end{equation}}
\def\beq{\begin{equation}}
\def\eeq{\end{equation}}
\def\bc{\begin{center}}
\def\ec{\end{center}}
\def\bea{\begin{eqnarray}}
\def\eea{\end{eqnarray}}

\def\nn{\nonumber}

\newcommand{\GeV}{\;\text{GeV}}

\newcommand{\fb}{\;\text{fb}}

\newcommand{\derp}{\partial}

\newcommand{\hc}{\mathrm{h.c.}}

\newcommand{\diag}{{\rm{\bf diag}}}
\newcommand{\UH}{\mathbf{U}}

\newcommand{\TL}{\mathbf{T}}
\newcommand{\VL}{\mathbf{V}}

\newcommand{\DLR}{\mathbf{D}}
\newcommand{\DLL}{\mathcal{D}}

\newcommand{\tr}{{\rm Tr}}

\renewcommand{\to}{\rightarrow}

\newcommand{\BBu}{B^{\mu\nu}}
\newcommand{\BBd}{B_{\mu\nu}}

\newcommand{\WWd}{W_{\mu\nu}}

\newcommand{\bmat}{\begin{pmatrix}}
\newcommand{\emat}{\end{pmatrix}}

\newcommand{\ie} {{\it i.e.}}

\newcommand{\blue}[1]{\color{blue} #1 \color{black}}

\begin{document}
\begin{titlepage}
\vspace*{-1cm}
\phantom{hep-ph/***} 
{\flushleft
{\blue{FTUAM-14-22}}
\hfill{\blue{IFT-UAM/CSIC-14-056}}\\
{\blue{YITP-SB-14-17}}
\hfill{\blue{DFPD-2014/TH/13}}}
\vskip 1cm
\begin{center}
\mathversion{bold}
{\LARGE\bf CP violation with a dynamical Higgs}\\
\mathversion{normal}
\vskip .3cm
\end{center}
\vskip 0.5  cm
\begin{center}
{\large M.B.~Gavela}~$^{a)}$,
{\large J.\ Gonzalez--Fraile}~$^{b)}$,
{\large M.\ C.\ Gonzalez--Garcia}~$^{b),c),d)}$,
{\large L.~Merlo}~$^{a)}$,
{\large S.~Rigolin}~$^{e)}$,
{\large J.~Yepes}~$^{a)}$
\\
\vskip .7cm
{\footnotesize
$^{a)}$~
Departamento de F\'isica Te\'orica and Instituto de F\'{\i}sica Te\'orica, IFT-UAM/CSIC,\\
Universidad Aut\'onoma de Madrid, Cantoblanco, 28049, Madrid, Spain\\
\vskip .1cm
$^{b)}$~
Departament d'Estructura i Constituents de la Mat\`eria and ICC-UB,\\ Universitat de Barcelona, 647 Diagonal, E-08028 Barcelona, Spain\\
\vskip .1cm
$^{c)}$~
C.N.~Yang Institute for Theoretical Physics and Department of Physics and Astronomy, \\SUNY at Stony Brook, Stony Brook, NY 11794-3840, USA\\
\vskip .1cm
$^{d)}$~
Instituci\'o Catalana de Recerca i Estudis Avan\c{c}ats (ICREA)\\
\vskip .1cm
$^{e)}$~
Dipartimento di Fisica e Astronomia ``G.~Galilei'', Universit\`a di Padova and \\
INFN, Sezione di Padova, Via Marzolo~8, I-35131 Padua, Italy
\vskip .3cm
\begin{minipage}[l]{.9\textwidth}
\begin{center} 
\textit{E-mail:} 
\email{belen.gavela@uam.es},
\email{fraile@ecm.ub.edu},
\email{concha@insti.physics.sunysb.edu},
\email{luca.merlo@uam.es},
\email{stefano.rigolin@pd.infn.it},
\email{ju.yepes@uam.es}
\end{center}
\end{minipage}
}
\end{center}
\vskip 0.5cm
\begin{abstract}
We determine the complete set of independent gauge and gauge-Higgs CP-odd effective operators for the generic case of a
dynamical Higgs, up to four derivatives in the chiral expansion. The relation with the linear basis of dimension six
CP-odd operators is clarified. Phenomenological applications include bounds inferred from electric dipole moment limits,
and from present and future collider data on triple gauge coupling measurements and Higgs signals.
\end{abstract}
\end{titlepage}
\setcounter{footnote}{0}

\tableofcontents

%
%

\newpage

\section{Introduction}

While charge conjugation (C) and parity (P) are not exact symmetries
of the Standard Model of particle physics, present
data~\cite{CMS:yva,ATLAScouplings,Aad:2013wqa,Chatrchyan:2013mxa} are
consistent with the Higgs particle being the Standard Model (SM)
scalar~\cite{Englert:1964et,Higgs:1964ia,Higgs:1964pj}, 
which is defined as a CP-even $SU(2)_L$ doublet scalar.
Nevertheless, in the plausible perspective that particle
physics is not at the end of the road and beyond the SM physics (BSM)
is awaiting discovery as an explanation of the electroweak hierarchy
problem, it is necessary to track the possible non-doublet and/or
CP-odd components of the observed resonance, in particular in view of
the sizeable present error bars. This is underway through different
complementary strategies: kinematical analysis, direct searches of new
resonances expected in particular BSM theories, or indirect signals
other than kinematic ones. Indirect searches may well give fruitful
results prior to the discovery of new resonances, and may allow to
explore and disentangle the two possible avenues of realisation of
electroweak symmetry breaking (EWSB):
linear~\cite{Buchmuller:1985jz,Grzadkowski:2010es,Hagiwara:1993ck,Hagiwara:1996kf,Hagiwara:1993qt,GonzalezGarcia:1999fq,Low:2012rj,Corbett:2012dm,Ellis:2012hz,Giardino:2012dp,Montull:2012ik,Espinosa:2012im,Carmi:2012in,Banerjee:2012xc,Bonnet:2012nm,Plehn:2012iz,Djouadi:2012rh,Batell:2012ca,Moreau:2012da,Cacciapaglia:2012wb,Azatov:2012qz,Masso:2012eq,Passarino:2012cb,Corbett:2012ja,Falkowski:2013dza,Giardino:2013bma,Ellis:2013lra,Djouadi:2013qya,Contino:2013kra,Dumont:2013wma,Elias-Miro:2013mua,Lopez-Val:2013yba,Jenkins:2013zja,Pomarol:2013zra,Banerjee:2013apa,Alloul:2013naa,Englert:2014uua}
--which is typical of BSM theories in which the Higgs particle is
elementary-- or
non-linear~\cite{Bagger:1993zf,Koulovassilopoulos:1993pw,Burgess:1999ha,Grinstein:2007iv,Contino:2010mh,Azatov:2012bz,Buchalla:2012qq,Alonso:2012px,Alonso:2012pz,Buchalla:2013rka,Brivio:2013pma,Brivio:2014pfa}
--as for instance in models in which the Higgs boson is a composite
pseudo-goldstone boson of some strong-interacting BSM theory or a dilaton.

Interesting past and new proposals to search for CP-odd anomalous
couplings of the Higgs boson to fermions and gauge
bosons~\cite{Dell'Aquila:1985ve,Dell'Aquila:1985vc,Dell'Aquila:1985vb,Soni:1993jc,Chang:1993jy,Arens:1994wd,Choi:2002jk,Buszello:2002uu,Godbole:2007cn,Cao:2009ah,Gao:2010qx,DeRujula:2010ys,Plehn:2001nj,Buszello:2006hf,Hankele:2006ma,Klamke:2007cu,Englert:2012ct,Odagiri:2002nd,DelDuca:2006hk,Andersen:2010zx,Englert:2012xt,Djouadi:2013yb,Dolan:2014upa,Christensen:2010pf,Desai:2011yj,Ellis:2012xd,Godbole:2013saa,Delaunay:2013npa,Voloshin:2012tv,Korchin:2013ifa,Bishara:2013vya,Chen:2014ona,Freitas:2012kw,Djouadi:2013qya,Belusca-Maito:2014dpa,Shu:2013uua,Cheung:2013kla} rank from purely phenomenological analysis to
the identification of effective signals expected assuming either a
linear or a non-linear realisation of EWSB.  In previous literature,
some of the CP-odd gauge and/or gauge-Higgs operators to be discussed
below had not been explored, but traded instead by fermionic ones via
the equations of motion\footnote{See for instance
Ref.~\cite{Buchalla:2013rka} for an analysis in the framework of
non-linear EWSB.}. Nevertheless, it is theoretically very
interesting to identify and analyse the complete set of independent
CP-odd bosonic operators, as they may shed a direct light
on the nature of EWSB, which takes place precisely in the bosonic
sector. Moreover, the present LHC data offer increasingly rich and
precise constraints on gauge and gauge-Higgs couplings, up to the
point of becoming competitive with fermionic bounds in constraining
BSM theories; this trend may be further strengthened with the post-LHC
facilities presently under discussion.

We discuss here the issue of CP-violation in the case of non-linear realisations of EWSB.
To be generic and model-independent, a non-linear (also dubbed ``chiral") effective Lagrangian
will be used to describe physics at energies lower than the characteristic BSM scale(s).
The complete and independent set -- that is, the basis -- of CP-odd bosonic  operators of
the non-linear expansion will be determined here for the first time, up to four derivative couplings. The differences
with the leading anomalous couplings and signals expected from linear realizations of BSM physics
will be also identified. Phenomenological constraints resulting from limits on electric dipole
moments (EDMs) and from present LHC data will be derived as well, and future prospects briefly discussed.
The structure of the paper can be easily inferred from the Table of Contents.

%
%
\boldmath
\section{Effective CP-odd chiral bosonic Lagrangian}
\label{Sect:Basis}
\unboldmath

Reference~\cite{Alonso:2012px} developed the effective Lagrangian for a light dynamical
Higgs, up to four derivative couplings and restricted to the CP-even bosonic sector, except for the
inclusion of Yukawa-like interactions\footnote{As usual, derivative is understood in the sense
of covariant derivative. That is, a gauge field and a momentum have
both chiral dimension one and their inclusion in non-renormalizable
operators is weighted down by the same high-energy strong-interaction scale
$\Lambda_s$.}. Its CP-odd counterpart will be studied below.

The most up-to-date analyses of Higgs data have established that
the couplings of $h$ to the gauge bosons and the absolute values of its
couplings to fermions are compatible with the SM ones\footnote{The sign of the couplings between $h$ and fermions is
still to be measured, although a slight preference for a positive value is indicated in some two parameter
fits (see for example~\cite{Espinosa:2012im,Montull:2012ik,Azatov:2012qz})
which take into account one-loop induced EW corrections; we will consider this option in what follows.}.
It is then justified from the phenomenological point of view to consider the SM as the leading-order Lagrangian
$\LL_{SM}$ and treat as corrections all possible departures due to the unknown high-energy strong dynamics. Here only the CP-odd sector will be explicitly addressed, while the CP-even sector has been already studied in Refs.~\cite{Alonso:2012px} and will be left implicit. The effective Lagrangian can then be written as
\beq
\LL_\text{chiral} = \LL_{SM}+\Delta\LL_{\text{\cancel{CP}}}\,,
\label{Lchiral}
\eeq 
where the first term reads 
\beq
\begin{split}
\LL_{SM} =& \frac{1}{2} (\derp_\mu h)(\derp^\mu h) -\dfrac{1}{4}\BBd\BBu-\dfrac{1}{4}\WWd^a W^{a\mu\nu}-
\dfrac{1}{4} G^a_{\mu\nu}G^{a\mu\nu}- V (h)\\
 &-\dfrac{(v+h)^2}{4}\tr[\VL_\mu\VL^\mu]+ i\bar{Q}\slashed{D}Q+i\bar{L}\slashed{D}L\\
 &-\dfrac{v+  h}{\sqrt2}\left(\bar{Q}_L\UH \cY_Q Q_R+\hc\right)-\dfrac{v+  h}{\sqrt2}\left(\bar{L}_L\UH \cY_L L_R+\hc\right)\,\\
 &-\frac{g_s^2}{16\pi^2}\theta_s\,G^a_{\mu\nu}\, \tilde{G}^a_{\rho\sigma}\,.
\end{split}
\label{LLO}
\eeq
In this expression $\UH\equiv \exp\left(i \pi\cdot\tau/v\right)$
--with $\tau$ denoting the Pauli matrices-- is a unitary matrix which
efficiently encodes the longitudinal degrees of freedom of the heavy
gauge bosons and transforms as a $(2,2)$ of the global $SU(2)_L\times
SU(2)_R$ symmetry group of the Lagrangian, and \mbox{$\VL_\mu\equiv
\left(\DLR_\mu\UH\right)\UH^\dagger$} is the vector chiral field
transforming in the adjoint of $SU(2)_L$. Furthermore, $v$ is the EW
scale, defined via the $W$ gauge boson mass $M_W=gv/2$, and $h$
denotes the Higgs particle. The covariant derivative reads 
\beq
\DLR_\mu \UH(x) \equiv \derp_\mu \UH(x) +igW_{\mu}(x)\UH(x) 
- \dfrac{ig'}{2} B_\mu(x) \UH(x)\tau_3 \,, 
\eeq
with $W_\mu\equiv W_{\mu}^a(x)\tau_a/2$ and $B_\mu$ denoting the
$SU(2)_L$ and $U(1)_Y$ gauge bosons, respectively. In Eq.~(\ref{LLO}),
the first line describes the $h$ and gauge boson kinetic terms, as well as 
the effective scalar potential $V(h)$, accounting for the breaking of
the electroweak symmetry. The second line describes the $W$ and $Z$
masses and their interactions with $h$, as well as the kinetic terms
for Goldstone bosons and fermions. The third line corresponds to the
Yukawa-like interactions written in the fermionic mass eigenstate
basis. A compact notation for the
right-handed fields has been adopted, gathering them into
doublets $Q_R$ and $L_R$. $\cY_Q$
and $\cY_L$ are two $6\times6$ block-diagonal matrices containing the
usual Yukawa couplings:
\beq
\cY_Q\equiv\diag\left(Y_U,\, Y_D\right)\,,\qquad\qquad
\cY_L\equiv  \diag\left(Y_\nu,\, Y_L\right)\,,
\label{Yukawas}
\eeq
where the Cabibbo-Kobayashi-Maskawa mixing is understood to be encoded
in the definition of $Q_L$, thus accounting for the SM CP-even fermionic
couplings. Finally, the last term in Eq.~(\ref{LLO}) corresponds to
the well-known total derivative CP-odd gluonic coupling, for which the
notation used is that in which the dual field-tensor of any field
strength $X_{\mu\nu}$ is defined as $\tilde{X}^{\mu\nu} \equiv
\frac{1}{2} \epsilon^{\mu\nu\rho\sigma}X_{\rho\sigma}$.

This description is  data-driven and, while being a consistent chiral expansion up to four derivatives, the particular division in Eq.~(\ref{Lchiral}) does not match that in number of derivatives, usually adopted by chiral Lagrangian practitioners. For instance, the usual custodial breaking term $\tr(\TL\VL_\mu)\tr(\TL\VL^\mu)$, being $\TL\equiv\UH\tau_3\UH^\dag$, is a two derivative operator and is often listed among the leading order set in the chiral expansion; however, it is not present in the SM at tree level and data strongly constrain its coefficient so that in practice it can be always considered~\cite{Contino:2010mh} a subleading operator. Moreover, in the phenomenological Lagrangian in Eq.~(\ref{LLO}) the Higgs couplings with gauge bosons and fermions have been taken SM-like, as suggested by data. However, in the non-linearly realized EWSB framework, this is not guaranteed by any symmetry and it should be considered as a phenomenological accident. A more general notation has been adopted in Ref.~\cite{Alonso:2012px} for the case of the CP-even chiral Lagrangian for a dynamical Higgs. This issue is irrelevant for the focus of this paper, as if the latter notation was adopted here, the complete four-derivative basis $\Delta\LL_{\text{\cancel{CP}}}$ (see below Eq.~(\ref{CPbasis})) would be exactly the same.

\subsection{Basis of CP-odd pure gauge and gauge-Higgs operators}
 The CP-odd corrections will be parametrised as
\begin{align}
\Delta \LL_{\text{\cancel{CP}}}&= c_{\widetilde{B}}\,\cS_{\widetilde{B}}(h) + c_{\widetilde{W}}\,\cS_{\widetilde{W}}(h) + c_{\widetilde{G}}\,\cS_{\widetilde{G}}(h) + c_{2D}\,\cS_{2D}(h) + \sum_{i=1}^{16}\,c_i\,\cS_{i}(h)\,,
\label{LCP4}
\end{align}
where $c_i$ are model-dependent constant coefficients and
\beq
\begin{aligned}
&\cS_{\widetilde{B}}(h)\equiv -\frac{1}{2}g'^2\,B^{\mu\nu}\, \widetilde{B}_{\mu\nu}\,\cF_{\widetilde{B}}(h)\,,
&&\cS_{7}(h)\equiv g\, \text{Tr}\left(\TL\,\left[W^{\mu\nu},\VL_\mu\right]\right)\,\partial_{\nu}\cF_{7}(h)\,, \\ 
&\cS_{\widetilde{W}}(h)\equiv -\frac{1}{2}g^2\text{Tr}\left(W^{\mu\nu}\widetilde{W}_{\mu\nu}\right)\cF_{\widetilde{W}}(h)\,,
&&\cS_{8}(h)\equiv 2\,g^2\text{Tr}\left(\TL\,\widetilde{W}^{\mu\nu}\right)\text{Tr}\left(\TL\,W_{\mu\nu}\right)\cF_{8}(h)\,, \\ 
&\cS_{\widetilde{G}}(h)\equiv -\frac{1}{2}g^2_s\,G^{a\mu\nu}\, \widetilde{G}^a_{\mu\nu}\,\cF_{\widetilde{G}}(h)\,,
&&\cS_{9}(h)\equiv 2\,i\,g\text{Tr}\left(\widetilde{W}^{\mu\nu}\,\TL\right)\text{Tr}\left(\TL\,\VL_\mu\right)\partial_{\nu}\cF_{9}(h)\,,\\
&\cS_{2D}(h)\equiv i\,\frac{v^2}{4}\,\text{Tr}\left(\TL\,\cD^\mu\VL_\mu\right)\,\cF_{2D}(h)\,,
&&\cS_{10}(h)\equiv i\,\text{Tr}\left(\VL^\mu\,\cD^\nu\VL_\nu\right)\,\text{Tr}\left(\TL\,\VL_\mu\right)\,\cF_{10}(h)\,, \\
&\cS_1(h)\equiv 2g\,g'\,\widetilde{B}^{\mu\nu}\text{Tr}\left(\TL W_{\mu\nu}\right)\,\cF_1(h)\,,
&&\cS_{11}(h)\equiv i\,\text{Tr}\left(\TL\,\cD^\mu\VL_\mu\right)\,\text{Tr}\left(\VL^\nu\,\VL_\nu\right)\,\cF_{11}(h)\,,  \\ 
&\cS_{2}(h)\equiv 2\,i\,g'\,\widetilde{B}^{\mu\nu}\,\text{Tr}\left(\TL\,\VL_\mu\right)\,\partial_{\nu}\cF_{2}(h)\,,
&&\cS_{12}(h)\equiv i\,\text{Tr}\left([\VL^\mu,\TL]\,\cD^\nu\VL_\nu\right)\,\partial_{\mu}\cF_{12}(h)\,, \\ 
&\cS_{3}(h)\equiv 2\,i\,g\,\text{Tr}\left(\widetilde{W}^{\mu\nu}\,\VL_\mu\right)\,\partial_{\nu}\cF_{3}(h)\,,
&&\cS_{13}(h)\equiv i\,\text{Tr}\left(\TL\,\cD^\mu\VL_\mu\right)\,\partial^{\nu}\partial_{\nu}\cF_{13}(h)\,, \\
&\cS_{4}(h)\equiv g\text{Tr}\left(W^{\mu\nu}\VL_\mu\right)\text{Tr}\left(\TL\,\VL_\nu\right)\cF_{4}(h)\,,
&&\cS_{14}(h)\equiv i\,\text{Tr}\left(\TL\,\cD^\mu\VL_\mu\right)\,\partial^{\nu}\cF_{14}(h)\,\partial_{\nu}\cF'_{14}(h) \,, \\ 
&\cS_{5}(h)\equiv i\text{Tr}\left(\VL^\mu\,\VL^\nu\right)\text{Tr}\left(\TL\,\VL_\mu\right)\partial_{\nu}\cF_{5}(h)\,,
&&\cS_{15}(h)\equiv i\,\text{Tr}\left(\TL\,\VL^\mu\right)\,\left(\text{Tr}\left(\TL\,\VL^\nu\right)\right)^2\,\partial_{\mu}\cF_{15}(h)\,,\\ 
&\cS_{6}(h)\equiv i\text{Tr}\left(\VL^\mu\,\VL_\mu\right)\text{Tr}\left(\TL\,\VL^\nu\right)\partial_{\nu}\cF_{6}(h)\,,
&&\cS_{16}(h)\equiv i\,\text{Tr}\left(\TL\,\cD^\mu\VL_\mu\right)\,\left(\text{Tr}\left(\TL\,\VL^\nu\right)\right)^2\,\cF_{16}(h)\,,
\end{aligned}
\label{CPbasis}
\eeq
with the $\cF_i(h)$-functions  for all operators\footnote{The Higgs-independent term in this functional is physically
irrelevant for operators $\cS_{\widetilde{B}}(h)$, $\cS_{\widetilde{W}}(h)$, $\cS_{2D}(h)$.} but $\cS_{\widetilde{G}}(h)$ 
being generic functions of the scalar singlet $h$ defined as~\cite{Alonso:2012px} 
\beq
\cF_i(h)\equiv1+2\,{a}_i\,\frac{h}{v}+{b}_i\,\frac{h^2}{v^2}+\ldots\, ,
\label{F}
\eeq
with dots standing for terms with higher powers in $h/v$ which will not be considered below. $\cF_{\widetilde{G}}(h)$ will
be understood to be also of this form but for the first term in Eq.~(\ref{F}), as the Higgs-independent part of
$\cS_{\widetilde{G}}(h)$ has already been included in the SM Lagrangian, Eq.~(\ref{LLO}). 

The Lagrangian in Eqs.~(\ref{Lchiral}) and (\ref{LCP4}) describes the CP-odd low-energy effects of a high-energy strong dynamics responsible for the electroweak GBs, coupled to a generic scalar singlet $h$. Note that the number of independent operators in the non-linear expansion turned out to be larger than for the analogous
basis in the linear expansion~\cite{Brivio:2013pma,Brivio:2014pfa}, a generic feature when comparing both type of
effective Lagrangians; see Appendix~\ref{App:Siblings}. The basis is also larger than that for chiral expansions developed in the past for the case of a very heavy Higgs particle (i.e. absent at low energies)~\cite{Appelquist:1980vg,Longhitano:1980iz,Longhitano:1980tm,Appelquist:1993ka}, as:
i) terms which in the absence of the $\cF_i(h)$ functions were shown to be equivalent via total derivatives, are now independent; ii) new terms including derivatives of $h$ appear. Some of the operators of the list in Eq.~(\ref{CPbasis}) also appear in Refs.~\cite{Buchalla:2012qq,Buchalla:2013rka}; a complete comparison, however, is not possible, as certain bosonic operators in Eq.~(\ref{CPbasis}) have been translated into operators containing fermions, by using $h$ equations of motion.

%
%
\section{Phenomenology}
\label{sec:pheno}

In what follows we analyse the physical impact of the operators in the
CP-odd bosonic basis determined above. Some phenomenological bounds
and future prospects are discussed as well.


\subsection{CP-odd two-point functions}
\label{sec:2point}

Only the operators $\cS_{2D}(h)$ and $\cS_{13}(h)$ among those defined in Eq.~(\ref{CPbasis}) may a
priori induce renormalisation effects on the fields and couplings of the SM Lagrangian.  $\cS_{2D}(h)$ is a two-derivative coupling and thus part of the leading order of  the chiral expansion; in contrast, note that it has no analogue in the leading order ($d=4$) of the linear expansion --in other words in the SM Lagrangian-- as its lower-dimensional linear sibling would be a dimension six ($d=6$) operator, see Appendix~\ref{App:Siblings}.

$\cS_{2D}(h)$ and $\cS_{13}(h)$ contain two-point functions
which explicitly break the CP symmetry and as a consequence the
Lagrangian eigenstates may not be CP-eigenstates. Those two couplings
result in a mixing of $h$ with the Goldstone bosons which in the SM give masses to the $W$ and $Z$
bosons, see below. Their physical impact 
reduces simply to anomalous CP-odd Higgs-fermion and Higgs-$Z$ couplings, as we show next in detail.

Consider the linear combination of the two operators $\cS_{2D}(h)$ and
$\cS_{13}(h)$, together with the $h$-kinetic term and the gauge-boson
mass term in the Lagrangian of Eq.~(\ref{Lchiral}), and let us focus first on their contribution to two-point functions:
\begin{align}
&\LL_{\text{chiral}} \supset \frac{1}{2}\,\partial^\mu h\,\partial_\mu
  h-\frac{(v+h)^2}{4}\,\text{Tr}\left(\VL^\mu\,\VL_\mu\right) +
  c_{2D}\,\cS_{2D}(h) + c_{13}\,\cS_{13}(h) \nn\\ 
& \supset \frac{1}{2}\,\partial^\mu h\,\partial_\mu h+\frac{v^2}{4}\,\text{Tr}\left(\derp^\mu\UH^\dag\,\derp_\mu\UH\right)
  +\frac{i}{2}\,v\,\text{Tr}\big(\TL\,\left(\partial_\mu\partial^\mu
    \UH\right)\UH^\dagger\big)\,( \hat{a}_{2 D}\, h +
  \frac{4}{v^2}\,\hat{a}_{13}\, \Box h)+ \label{Lchibefore} \\
&\phantom{\supset} +
  \frac{i}{2}\,g'\,B^\mu\,\left\{\frac{v^2}{4}\text{Tr}\bigg(\left(\partial_\mu
    \UH\right)\,\tau_3\,\UH^\dagger-\UH\,\tau_3 \left(\partial_\mu
    \UH^\dagger\right)\bigg) + i\,v\left[ \hat a_{2D}\,\partial_\mu
    h + \frac{4}{v^2}\,\hat a_{13}\,\partial_\mu\left(\Box
    h\right)\right]\right\} \nn\\
&\phantom{\supset}+ \frac{i}{2}\,g\,W^i_\mu\left\{\frac{v^2}{4}\text{Tr}\bigg(\left(\partial^\mu
    \UH^\dagger\right)\tau^i\UH-\UH^\dagger\tau^i \left(\partial^\mu
    \UH\right)\bigg) - \frac{iv}{2}\left[\hat{a}_{2D}\,\partial_\mu h
    + \frac{4}{v^2}\, \hat{a}_{13}\,\partial_\mu\left(\Box
    h\right)\right]\text{Tr}\left(\TL\,\tau^i\right)\right\}\,,\nn
\end{align}
where for simplicity the definitions
\begin{equation}
\hat{a}_i \equiv c_i a_i
\label{hata}
\end{equation}
have been implemented, with $c_i$ being the operator coefficients in Eq.~(\ref{LCP4}) and $a_i$ the coefficients of the terms linear
in the Higgs field in Eq.~(\ref{F}).

In what concerns the Lagrangian two-point functions, the dependence on $\hat{a}_{2D}$ and $\hat{a}_{13 }$ in
Eq.~(\ref{Lchibefore}) can be reabsorbed via a phase redefinition of the Goldstone boson $\UH$ matrix in Eq.~(\ref{LLO}) of the form 
\begin{align}
\UH&=\tilde{\UH}\, \exp{\left[-\frac{i}{v}\left(\hat{a}_{2D}\,h +  4\,\hat{a}_{13}\,\frac{\Box h}{v^2}\right)\tau_3 \right]}\,,
\label{Uredefinition}\,
\end{align}
at first order in the $\hat{a_i}$ coefficients.  
This redefinition  is a
non-linear version of the simple Higgs-field redefinition proposed in
Ref.~\cite{Georgi:1986df} when analysing the effective linear axion Lagrangian.
 $\tilde{\UH}$ is then  the resulting physical matrix of the Goldstone bosons eaten by the $W$
and $Z$ bosons, to be identified with the identity in the unitary
gauge.  The gauge-fixing terms can now be written in the standard form,
\beq
\begin{aligned}
\LL^{\text{GF}}_{B}&=-\frac{1}{4\,\eta}\,\text{Tr}\left(
\left[\partial_\mu
 B^\mu - \frac{i}{4}\,\eta\,g'\,v^2\left(\widetilde{\UH}\tau_3-\tau_3
 \widetilde{\UH}^\dagger\right)\right]^2\right)  \\
\LL^{\text{GF}}_{W}&=-\frac{1}{\eta}\,\text{Tr}\left(\left[\partial_\mu
 W^\mu +
 \frac{i}{8}\,\eta\,g\,v^2\left(\widetilde{\UH}
-\widetilde{\UH}^\dagger\right)\right]^2\right)\,,
\end{aligned}
\eeq
 removing all mixed gauge
boson-Goldstone bosons and gauge boson-$h$ two-point couplings. 

After the redefinition in Eq.~(\ref{Uredefinition})
and at first order on the operator coefficients,  
 the SM Lagrangian Eq.~(\ref{LLO}) gets physical corrections  
  given by 
 \begin{align}
 \Delta\LL_{\text{Yuk}}\, + \, \Delta\LL_{\text{Bos}}\, ,
 \end{align}
 with
\begin{align}
\Delta \LL_{\text{Yuk}} &= \frac{i}{v}\left(\hat{a}_{2 D}\,h +
4\,\hat{a}_{13}\,\frac{\Box h}{v^2}
\right)\,\frac{(v+h)}{\sqrt{2}}\left(\bar{Q}_L\, \widetilde\UH\,\cY_Q
\,\tau_3\,Q_R-\mbox{h.c.}\right)\,+\,\left[Q_{L,R} \Longrightarrow L_{L,R}\right]\,,
\label{eq:cpoddyuk}
\end{align}
and
\beq
\begin{aligned}
\Delta \LL_{\text{Bos}}=
& -i\left(1+\frac{h}{v}\right)\partial_\mu h \text{Tr}
\left(\TL\left(\partial^\mu \widetilde\UH\right) \widetilde \UH^\dagger\right)
\left(\hat a_{2D}\, h +4\,\hat a_{13} \frac{\square h}{v^2}\right)
\\
& -i\,\text{Tr}\left(\TL\,\left(\partial_\mu\partial^\mu
  \widetilde\UH\right)\widetilde\UH^\dagger\right)\,
\left[\left(\hat{a}_{2 D}-\dfrac{\hat{b}_{2D}}{4}\right)h^2
+4\left(\hat{a}_{13}-\dfrac{\hat{b}_{13}}{2}\right)
\dfrac{h\,\square h}{v^2} \right. \\
&\hspace{4.8cm}\left. -2\hat{b}_{13}\frac{\partial_\nu h\partial^\nu h}{v^2}
+\,\dfrac{h^2}{2v}
\left(\hat{a}_{2 D}\,h+4\,\hat{a}_{13}\dfrac{\square h}{v^2}
\right) \right]\\
&-\bigg[g\,\text{Tr}\left(\TL W^\mu\right)-g'B^\mu\bigg]
\left[\left(\hat{a}_{2 D} -\dfrac{\hat{b}_{2D}}{2}\right)h \,\derp_\mu h
+4\left(\hat{a}_{13}-
\dfrac{\hat{b}_{13}}
{2}\right)
\dfrac{h\,\partial^\mu\square h}{v^2}  \right. \\
&\hspace{4.6cm}
 -2\hat{b}_{13}\left(\frac{\square h\,\partial_\mu h 
}{v^2}+2
\frac{\partial_\nu h\,\partial_\mu\partial^\nu h}{v^2}\right)
\\
& \hspace{4.6cm}\left.
+\dfrac{h^2}{2v}
\left(\hat{a}_{2 D}\,\derp_\mu h
+4\,\hat{a}_{13}\dfrac{\partial_\mu\square h}{v^2}\right)
\right]
\end{aligned}
\eeq
where $\hat b_i\equiv c_i b_i$. The ``tilde" over $\widetilde\UH$ will be dropped from now on.

Anomalous $qqh$, $\ell\ell h$ and $Zhh$ vertices follow; the
corresponding Feynman rules can be found in Appendix~\ref{AppFR}. It
is worth to remark that if a generic $\cF_i(h)$ function is considered
also for the Yukawa terms instead of the SM-like dependence in
Eq.~(\ref{LLO}), further quartic $qqhh$ and $\ell\ell hh$ anomalous
vertices will be revealed in addition to those shown in
Eq.~(\ref{eq:cpoddyuk}); we postpone the analysis of these two-Higgs 
exotic interactions to a future publication.

In addition to the tree-level impact discussed,
$\cS_{2D}(h)$ and $\cS_{13}(h)$ induce one-loop corrections to the Higgs gauge-boson
couplings, see Sec.~\ref{CP-HVV}, which in turn can be bounded from the strong experimental
limits on fermionic EDMs, see Eq.~(\ref{moreEDMbounds}).


\subsection{Triple gauge boson couplings}
\label{subsec:tgv}

The operators in Eq.~(\ref{CPbasis}) induce tree-level modifications of 
the self-couplings of the electroweak gauge bosons as well as of the Higgs-gauge
boson vertices involving three or more particles: their impact on the Feynman
rules of the theory are given in Appendix~\ref{AppFR}. 

We first focus on the CP-violating triple gauge boson
couplings $W^+ W^-\gamma $ and $W^+W^-Z$, originated from the operators in Eq.~(\ref{CPbasis}). Following
Ref.~\cite{Hagiwara:1986vm}, the CP-odd sector of the Lagrangian that
describes triple gauge boson vertices (TGVs) can be parametrised as:
\begin{eqnarray} 
{\mathcal L}^{WWV}_{\text{eff, \cancel{CP}}} = g_{WWV} \Bigg( &
g_4^V W^\dagger_\mu W_\nu (\partial^\mu V^\nu + \partial^\nu V^\mu)
 - i \tilde{\kappa}_V W^\dagger_\mu W_\nu \tilde{V}^{\mu\nu}
- i \frac{\tilde{\lambda}_V}{M_W^2} W^\dagger_{\sigma\mu} W^\mu_\nu 
\tilde{V}^{\nu\sigma}\nonumber\\
&+  \tilde{g}_{6}^V ( W^\dagger_\nu\partial_\mu W^\mu + W_\nu \partial_\mu W^{\dagger\mu} )V^\nu
+  \tilde{g}_{7}^V W^\dagger_\mu W^\mu \partial^\nu V_\nu
\Bigg)\,, 
\label{eq:wwv}
\end{eqnarray}
where $V \equiv \{\gamma, Z\}$ and $g_{WW\gamma} \equiv e=g
\sin\theta_W$, $g_{WWZ} = g \cos\theta_W$. 
In this equation $W^\pm_{\mu\nu}$ and
$V_{\mu\nu}$ stand exclusively for the kinetic part of the corresponding gauge field
strengths, and the dual tensor $\tilde{V}_{\mu\nu}$ has been defined in Sect.~\ref{Sect:Basis}.
In writing Eq.~(\ref{eq:wwv}) we have introduced the
coefficients $\tilde g_{6}^V$ and $\tilde g_7^V$ associated to operators
that contain the
contraction $\DLL_\mu\VL^\mu$; its $\partial_\mu\VL^\mu$ part vanishes
only for on-shell gauge bosons; in all generality $\DLL_\mu\VL^\mu$
insertions could only be disregarded in
the present context when fermion masses are neglected.
In the SM all couplings in Eq.~(\ref{eq:wwv}) vanish.

Electromagnetic gauge invariance requires $g_{4}^{\gamma} =0$, while
the CP-odd bosonic operators in Eq.~(\ref{CPbasis}) give the following contributions
to the phenomenological coefficients in Eq.~(\ref{eq:wwv}): 
\beq
\begin{aligned}
 \tilde{\kappa}_\gamma &=-\frac{4e^2}{s_\theta^2}\left( c_1+ 2c_8\right)\, , \qquad\qquad
&&\tilde{\kappa}_Z =\frac{4e^2}{c_\theta^2}
\left( c_1- 2\frac{c_\theta^2}{s_\theta^2}c_8\right) \, ,  \\
g_4^Z &=\frac{e^2}{2c_\theta^2s_\theta^2} c_4 \, , \qquad\qquad
&&\tilde{g}_{6}^Z =\frac{e^2}{2c_\theta^2s_\theta^2}\left( c_4 + c_{10} \right)\, ,\\
\tilde{g}_{7}^Z &=-\frac{e^2}{2c_\theta^2s_\theta^2} \,\left( c_4 - 2c_{11} \right),\qquad\qquad
&&\tilde{g}_{6}^\gamma  =\tilde{g}_{7}^\gamma=\tilde \lambda_\gamma=\tilde\lambda_Z=0 \,.
\label{eq:coeftgv}
 \end{aligned}
 \eeq

For completeness, note that there is an additional contribution  
to the $ZZZ$ vertex of the form:
\begin{equation} 
{\mathcal L}^{3Z}_{\text{eff, \cancel{CP}}}\, = 
\tilde g_{3Z} \, Z_\mu Z^\mu \partial_\nu Z^\nu \, , 
\label{eq:zzz}
\end{equation}
with 
\begin{equation}
\tilde g_{3Z}=\frac{e^3}{2c_\theta^3s_\theta^3} 
\left( c_{10} + c_{11} + 2c_{16} \right) \, ,
\label{eq:g3z}
\end{equation}
which, alike to the phenomenological couplings 
$\tilde g_{6}^V$ and $\tilde g_7^V$ in Eq.~(\ref{eq:wwv}),
vanishes for on-shell $Z$ bosons and in general can be disregarded in
the present context when the masses of fermions coupling to the $Z$
are neglected.

It is interesting to compare the expected signals from the chiral Lagrangian presented here and the $d=6$ linear realization. At the level of TGVs, there are six independent four-derivatives chiral CP-odd operators contributing to Eqs.~(\ref{eq:coeftgv}) and (\ref{eq:zzz}), while only two are present in the set of $d=6$ linear ones. Furthermore the nature of the phenomenological couplings involved is different, as some of the former correspond to $d=8$ linear operators, while one of the latter set is a six derivatives. More explicitly, the two linear CP-odd operators at $d=6$ are~\cite{Buchmuller:1985jz,Grzadkowski:2010es}
\beq
{\cal O}_{\widetilde WB}=g\,g'\,\epsilon^{\mu\nu\rho\sigma}\,B_{\mu\nu}\,W_{\rho\sigma}^j\,\left(\Phi^{\dagger}\,\tau_j\,\Phi\right)\,, \qquad\qquad 
{\cal O}_{\widetilde W WW}=i\epsilon_{ijk}\tilde {W^i}^\nu_\mu {W^j}^\lambda_\nu {W^k}_\lambda^\mu\,,
\eeq
where the first one is the sibling of $\cS_{1}(h)$ (see Appendix \ref{App:Siblings}) while ${\cal O}_{\tilde W WW}$
does not have an equivalent operator in the chiral expansion up to four derivatives. Thus in this case the effective couplings in Eq.~(\ref{eq:wwv}) verify:
\beq
\begin{gathered}
\tilde\kappa^{(lin,d=6)}_\gamma= -\frac{c_\theta^2}{s_\theta^2} \tilde\kappa^{(lin,d=6)}_Z\,,\qquad\qquad\qquad
\tilde\lambda^{(lin,d=6)}_\gamma=\tilde\lambda^{(lin,d=6)}_Z\,,  \\
g_4^{Z,(lin,d=6)}=\tilde{g}_{6}^{Z,(lin,d=6)} =\tilde{g}_{7}^{Z,(lin,d=6)} =
\tilde{g}_{6}^{\gamma,(lin,d=6)}  =\tilde{g}_{7}^{\gamma,(lin,d=6)}=0 \,.
\end{gathered}
\label{eq:tgvlin}
\eeq

Hence, generically, if $g_4^Z$ is found larger than $\tilde\lambda_\gamma$ or $\tilde\lambda_Z$, it would point out towards a chiral realization of the EWSB, while the contrary would signal towards the linear realization. Furthermore if  non-zero $\tilde\kappa_\gamma$  and $\tilde\kappa_Z$ are observed, the relation in Eq.~(\ref{eq:tgvlin}) could be also tested.

The strongest constraints on CP violation in the $W^+ W^-\gamma$ vertex arise from its contributions to fermionic EDMs that they can induce at one-loop, while constraints on CP-violating $W^+ W^-Z $ couplings can be obtained from the study of gauge-boson production at colliders. We further elaborate below in these two types of signals.


\boldmath 
\subsubsection{CP violation in $WW\gamma$: fermionic EDMs}
\label{subsubsec:edm}
\unboldmath

Electric dipole moments for quarks and leptons are generically
the best windows on BSM sources of CP-violation, due to the
combination of the very stringent experimental bounds with the fact
that they tend to be almost free from SM background contributions:  fermionic EDMs are suppressed in
the SM beyond two electroweak boson exchange,
while in most BSM theories they are induced at one-loop level.
 
\begin{figure}[h!]
\begin{center}
\includegraphics[scale=0.6]{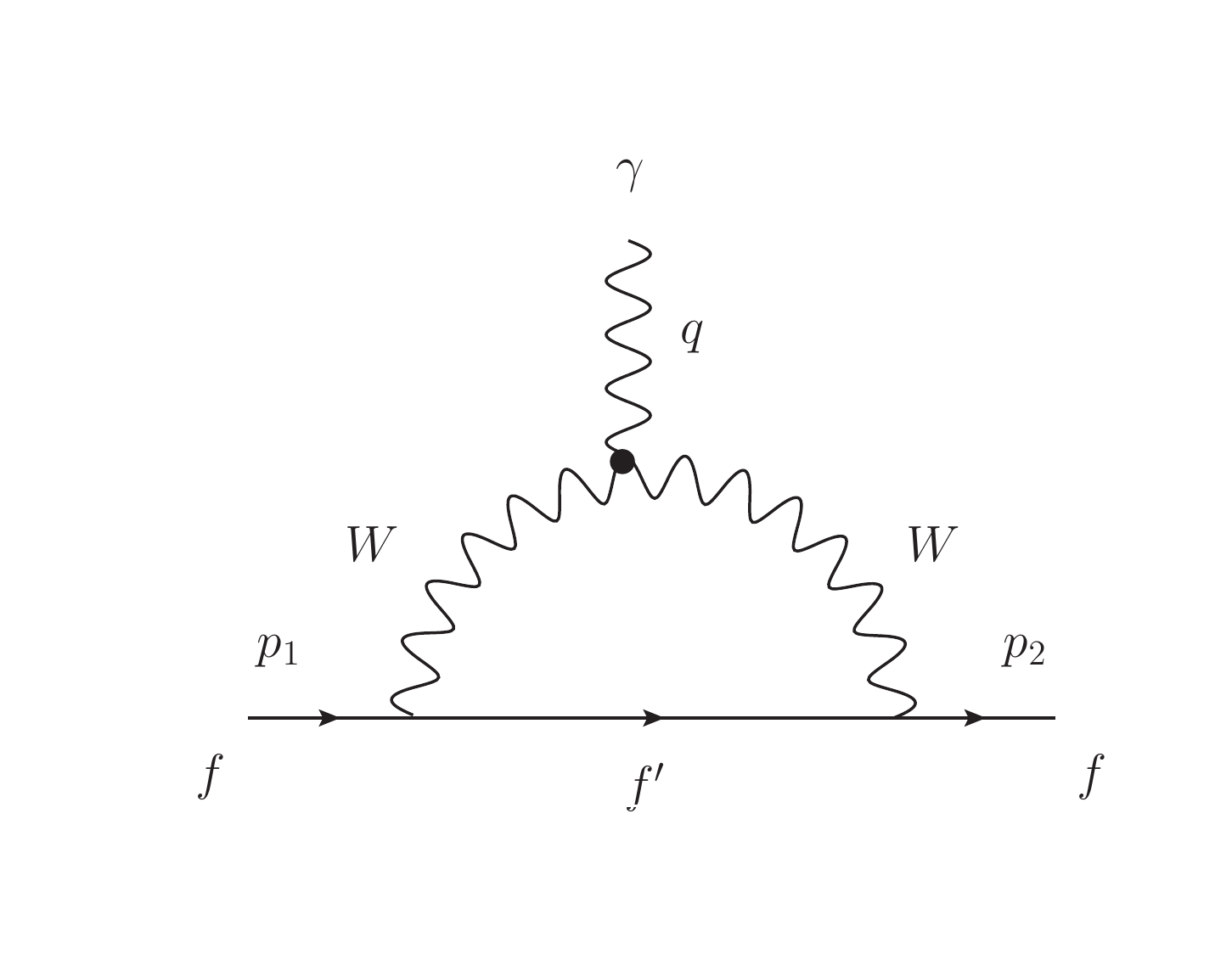}
\caption{\emph{A CP-odd TGV coupling inducing a fermionic EDM interaction.}}
\label{fig:TGV}
\end{center}
\end{figure}

Although none of the operators in the chiral basis above --
Eq.~(\ref{CPbasis}) -- induces tree-level contributions to EDMs, two of
them, $\cS_1(h)$ and $\cS_8(h)$, contain gauge boson couplings
involving the photon, of the form
\be
 + \frac{i}{2} \epsilon_{\mu\nu\rho\sigma }W^+_{\mu } W^-_{\nu } A^{\rho\sigma} \,,
\ee
where $ A^{\rho\sigma}$ denotes the photon field strength, see Eqs.(\ref{eq:wwv})
and (\ref{eq:coeftgv}) and 
Appendix~\ref{AppFR}. This coupling induces in turn a one-loop contribution
to fermion EDMs, see Fig.~\ref{fig:TGV}.

The amplitude corresponding to this Feynman diagram can be parametrised as
\begin{align}
\mathcal{A}_{f}&\equiv -i\,d_f\,\overline{u}\left(p_2\right)\,\sigma_{\mu \nu}q^{\nu }\gamma^5 \,u\left(p_1\right)\,,
\end{align}
where $d_f$ denotes the fermionic EDM strength. The corresponding
integral diverges logarithmically\footnote{For a specific UV model which does not lead to logarithmic diverging EDM see \cite{Appelquist:2004mn}.}; assuming a physical cut-off
$\Lambda_s$ for the high energy BSM theory and following the generic
computation in Ref.~\cite{Marciano:1986eh}, we obtain for the
contribution from $\cS_1(h)$ and $\cS_8(h)$:
\be
d_f=\left(c_1 + 2\,c_8\right)\frac{e^3\,G_F\,T_{3 L}\,\csc ^2\theta
  _W}{ \sqrt{2}\,\pi ^2}\,\,m_f\,\left[\log \left(\frac{\Lambda_s
    ^2}{M_W^2}\right)+ \mathcal{O}(1)\right]\,,
\ee
where $T_{3L}$ stands for the fermion weak isospin, $\theta_W$ denotes
the Weinberg angle and $G_F$ the Fermi coupling constant.
The present experimental bound on the electron  EDM~\cite{Baron:2013eja}, 
\be
\left|\frac{d_e}{e}\right|<8.7\times 10^{-29}\,\text{cm}\,,\qquad \text{at 90\% CL}\,,
\ee
implies then a limit 
\be
\left|\left(c_1 + 2\,c_8\right)\left[\log \left(\frac{\Lambda_s
    ^2}{M_W^2}\right) + \mathcal{O}(1)\right]\right|<5.2\times 10^{-5}\,.
    \label{c1c8eEDM}
\ee
Using as values for the constituent quark masses $m_u=m_d=m_N/3$, the
experimental limit on the neutron EDM~\cite{Baker:2006ts},
\be
\left|\frac{d_n}{e}\right|<2.9\times 10^{-26}\,\text{cm}\,,\qquad
\text{at 90\% CL}\,,
\label{eq:eedm}
\ee
allows to set an even stronger limit on the combination of $\cS_1(h)$
and $\cS_8(h)$ operator coefficients:
\be
\left|\left(c_1 + 2\,c_8\right)\left[\log
  \left(\frac{\Lambda_s^2}{M_W^2}\right) + \mathcal{O}(1)
  \right]\right|<2.8\times 10^{-5}\,.  
 \label{c1c8nEDM}
\ee

Weaker but more direct bounds on these operators can be imposed
from the study of $W\gamma$ production at colliders. For example
the recent study in Ref.~\cite{Dawson:2013owa} concluded that the future 14 TeV LHC data
with 10 fb$^{-1}$ can place  a 95\% CL bound
\begin{equation}
|\tilde\kappa_\gamma|\leq 0.05\quad \Longrightarrow \quad |c_1+2 c_8|\leq 0.03 \, .
\label{eq:c1c8edm}
\end{equation}


\boldmath
\subsubsection{CP violation in $WWZ$: Collider bounds and signatures}
\label{WWZcollider}
\unboldmath

At present the strongest direct constraints on CP-violating effects
in the $WWZ$ vertex are imposed by the combination of results using
the LEP collaboration studies on the observation of the angular
distribution of $W's$ and their decay products in $WW$ production at
LEPII~\cite{Abbiendi:2000ei,Abdallah:2008sf,Schael:2004tq}. The combination
yields the following 1$\sigma$ (68\% CL) constraints~\cite{PDG}
\beq
-0.47 \leq g_4^Z \leq -0.13\,,\qquad 
-0.14\leq \tilde{\kappa}_Z \leq -0.06 \,,\qquad
-0.16\leq \tilde{\lambda}_Z\leq -0.02\,,
\label{eq:wwzlep}
\eeq
which in terms of the coefficients of operators in Eq.~(\ref{CPbasis})
implies
\beq
-1.8 \leq c_4 \leq -0.50 \,,\qquad\qquad
-0.29\leq \left(c_1 - 2\frac{c_\theta^2}{s_\theta^2}c_8\right) \leq -0.13\,.
\label{eq:wwzlep2}
\eeq
Note that the bounds in Eq.~(\ref{eq:wwzlep}) are obtained assuming one
effective coupling in Eq.~(\ref{eq:wwv}) being different from zero at a time,
which is consistent with the predictions
from the dynamical Higgs Lagrangian, Eq.~(\ref{eq:coeftgv}), since different
operators lead to independent modifications of the effective couplings
$g^Z_4$ and $\tilde \kappa_Z$.

In what concerns Tevatron and LHC data, anomalous CP-odd TGV interactions 
have not been studied in detail yet. To fill this gap we present
in what follows our analysis of the LHC potential to measure deviations
or set exclusion bounds on CP-odd $WWZ$ anomalous TGVs, extending
our preliminary study~\cite{Eboli:2010qd}. At LEP the experimental
analyses which lead to the bounds in Eq.~(\ref{eq:wwzlep})
were based on the study of the angular
distributions of the final state particles in the event.
In contrast, at the LHC, the higher collision energy -- well above the 
$WW$ and $WZ$ thresholds --  makes the use of kinematic variables related 
to the energy of the event more suitable for the measurement of TGV.

The study in Ref.~\cite{Eboli:2010qd} concluded that the $pp\rightarrow W^\pm Z$
process has higher potential to observe $g_4^Z$ than  the $pp\rightarrow
W^⁺W^⁻$ channel, while both channels have a similar power to study
$\tilde{\kappa}_Z$ and $\tilde{\lambda}_Z$. Furthermore, it was also
discussed the use of several kinematic distributions to characterize the
presence of a non-vanishing CP-violating coupling
and the use of some asymmetries to characterize its CP nature.
So far the LHC has already
collected almost $25$ times more data than the luminosity considered
in that preliminary study which we update here. In addition, in this
update we take advantage of a more realistic background evaluation, by
using the results of the experimental LHC analysis on other anomalous
TGV interactions~\cite{Aad:2012twa}\footnote{This strategy was also 
the starting point for the study of the CP conserving, but C and P
violating coupling $g_5^Z$ presented in Ref.~\cite{Brivio:2013pma}.}.

In this section we study the process
\begin{equation}
 pp\rightarrow \ell^{\prime\pm}\ell^+\ell^-E_T^{{\rm miss}}\,,
\label{eq:3lmiss}
\end{equation}
where $\ell^{(\prime)}=e$ or $\mu$. The main background for the
detection of anomalous TGV interactions is the irreducible SM production 
of $W^\pm Z$ pairs. In addition
there are further reducible backgrounds like $W$ or $Z$ production
with jets, $ZZ$ production followed by the leptonic decay of the $Z$'s
with one charged lepton escaping detection, and $t\bar{t}$ pair
production.

We simulate the signal and the SM irreducible background using an
implementation of the anomalous vertices $g_4^Z$, $\tilde\kappa_Z$, and
$\tilde \lambda_Z$ in
FeynRules~\cite{Christensen:2008py} interfaced with MadGraph
5~\cite{Alwall:2011uj} for event generation. We account for the
different detection efficiencies by rescaling our simulation of
the SM production of $W^\pm Z$ pairs to the values quoted
by ATLAS~\cite{Aad:2012twa} for the study of $\Delta
\kappa_Z$, $g_1^Z$ and $\lambda_Z$. However, we have also cross-checked the
results using a setup where the signal simulation is based on the same
FeynRules~\cite{Christensen:2008py} and MadGraph5~\cite{Alwall:2011uj}
implementation, interfaced then with PYTHIA~\cite{Sjostrand:2006za}
for parton shower and hadronization, and with PGS 4~\cite{pgs} for
detector simulation. Finally, the reducible backgrounds for the 7 TeV
data analysis are obtained from the simulations presented in the ATLAS
search~\cite{Aad:2012twa}, and they are properly rescaled for the 8
and 14 TeV runs.
 
In order to make our simulations more realistic, we closely follow the
TGV analysis performed by ATLAS~\cite{Aad:2012twa}. The
kinematic study of the $W^\pm Z$ production starts with the usual detection
and isolation cuts on the final state leptons. Muons and electrons are considered
if their transverse momentum with respect to
the collision axis $z$, $p_T\equiv\sqrt{p_x^2+p_y^2}$, and their pseudorapidity 
$\eta\equiv\frac{1}{2}\ln\frac{|\vec p| +p_z}{|\vec p| -p_z}$, satisfy
\beq
\begin{gathered}
 p_T^\ell>15\GeV\,,\qquad  \quad
 |\eta^\mu|<2.5\,,\\  
 |\eta^e|<1.37\qquad \text{or}\qquad 1.52< |\eta^e|<2.47\, .
\label{eq:cut0}
\end{gathered}
\eeq
To guarantee the isolation of muons (electrons), we require that the
scalar sum of the $p_T$ of the particles within
$\Delta R\equiv\sqrt{\Delta\eta^2+\Delta\phi^2}=0.3$ of the
muon (electron), excluding the muon (electron) track, is smaller than
15$\%$ (13\%) of the charged lepton $p_T$. In the cases when the final
state contains both muons and electrons, a further isolation
requirement has been imposed:
\begin{eqnarray}
 \Delta R_{e\mu} > 0.1\,.
\label{eq:cut1}
\end{eqnarray}

It is also required that at least two leptons with the same flavour and
opposite charge are present in the event and that their invariant mass
is compatible with the $Z$ mass, \ie
\beq
 M_{\ell^+\ell^-}\in \left[M_Z-10,\ \ M_Z+10\right]\GeV\,.
\label{eq:cut2}
\eeq
In what follows we refer to $p^Z$ as the momentum of this 
$\ell^+\ell^-$ pair, $p^Z\equiv p^{\ell^+}+p^{\ell^-}$.  
We further impose that a third lepton is present which
passes the above detection requirements and whose transverse momentum
satisfies in addition
\beq
 p_T^{\ell'}>20\GeV\,.
\label{eq:cut3}
\eeq
Moreover, with the purpose of suppressing most of the
$Z+\rm{jets}$ and other diboson production backgrounds, we require
\beq
E_T^{{\rm miss}}>25\GeV
\qquad \text{and} \qquad 
M_T^W>20\GeV\,,
\label{eq:cut4}
\eeq
where $ E_T^{{\rm miss}}$ is the missing transverse energy and the
transverse mass is
\linebreak
\mbox{$M_T^W=\sqrt{2p_T^\ell E_T^{{\rm miss}}\left(1-\cos(\Delta\phi)\right)}$},
with $p_T^{\ell'}$ being the transverse momentum of the third lepton, and
$\Delta\phi$ the azimuthal angle between the missing
transverse momentum and the third lepton. Finally, it is required that
at least one electron or one muon has a transverse momentum complying
with
\beq
p_T^{e(\mu)}>25\ \  (20)\GeV\,.
\eeq

Our Monte Carlo simulations have been tuned to the ATLAS
ones~\cite{Aad:2012twa}, so as to incorporate more realistic detection
efficiencies. Initially, a global $k$-factor is introduced to account
for the higher order corrections to the process in Eq.~(\ref{eq:3lmiss}) by
comparing our leading order predictions to the NLO ones used in the
ATLAS search~\cite{Aad:2012twa}, leading to $k\sim 1.7$. Next, we
compare our results after cuts with the ones quoted by ATLAS in Table
1 of Ref.~\cite{Aad:2012twa}. We tune our simulation by applying a
correction factor per flavour channel ($eee$, $ee\mu$, $e\mu\mu$ and
$\mu\mu\mu$) that is almost equivalent to introducing a detection efficiency of
$\epsilon^e=0.8\ (0.95)$ for electrons (muons).  
These efficiencies have been employed in our simulations for signal and
backgrounds.

After the selection procedure,
in the presence of anomalous TGVs the cross section for the process
$pp\rightarrow \ell^{\prime\pm}\ell^+\ell^-E_T^{{\rm miss}}$ can be
qualitatively described by:
\beq
 \sigma\ =\  \sigma_{{\rm bck}}\ + \sigma_{SM}\ +
 \sum_{i,j\geqslant i}\sigma_{{\rm ano}}^{ij}g_{{\rm ano}}^ig_{{\rm ano}}^j\,.
\eeq
Here $\sigma_{SM}$ corresponds to the irreducible SM $W^\pm Z$ background,
while $\sigma_{{\rm bck}}$ stands for all background
sources except for the SM EW $W^\pm Z$ production. Additionally
$\sigma_{{\rm ano}}^{ij}$ are the pure anomalous contributions. Notice 
that because of the CP-violating nature of the anomalous couplings there
is no interference between those and the SM contributing to the total cross
section. Furthermore
in the present study we assume only one coupling departing from its 
SM value at a time (\ie\ always $i=j$) which, as mentioned above,
is consistent with the expectations from the dynamical Higgs 
effective operators, Eq.~(\ref{eq:coeftgv}), since they lead to
independent modifications of the two relevant effective couplings 
$g^Z_4$ and $\tilde \kappa_Z$.
We present in
Table~\ref{tab:cs} the values of $\sigma_{SM}$, $\sigma_{{\rm bck}}$
and $\sigma_{{\rm ano}}$ for center--of--mass energies of 7, 8 and 14
TeV\footnote{For completeness we make our study for the most general
CP-violating WWZ vertex in Eq.~(\ref{eq:wwv}) and evaluate the sensitivity
to $\tilde{\lambda}^Z$ as well, even though
this coupling is generated at higher order in the chiral expansion
as shown in Eq.~(\ref{eq:coeftgv}).}.
\begin{table}[htb!]
\centering
\begin{tabular}{|c|c|c|c|c|c|}
\hline
 COM Energy  & $\sigma_{SM}$ (fb) & $\sigma_{{\rm bck}}$ (fb) & $\sigma_{{\rm ano}}^{g_4^z}$ (fb)
 & $\sigma_{ {\rm ano}}^{\tilde{\kappa}_z}$ & $\sigma_{{\rm ano}}^{\tilde{\lambda}_z}$
\\
\hline
7 TeV & 47.7 & 14.3 & 846 & 56.0 & 1914
\\
\hline
8 TeV & 55.3 & 16.8 & 1117 & 67.7 & 2556 
\\
\hline
14 TeV & 97.0 & 29.0 & 3034 & 134 & 7471
\\
\hline
\end{tabular}
\caption{\em Values of the cross section predictions for the process
$pp\rightarrow \ell^{\prime\pm}\ell^+\ell^-E_T^{{\rm miss}}$ after
applying all the cuts described in the text.
$\sigma_{SM}$ is the SM contribution coming
from EW $W^\pm Z$ production, $\sigma_{{\rm ano}}^{i}$ are the pure anomalous
contributions and $\sigma_{{\rm bck}}$ corresponds to all the background
sources except for the electroweak SM $W^\pm Z$ production.}
\label{tab:cs}
\end{table}

\begin{figure}[htb!]
  \centering
  \includegraphics[width=0.48\textwidth]{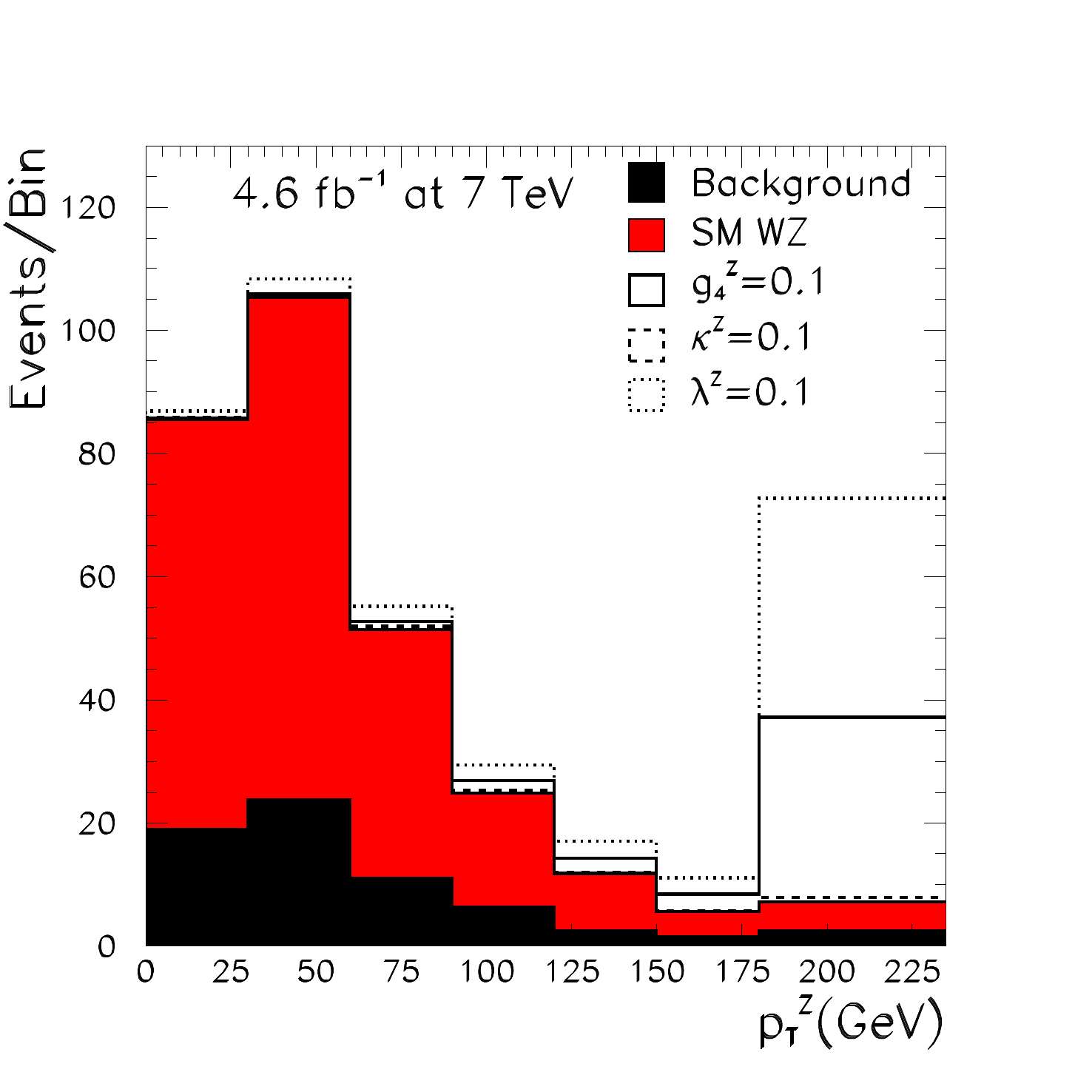}
  \includegraphics[width=0.48\textwidth]{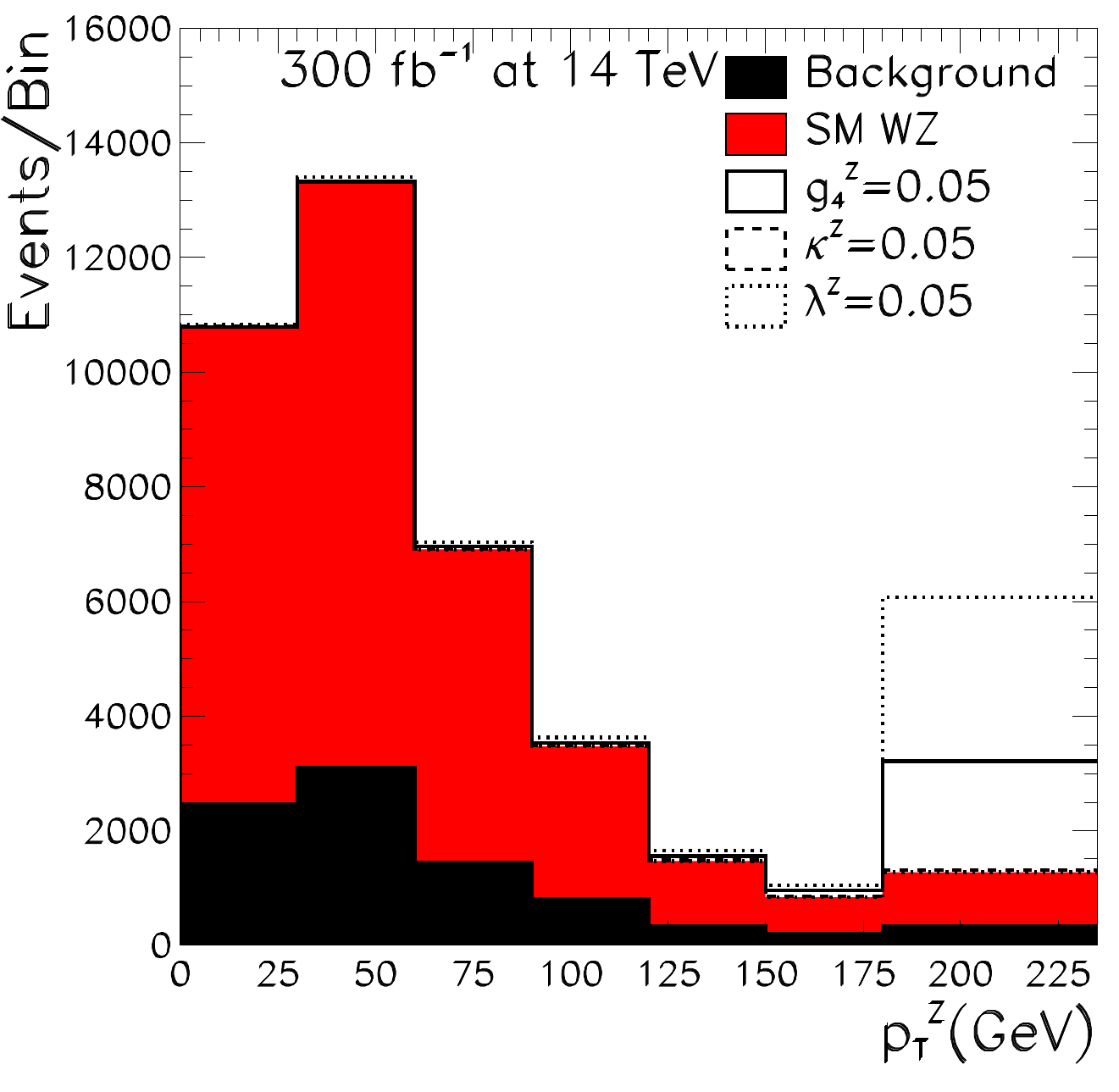}
 \caption{\em In the left (right) panel we show the distribution of events
with respect to $p_T^Z$ for the 7 (14) TeV run
assuming ${\cal L}=4.64$ ($300$) fb$^{-1}$ of integrated luminosity.
The black histogram contains all background sources, except for
the SM $W^\pm Z$ production, the red histogram represents the sum of all
the backgrounds and finally the solid (dashed) [dotted] distribution
corresponds to the addition of the contribution of an anomalous TGV
with a value $g_4^Z=0.1$ ($\tilde{\kappa}_Z=0.1$) [$\tilde{\lambda}_Z=0.1$]
for the 7 TeV run and $g_4^Z=0.05$ ($\tilde{\kappa}_Z=0.05$)
[$\tilde{\lambda}_Z=0.05$] for the 14 TeV run.
The last bin contains all the events with $p_T^Z>180$ GeV.}
\label{fig:wwz}
\end{figure}

In order to quantify the reachable sensitivity on the determination of
the different anomalous TGVs,
advantage has been taken in this analysis 
of the fact that anomalous TGVs enhance the cross sections at high
energies. Ref.~\cite{Eboli:2010qd} shows that the variables
$M_{WZ}^{{\rm rec}}$ (the reconstructed $WZ$ invariant mass),
$p_T^{\ell\ {\rm max}}$ and $p_T^Z$ are able
to trace well this energy dependence, leading to similar sensitivities
to the anomalous TGVs. Here, we chose $p_T^Z$  because
this variable is strongly correlated with the subprocess
center--of--mass energy ($\hat{s}$), and, furthermore, it can be
directly reconstructed with good precision from the measured lepton
momenta.
In the left (right) panel of Fig.~\ref{fig:wwz} we show the number
of expected events with respect to the transverse momentum of the $Z$
candidate for the 7 (14) TeV run, assuming an integrated luminosity of
${\cal L}=4.64$ ($300$) fb$^{-1}$.  
The figure captures the enhancement of events at the higher
values of $p_T^Z$ that the presence of anomalous TGV interactions causes.
We can also observe how the effect of $\tilde{\kappa}_Z$ is
weaker than the effect of introducing $g_4^Z$ or
$\tilde{\lambda}_Z$.

We have followed two procedures to estimate the LHC potential to probe
anomalous CP-violating couplings. In a more conservative approach, 
we have performed a simple event counting analysis assuming that the number
of observed events corresponds to the SM prediction, and we look for the
values of the corresponding anomalous couplings which are inside
the $68\%$ and $95\%$ CL allowed
regions. In this case an additional cut $ p_T^Z>90\GeV$ 
was applied in the analysis to enhance the sensitivity~\cite{Eboli:2010qd}.
On a second analysis, a simple $\chi^2$ has been built based on the contents
of the different bins of the $p_T^Z$ distribution, with
the binning shown in Fig.~\ref{fig:wwz}.
Once again, it is assumed that the observed $p_T^Z$ spectrum corresponds
to the SM expectations and we seek the values of the corresponding
anomalous couplings that
are inside the $68\%$ and $95\%$ allowed regions. In general
the binned analysis yields $10\%-30\%$ better sensitivity.
The results of the binned analysis are presented in
Table~\ref{tab:results1}.

\begin{table}[htb!]
\centering
\begin{tabular}{|c|c|c|c|c|}
\hline
   & \multicolumn{2}{ c}{$68\%$ C.L. range} & \multicolumn{2}{|c|}{$95\%$ C.L. range}
\\
\hline
    & 
{7+8 TeV} & {7+8+14 TeV} &{7+8 TeV} & {7+8+14 TeV} 
 \\
\hline
$g_4^Z$ & $(-0.019,\ 0.019)$ & $(-0.007,\ 0.007)$  & $(-0.027,\ 0.027)$
& $(-0.010,\ 0.010)$
\\
\hline
$\tilde{\kappa}_Z$ & 
$(-0.12,\ 0.12)$ & $(-0.047,\ 0.047)$   & $(-0.17,\ 0.17)$
& $(-0.067,\ 0.067)$
\\
\hline
$\tilde{\lambda}_Z$ & $(-0.012,\ 0.012)$ & $(-0.004,\ 0.004)$ 
& $(-0.018,\ 0.018)$ & $(-0.006,\ 0.006)$
\\
\hline\hline
$c_4$ & $(-0.074,\ 0.074)$ & $(-0.027,\ 0.027)$  & $(-0.10,\ 0.10)$
& $(-0.039,\ 0.039)$
\\
\hline
$c_1 - 2\frac{c_\theta^2}{s_\theta^2}c_8$ &
$(-0.25,\ 0.25)$ & $(-0.099,\ 0.099)$   & $(-0.36,\ 0.36)$
& $(-0.14,\ 0.14)$\\ \hline
\end{tabular}
\caption{\em Expected sensitivity on $g_4^Z$, $\tilde{\kappa}_Z$ and
$\tilde{\lambda}_Z$ at the LHC, and the corresponding precision
reachable on the non-linear operator coefficients. We assume
${\cal L}=4.64$ fb$^{-1}$ for the 7 TeV run, ${\cal L}=19.6$
fb$^{-1}$ for the 8 TeV one and ${\cal L}=300$ fb$^{-1}$
for the future 14 TeV expectations.}
\label{tab:results1}
\end{table}

From Table~\ref{tab:results1} we read that the 7 and 8 TeV data sets
could clearly increase the existing limits on $g_4^Z$, and
consequently on $c_4$, and the future 14 TeV run would rapidly approach
the few per cent level. Conversely, as it was expected, the reachable
sensitivity on $\tilde{\kappa}_Z$ is weaker. Nevertheless, the future
14 TeV run has the potential to improve the direct bounds that LEP was
able to derive, and settle consequently the strongest direct available limits
on the corresponding combination of $c_1$ and $c_8$ couplings. Notice
that this combination is different from the $c_1$ and $c_8$
combination contributing to $\tilde{\kappa}_\gamma$, which is
bounded from EDM measurements, see Eqs.~(\ref{eq:coeftgv})
and~(\ref{eq:c1c8edm}). Thus, both measurements are complementary.

\begin{figure}[h!]
\centering
\includegraphics[width=0.5\textwidth]{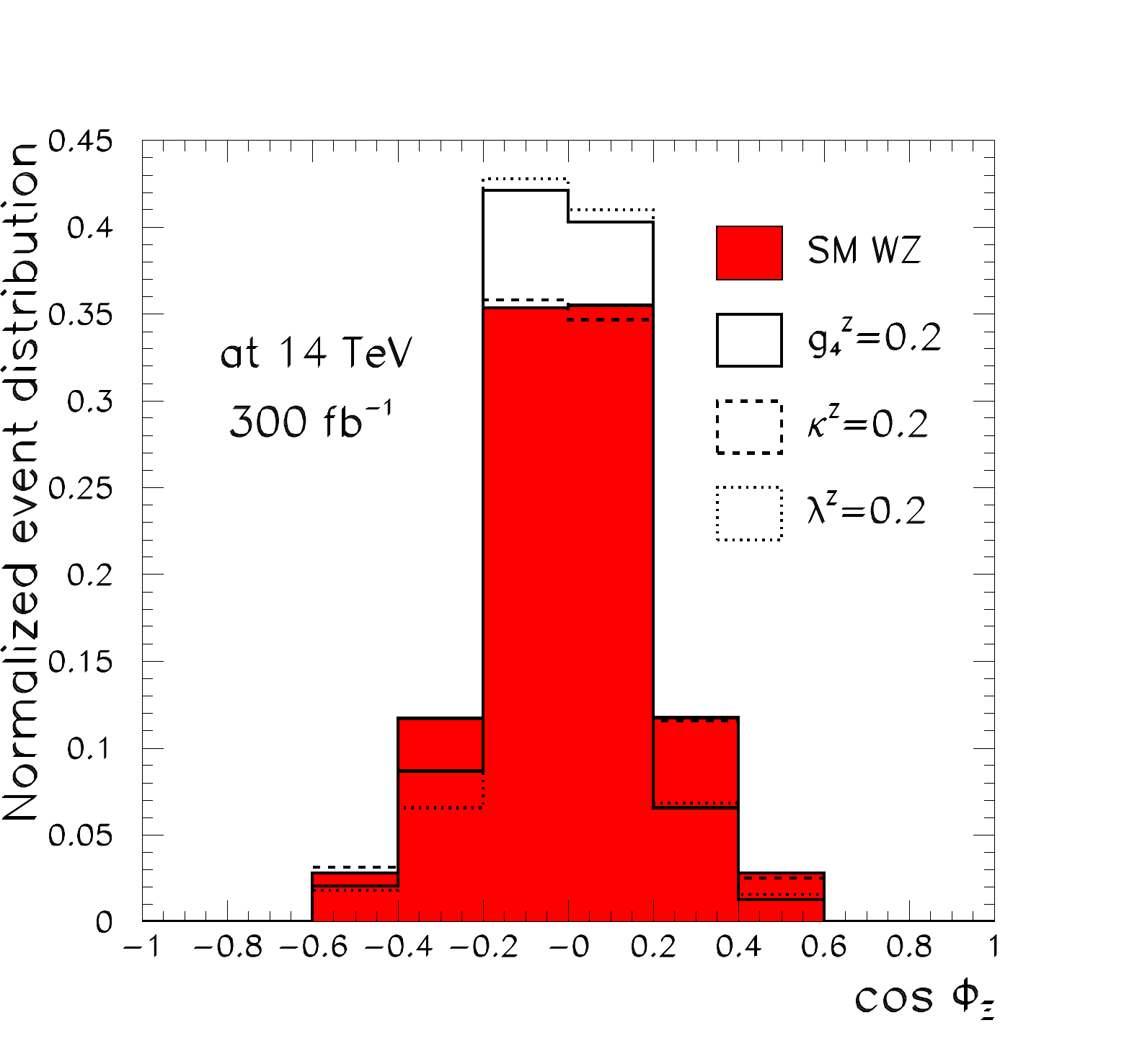}
\caption{\em Distribution of $pp\rightarrow \ell^{\prime\pm}\ell^+\ell^-E_T^{{\rm miss}}$
contributions with respect to $\cos\theta_\Xi$, after the cuts
described from Eqs.~(\ref{eq:cut0})--(\ref{eq:cut4}) are applied, considering
$300 \fb^{-1}$ of integrated luminosity collected at 14 TeV. The sign-symmetric
electroweak SM $W^\pm Z$  distribution  is shown as the red histogram and 
the distribution for the SM plus the contribution
of $g_4^Z=0.2$ ($\tilde{\kappa}_Z=0.2$) [$\tilde{\lambda}_Z=0.2$]
is shown as the solid (dashed) [dotted] line. All the distributions
are normalized to one for an easier comparison.}
\label{fig:cpsign}
\end{figure}

Up to this point the analysis that we have performed has not benefitted 
from the CP-odd nature of the TGV interactions.
Different studies~\cite{Dawson:1996ge,Kumar:2008ng,Han:2009ra,Eboli:2010qd}
have addressed the CP-odd nature of the anomalous TGVs by 
constructing some CP-odd or $\hat{T}$-odd observable.
In particular, in Ref.~\cite{Kumar:2008ng} it was shown that 
ideally in $pp\rightarrow W^\pm Z$ an asymmetric observable based
on the sign of the cross--product $p_q\cdot (p_Z\times p_{\ell^\prime})$
could be a direct probe of CP-violation, where here $p_q$ is the four-momentum
of the incoming quark. At the LHC, however, $p_q$ cannot be fully determined
and for this reason we build instead as a reconstructable correlated sign variable
\beq
\Xi_\pm \equiv\mbox{sign}(p^{\ell^\prime})_z\, \mbox{sign}
(p^{\ell^\prime}\times p^Z)_z\, ,
\label{ju}
\eeq
where $z$ is the collision axis. We define the  
sign-weighted cross section as 
\beq
\Delta \sigma \equiv \int d\sigma  \,\Xi_\pm \equiv   
\sum_i g^i_{\text{ano}} \, \Delta \sigma^{i}_{\text{ano}} \, .
\label{deltasigCP}
\eeq
A CP-odd TGV gives a measurable contribution to this sign-weighted
cross section which is linearly dependent on the coupling. 
On the contrary  the SM background is symmetric with respect to
$\Xi_\pm$ and it gives a null contribution to the sign-weighted
cross section in Eq.~(\ref{deltasigCP}). This behaviour is illustrated
in Fig.~\ref{fig:cpsign} where we show the distribution of events
at 14 TeV, assuming $300\fb^{-1}$ of integrated luminosity,
with respect to  the related variable
\beq	
\cos\theta_\Xi \equiv\cos\theta^{\ell^\prime}\cos\theta^{Z\times\ell^\prime}\; ,
\label{cos}
\eeq
where the angles are defined with respect to the $z$ axis. In this form
$\mbox{sign}(\cos\theta_\Xi)=\Xi_\pm$.

The corresponding sign-weighted cross sections at 14 TeV are
\beq
\Delta\sigma_{{\rm ano}}^{g^Z_4}=-59\fb\, , \qquad\qquad
\Delta\sigma_{{\rm ano}}^{\tilde\kappa_Z}=-9.7\fb\,,\qquad\qquad
\Delta\sigma_{{\rm ano}}^{\tilde\lambda_Z} =-137 \fb\,. 
\eeq
With a luminosity of 300 fb$^{-1}$ this CP-violation induced asymmetry
could be observed with 95\% CL above the
statistical fluctuations of the SM background for
\beq
|g^Z_4| \geq 0.02\, ,  \qquad\qquad  
|\tilde\kappa_Z|\geq 0.13  \, ,  \qquad\qquad
|\tilde \lambda_>|\geq 0.01 \, ,
\label{tired1}
\eeq
or what is equivalent for 
\beq
|c_4| \geq 0.08\,, \qquad\qquad  
\left|\left(c_1 - 2\frac{c_\theta^2}{s_\theta^2}c_8\right)\right|\geq 0.27\, .
\label{tired2}
\eeq


\subsection{CP violation in Higgs couplings to gauge-boson pairs} 
\label{CP-HVV}
The effective operators described in Eq.~(\ref{CPbasis})
also give rise to CP-odd interactions involving the Higgs particle and two gauge
bosons, to which we refer as HVV couplings. The CP-odd interactions
can be phenomenologically parametrized as
\beq
\begin{aligned}
{\cal L}_{{\rm eff, \cancel{CP}}}^{{\rm HVV}} =& 
\tilde{g}_{Hgg} \, h G^a_{\mu\nu} \tilde{G}^{a\mu\nu} +
\tilde{g}_{H \gamma \gamma}\,  h A_{\mu \nu} \tilde{A}^{\mu \nu} 
+ \tilde{g}_{H Z \gamma} \, h  A_{\mu \nu} \tilde{Z}^{\mu \nu}\\
&+ \tilde{g}^{(2)}_{H Z Z}  \, h Z_{\mu \nu} \tilde{Z}^{\mu \nu} +
\tilde{g}^{(2)}_{H W W}  \, h W^+_{\mu \nu} \tilde{W}^{- \, \mu \nu}\\
&+ \left[\tilde{g}^{(1)}_{H W W}  \left (W^+_{\mu \nu} W^{- \, \mu} \partial^{\nu} h \right) +\hc\right] 
+ \left[\tilde{g}^{(5)}_{H W W} \left(\partial_\mu W^{+\mu} W^-_\nu \partial^\nu h\right)+\hc\right] \,,
\end{aligned}
\label{eq:hvv}
\eeq
with tree level contributions
\beq
\begin{aligned}
\tilde{g}_{Hgg}&=-\frac{g_S^2}{v} \hat{a}_{\tilde{G}}\,,
\qquad\qquad\qquad\tilde{g}_{H \gamma \gamma}=\frac{4e^2}{v} \left(
-\frac{1}{4}\hat{a}_{\tilde{B}}+ \hat{a}_8 + \hat{a}_1 -
\frac{1}{8}\hat{a}_{\tilde{W}}\right)\,,  \\ 
\tilde{g}_{H Z \gamma}&=-\frac{8e^2s_\theta}{vc_\theta} \left(
-\frac{1}{4}\hat{a}_{\tilde{B}}-\frac{c_\theta^2}{2s_\theta^2}
(-\frac{1}{4}\hat{a}_{\tilde{W}}+2\hat{a}_8)+\frac{1}{8s_\theta^2}(
2\hat{a}_2+\hat{a}_3+2\hat{a}_9)-\frac{c_{2\theta}}{2s_\theta^2}\hat{a}_1\right) \,, \\ 
\tilde{g}_{H ZZ}^{(2)}&=\frac{4e^2s_\theta^2}{vc_\theta^2} \left(
-\frac{1}{4}\hat{a}_{\tilde{B}} +
\frac{c_\theta^4}{s_\theta^4}\hat{a}_8-\frac{c_\theta^2}{s_\theta^2}
\hat{a}_1 + \frac{1}{2s_\theta^2} \hat{a}_2 -
\frac{c_\theta^4}{8s_\theta^4}\hat{a}_{\tilde{W}}-\frac{c_\theta^2}{2s_\theta^4}
\hat{a}_9-\frac{c_\theta^2}{4s_\theta^4} \hat{a}_3 \right)\,,\\ 
\tilde{g}_{H WW}^{(2)}&=-\frac{2e^2}{vs_\theta^2} \left(\frac{1}{2}\hat{a}_{\tilde{W}}+\hat{a}_3\right)\,,
\qquad\quad\tilde{g}^{(1)}_{H WW}=\frac{2e^2}{vs_\theta^2} i \hat{a}_{7}\,, 
\qquad\quad\tilde{g}^{(5)}_{H WW}=-\frac{2e^2}{vs_\theta^2} i \hat{a}_{12}\,,
\label{eq:coefhvv}
\end{aligned}
\eeq
and where the $\hat{a_i}$ coefficients have been defined in Eq.~(\ref{hata}).
Additionally, the effective CP-odd Higgs-fermion couplings induced by
the mixing effects described in Sec.~\ref{sec:2point} generate one-loop
induced HVV couplings such as
\begin{eqnarray}
\tilde{g}_{Hgg}=\frac{\alpha_S}{8\pi v } \left(\hat{a}_{2D}-\frac{4 p_h^2}{v^2}\hat{a}_{13}\right) F_{{\rm odd}}^{{\rm CP}}(x_f) =
\frac{3}{8}\frac{\alpha_S}{\alpha_{{\rm em}}} \tilde{g}_{H\gamma\gamma}\; ,
\end{eqnarray}
where $F_{{\rm odd}}^{{\rm CP}}(x_f)$ is the form factor from the
fermionic one-loop processes~\cite{Spira:1995rr}, that in the limit of
high fermion masses ($x_f\equiv 4M_f^2/ M_h^2\gg1$) is approximately
$F_{{\rm odd}}^{{\rm CP}}=1$, almost equal to the form factor for the
CP-even Yukawa-fermion contribution to $h G^a_{\mu\nu} G^{a\mu\nu}$
and $h A_{\mu\nu} A^{\mu\nu}$ in the same limit,
$F_{{\rm even}}^{{\rm CP}}(x_f)$.
In addition to effects on the Higgs signals, these operators,
together with those giving direct contributions to
$\tilde{g}_{H \gamma \gamma}$ in Eq.~(\ref{eq:coefhvv}) give also a contribution
to the fermion EDMs~\cite{McKeen:2012av} of the
form\footnote{In writing Eq.~(\ref{eq:edm2loop}) we have only considered the relevant
loop of top quarks in the loop-induced part of the $h\gamma\gamma$ vertex
(both CP-odd and CP-even) generated by $\cS_{2D}(h)$ and  we have
neglected the corresponding $\mathcal{O}(m_f^2/m_H^2)$ contribution
from $\cS_{13}(h)$.}
\beq
\begin{split}
d_f=\frac{e^3m_f}{\pi^2 v^2} 
\left[-\frac{1}{4}\hat{a}_{\tilde{B}}+ \hat{a}_8 + \hat{a}_1 
- \frac{1}{8} \hat{a}_{\tilde{W}}+\frac{1}{48\pi^2}
\hat{a}_{2D}\left(F_{{\rm odd}}^{{\rm CP}}(x_{\rm top})
+\frac{2}{3}F_{{\rm even}}^{{\rm CP}}(x_{\rm top})\right)
\right]\times\\
\times \left[\log\frac{\Lambda_s^2}{m_H^2}+ \mathcal{O}(1)\right]\,,
\end{split}
\label{eq:edm2loop}
\eeq
whose size can be constrained, for example, from the present bound on the
electron EDM in Eq.(\ref{eq:eedm}):  
\beq
\begin{split}
\left|
\left(
-\frac{1}{4}\hat{a}_{\tilde{B}}+ \hat{a}_8 + \hat{a}_1 
- \frac{1}{8} \hat{a}_{\tilde{W}}+\frac{1}{48\pi^2}
\hat{a}_{2D}\left(F_{{\rm odd}}^{{\rm CP}}(x_{\rm top})+\frac{2}{3}F_{{\rm even}}^{{\rm CP}}(x_{\rm top})\right)
\right)\times\right.\phantom{\hspace{1cm}}\\
\left.\times\left[
\log\left(\frac{\Lambda_s^2}{m_H^2}\right) + \mathcal{O}(1)
\right]
\right|<5.6\times 10^{-5}\,.  
\end{split}
\label{moreEDMbounds}
\eeq

While twelve chiral CP-odd operators affect HVV vertices, in the case of a linear realization of EWSB, at $d=6$ only six operators contribute to Eq.~(\ref{eq:hvv}), which are the siblings of chiral operators $\cS_1(h)$, $\cS_2(h)$, $\cS_3(h)$, $\cS_G(h)$, $\cS_B(h)$, $\cS_W(h)$ listed in Appendix~\ref{App:Siblings}. Hence $\tilde{g}_{Hgg}$, $\tilde{g}_{H \gamma \gamma}$, $\tilde{g}_{H Z \gamma}$, $\tilde{g}_{H ZZ}^{(2)}$, and $\tilde{g}_{H WW}^{(2)}$ can be generated at $d=6$ with independent coefficients, while $\tilde{g}^{(1),(lin,d=6)}_{H WW}=\tilde{g}^{(5) (lin,d=6)}_{H WW}=0$. Also one-loop HVV couplings will be induced by the $d=6$ operator sibling of $\cS_{2D}(h)$.

Generically because larger number of operators contribute to a given coupling in the chiral expansion, cancellations
between their contributions can occur which are not possible in the case of the linear expansion at $d=6$. However, we notice that for the HVV couplings in Eq.~(\ref{eq:coefhvv}) enough independent operators contribute in linear EWSB at $d=6$ such that cancellations are also possible in this case.

Measuring the CP properties of the Higgs couplings is a subject
with an extensive literature before and after the Higgs discovery.
For the sake of concreteness we focus here on the experimental
results on the most studied channel,
$h\rightarrow ZZ\rightarrow \ell^+\ell^-\ell^{\prime+}\ell^{\prime-}$,
for which combined results of the full 7+8 TeV LHC runs have been presented
both by CMS~\cite{Chatrchyan:2012jja,Chatrchyan:2013mxa} and
ATLAS~\cite{Aad:2013xqa,ATLAS:2013nma} collaborations.

Historically the key observables for measuring the CP properties of
the Higgs in this channel were established in the seminal
works in Refs~\cite{Dell'Aquila:1985ve,Dell'Aquila:1985vc,Dell'Aquila:1985vb},
that were followed by an abundant literature on their applications to
the
LHC~\cite{Soni:1993jc,Chang:1993jy,Arens:1994wd,Choi:2002jk,Buszello:2002uu,Godbole:2007cn,Cao:2009ah}. Most
of these early phenomenological studies were based on the study of
single variable observables. Most recently, an almost together with
the first LHC collisions, two different new multivariable
methods~\cite{Gao:2010qx,DeRujula:2010ys} were proposed to use all
the kinematic information of the event as input into the likelihood, to
compare and exclude between different Higgs spin and parity
hypothesis. These phenomenological studies set the roots of the first
LHC experimental analyses of spin and CP properties of the Higgs in
this
channel~\cite{Chatrchyan:2012jja,Chatrchyan:2013mxa,Aad:2013xqa,ATLAS:2013nma}.

In particular the results of the experimental constraints from the CMS
analysis~\cite{Chatrchyan:2012jja,Chatrchyan:2013mxa} 
can be translated into the language of the effective operators of a light 
dynamical Higgs in Eq.~(\ref{CPbasis}). With this purpose we notice
that in Ref.~\cite{Chatrchyan:2013mxa} the $h\rightarrow ZZ$ vertex is described
using the notation in~\cite{Gao:2010qx}:
\begin{eqnarray}
A(h\rightarrow ZZ)=v^{-1}\left(d_1 m_Z^2 \epsilon_1^* \epsilon_2^*
+ d_2 f_{\mu\nu}^{*(1)} f^{\mu\nu *(2)}
+ d_3 f_{\mu\nu}^{*(1)} \tilde{f}^{\mu\nu *(2)}\right)\, ,
\label{cmsampl}
\end{eqnarray}
where $f_{\mu\nu}^{(i)}=\epsilon_\mu^iq_\nu^i-\epsilon_\nu^iq_\mu^i$,
$\tilde{f}_{\mu\nu}^{(i)}=\frac{1}{2}\epsilon_{\mu\nu\alpha\beta}f^{\alpha\beta(i)}
=\epsilon_{\mu\nu\alpha\beta}\epsilon_i^\alpha q_i^\beta$, with
$\epsilon^{1,2}$ being the polarization vectors of the $Z$ bosons and
$q_{1,2}$ the corresponding four-momenta. In the SM $d_1=2i$, while
$d_2$ only receives marginally contributions from high order diagrams, that
can be safely neglected leading to $d_2=d_3=0$.
The $d_3$ term is CP-odd and its interference with the
CP-conserving terms $d_1$ or $d_2$ leads to the CP-violating signals 
that are analyzed.

The effective operators in Eq.~(\ref{CPbasis}) give a non-vanishing
contribution to $d_3$ which,  from Eqs.~(\ref{eq:hvv})
and~(\ref{eq:coefhvv}), reads 
\begin{equation}
d_3=-2 i v \tilde g^{(2)}_{HZZ}\, ,
\end{equation}
while as long as no CP-conserving operators are considered $d_2=0$
and $d_1=d_{1,SM}$.

In Ref.~\cite{Chatrchyan:2013mxa} a measure of  CP-violation in the
$h\rightarrow ZZ^*\rightarrow 4l$  observables was defined as
\begin{eqnarray}
 f_{d_3}=\frac{|d_3|^2\sigma_3}{|d_1|^2\sigma_1+|d_3|^2\sigma_3}\;\; ,
\end{eqnarray}
where $\sigma_1$ ($\sigma_3$) corresponds to the cross section
for the process $h\rightarrow ZZ$ when $d_1=1$ ($d_3=1$) and $d_3=0$ ($d_1=1$).
For $M_h=125.6$ GeV, $\frac{\sigma_1}{\sigma_3}=6.36$. In
Ref.~\cite{Chatrchyan:2013mxa}
$f_{d_3}$ was fitted as one of the parameters of the multivariable analysis,
obtaining the measured value
\beq
 f_{d_3}=0.00^{+0.17}_{-0.00} \qquad\Longrightarrow \qquad
\frac{|d_3|}{|d_1|}=0.00^{+1.14}_{-0.00}\,,
 \label{68}
\eeq
pointing to the CP-even nature of the state. Furthermore, $95\%$ CL exclusion
bounds on $f_{d_3}$ were derived,
\beq
 f_{d_3}<0.51
 \qquad \Longrightarrow \qquad
 \frac{|d_3|}{|d_1|}<2.57\,.
\label{eq:CMSbound}
\eeq
We can directly translate the bounds in Eq.~(\ref{eq:CMSbound}) to
$68(95)\%$ CL constraints on the coefficients of the relevant CP-violating operators,
\begin{equation}
\left|-\frac{1}{4}\hat{a}_{\tilde{B}} + \frac{c_\theta^4}{s_\theta^4}\hat{a}_8
-\frac{c_\theta^2}{s_\theta^2} \hat{a}_1 + \frac{1}{2s_\theta^2} 
\hat{a}_2 - \frac{c_\theta^4}{8s_\theta^4}\hat{a}_{\tilde{W}}
-\frac{c_\theta^2}{2s_\theta^4} \hat{a}_9-\frac{c_\theta^2}{4s_\theta^4} 
\hat{a}_3
\right| \leq 10.3\ (23.3)\,. 
\label{tired3}
\end{equation}

In Ref.~\cite{CMS:2013xfa} the same analysis was applied to derive the future
expectations when $300 (3000) \fb^{-1}$ are collected at 14 TeV. The
corresponding expected sensitivities at $95\%$ CL are
\beq 
f_{d_3}\leq 0.13\, (0.04) \qquad \text{for}\qquad300\,(3000)\fb^{-1}\,.
\eeq
They can be translated into the following sensitivity at $95\%$ CL to the relevant
combination of operators:
\begin{eqnarray}
 &&\left|-\frac{1}{4}\hat{a}_{\tilde{B}} +
  \frac{c_\theta^4}{s_\theta^4}\hat{a}_8
  -\frac{c_\theta^2}{s_\theta^2} \hat{a}_1 + \frac{1}{2s_\theta^2}
  \hat{a}_2 - \frac{c_\theta^4}{8s_\theta^4}\hat{a}_{\tilde{W}}
  -\frac{c_\theta^2}{2s_\theta^4}
  \hat{a}_9-\frac{c_\theta^2}{4s_\theta^4} \hat{a}_3 \right|\leq 8.8\ \ (4.6)\, ,
\label{expcms}
\end{eqnarray}
for 300 (3000) fb$^{-1}$.

Observables to study the CP properties of the Higgs couplings have
also been proposed in the production channel $pp\rightarrow h jj$
followed by the Higgs decay into $\tau^+\tau^-$, $W^+W^-$, or
$\gamma\gamma$~\cite{Plehn:2001nj,Buszello:2006hf,Hankele:2006ma,Klamke:2007cu,
Englert:2012ct,Odagiri:2002nd,DelDuca:2006hk,Andersen:2010zx,Englert:2012xt,
Djouadi:2013yb,Dolan:2014upa}. Depending on the kinematic cuts
imposed, the study is most sensitive to CP-violating effects in the
$hWW$ (from $\cS_{\tilde{W}}(h)$, $\cS_3(h)$ and/or $\cS_7$(h)) and $hZZ$ (from
$\cS_{\tilde{B}}(h)$, $\cS_{\tilde{W}}(h)$, $\cS_1(h)$, $\cS_2(h)$, $\cS_3(h)$, $\cS_8(h)$ and/or
$\cS_9(h)$) vertices contributing to Higgs production through vector boson
fusion, or in the $hgg$ vertex (from $\cS_{\tilde{G}}(h)$, and from loop
induced $\cS_{2D}(h)$ and $\cS_{13}(h)$) contributing to production by gluon
fusion. The sensitivity to CP violating observables in associated
production processes $pp\rightarrow hZ\rightarrow
b\bar{b}\ell^+\ell^-$ and $pp\rightarrow hW\rightarrow
\ell^+jjE_T^{{\rm miss}}$ has also been studied in
Refs.~\cite{Christensen:2010pf,Desai:2011yj,Ellis:2012xd,Englert:2012xt,
Godbole:2013saa,Delaunay:2013npa}, and in pure gluon fusion
production followed by Higgs decay into $\gamma\gamma$ or to
$Z\gamma$~\cite{Voloshin:2012tv,Korchin:2013ifa,Bishara:2013vya,Chen:2014ona}.

Finally, it is also possible to quantify the potential to observe or
bound CP-odd interactions from global analyses of the Higgs signal
strengths~\cite{Freitas:2012kw,Djouadi:2013qya,Belusca-Maito:2014dpa}. However
in this case the analysis does not contain any genuinely CP-violating
observable and consequently it is always sensitive to combinations of
CP-even and CP-odd interactions.

%
%
\section{Conclusions}

Charge conjugation and parity are not exact symmetries of the Standard
Model of particle physics, and furthermore electroweak interactions
have proven that neither their product is a symmetry of
nature. In addition, new sources of CP-violation are likely needed to
explain the matter-antimatter asymmetry of the universe.  More
importantly, the extreme fine-tuning of the SM parameters implied by
the strong CP problem suggests as well new sources of CP violation. On
the other hand, the questions of whether the Higgs is elementary or
composite, and of whether EWSB is realised linearly or
non-linearly are still open.

We have focused here in the non-linear option for EWSB, approaching
the issue through the model-independent tool of effective
Lagrangians. We have established here for the first time the complete set of independent gauge
and gauge-Higgs CP-odd effective operators for the generic case of a
light dynamical Higgs, up to four derivatives in the chiral expansion,
see the basis in Eq.~(\ref{CPbasis}). The relation with the ensemble
of $d=6$ CP-odd linear operators has been clarified as
well. 

One interesting result is that an anomalous CP-odd coupling
$\cS_{2D}(h)$ is shown to be present already at the leading order of the
chiral Lagrangian, that is, at the two-derivative level. It affects
the renormalization of the SM parameters inducing a CP-odd component in
fermion-Higgs and fermion-$Z$ interactions. That coupling is instead not present at the
leading order ($d=4$) of the linear expansion, in other words in the
SM Lagrangian, as its would-be linear sibling turns out to be a $d=6$
operator. A similar contribution to two-point functions and with a similar physical
impact stems as well from a four-derivative operator, $\cS_{13}(h)$. Furthermore, focussing to triple gauge boson vertices, there are six independent four-derivatives chiral CP-odd operators contributing to these couplings, while only two are present in the set of $d=6$ linear ones. Moreover, the nature of the phenomenological couplings involved is different, as some of the former correspond to $d=8$ linear operators, while one of the latter set is a six derivatives. Considering instead the Higgs to two gauge boson vertices, there are in total twelve independent four-derivatives chiral CP-odd operators contributing to these interactions, while only seven are present for the $d=6$ linear case.

We have established bounds on the CP-odd non-linear operator
coefficients, mainly from anomalous triple vertices versus two types
of experimental data: i) limits on fermionic EDMs
which, not surprisingly given the very fine experimental precision,
set some of the quantitatively tightest constraints; ii) present and
future LHC data, in particular from the impact of TGV and Higgs-gauge
boson triple couplings.

More precisely, among the TGV we have evaluated the one-loop
contribution to fermionic EDMs from the anomalous CP--odd
$WW\gamma$ vertex, and derived then the corresponding bounds on
the relevant non-linear operator coefficients, see Eqs.~(\ref{c1c8eEDM}) and
(\ref{c1c8nEDM}).

The bounds on the strength of anomalous CP--odd $WWZ$ vertices
have been explored here from both CP-blind and from CP-sensitive
observables. The strongest limits are still coming from LEP
analyses, and we have translated them into bounds for the non-linear
operator coefficients, see Eq.~(\ref{eq:wwzlep2}). Furthermore, the
direct measurement of this vertex through CP-blind signals in gauge
boson single or pair production at colliders has been addressed. In
Sec.~\ref{WWZcollider} we have thus estimated the present and
future potential of LHC to measure anomalous CP--odd TGVs performing a
realistic collider analysis of $WZ$ pair production. In doing so
we have exploited that anomalous TGVs enhance the
cross sections at high energies by quantifying the dependence of
the expectations on kinematic variables which trace well this energy
behaviour. The conclusion
is that the LHC has the potential to improve the LEP bounds using the 7
and 8 TeV collected data sets, as shown in Table~\ref{tab:results1},
while the precision reachable in the future 14 TeV run will approach
the per cent level on the anomalous coefficients.

Furthermore, on the realm of CP-odd observables, we have presented
the LHC potential to decipher the CP nature of an hypothetical
anomalous TGV observation by defining CP--odd sensitive asymmetries.
Through the asymmetry defined in Eq.~(\ref{deltasigCP}), it has been shown
that the future LHC run will have the capability to establish the CP
nature of the $WWZ$ vertex for a large range of the parameter space
that can be covered in that run, see Eqs.~(\ref{tired1}) and
(\ref{tired2}).

For CP-odd observables sensitive to anomalous Higgs--gauge boson
trilinear vertices, the focus has been set on the limits than can be
obtained from the existing 7 and 8 TeV LHC experimental Higgs searches
that benefit from genuinely CP--odd observables. We have translated
the bounds from the CMS study of the Higgs boson properties on the
leptonic $h\rightarrow ZZ$ channel to the relevant combination of
non-linear operator coefficients, see Eq.~(\ref{tired3}). The future
sensitivity estimated by CMS in the same framework has also been
translated into the future reachable sensitivity on the same
combination of coefficients, Eq.~(\ref{expcms}). Finally, we have also
noticed that those combinations of non-linear operators contributing
to the $h\gamma\gamma$ vertex can be constrained from the contribution
of this trilinear coupling to fermionic EDMs, as illustrated in
Eq.~(\ref{moreEDMbounds}).

The quest of new sources of CP-violation is well justified if not
mandated by present observations and SM puzzles, while the elementary
or composite nature of the Higgs and the maybe related nature --linear or non-linear-- of the EWSB mechanism are other fundamental and urgent issues in particle physics. The model-independent theoretical analysis
of CP violation performed in this paper for the case of a light
dynamical Higgs, as well as the new limits established and the new
phenomenological tools developed, should be useful in shedding light on
these fundamental issues.

%
%

\section*{Acknowledgements}
We thank O. Eboli for interesting discussions. We acknowledge partial
support of the European Union network FP7 ITN INVISIBLES (Marie Curie
Actions, PITN-GA-2011-289442), of CiCYT through the project
FPA2009-09017, of the European Union FP7 ITN UNILHC (Marie Curie
Actions, PITN-GA-2009-237920), of MICINN through the grant
BES-2010-037869, of the Spanish MINECO Centro de Excelencia Severo
Ochoa Programme under grant SEV-2012-0249, and of the Italian
Ministero dell'Uni\-ver\-si\-t\`a e della Ricerca Scientifica through
the COFIN program (PRIN 2008) and the contract
MRTN-CT-2006-035505. M.C.G-G is supported by USA-NSF grant PHY-09-6739,
and together with J.G-F by MICINN FPA2010-20807 and
consolider-ingenio 2010 program CSD-2008-0037. J.G-F is further
supported by ME FPU grant AP2009-2546.
The work of L.M. is supported by the Juan de la Cierva programme
(JCI-2011-09244). The work of J.Y. is supported by the Spanish MINECO Centro de Excelencia Severo
Ochoa Programme under grant SEV-2012-0249.

 %
%
 \newpage
  
\appendix
\small

%
%
\section{Linear siblings of chiral operators}
\label{App:Siblings}

The interactions described by the chiral operators in Eq.~(\ref{CPbasis}) can also be described in the context of a linearly realised EWSB, through linear operators built in terms of the SM Higgs doublet. In this Appendix, the connection between the two expansions is discussed. As the number and nature of the leading order operators in the chiral and linear expansions are not the same, an exact  correspondence between the two kind of operators can be found only in the cases when $d=6$ linear operators are involved, as only for them complete bases of independent terms have been defined. Otherwise, it will be indicated which chiral operators should be combined in order to generate the gauge interactions contained in specific $d>6$ linear operators.

\begin{description}
\boldmath
\item[For chiral operators connected to $d=6$ linear operators:]
\unboldmath
\beq
\begin{aligned}
\cS_{\widetilde B}(h)&\to g'^2\,\epsilon^{\mu\nu\rho\sigma}B_{\mu\nu}\, B_{\rho\sigma} \left(\Phi^{\dagger}\,\Phi\right)\\
\cS_{\widetilde W}(h)&\to g^2\,\epsilon^{\mu\nu\rho\sigma}\,\text{Tr}\left(W_{\mu\nu}\,W_{\rho\sigma}\right) \left(\Phi^{\dagger}\,\Phi\right)\\
\cS_{\widetilde G}(h)&\to g^2_s\,\epsilon^{\mu\nu\rho\sigma}G^a_{\mu\nu}\, G^a_{\rho\sigma} \left(\Phi^{\dagger}\,\Phi\right)\\
\cS_{2D}(h)&\to\left(\Phi^{\dagger}\,\overleftrightarrow{D^\mu\,D_{\mu}}\Phi\right)\left(\Phi^{\dagger}\,\Phi\right)\\
\cS_{1}(h)
&\to 
g\,g'\,\epsilon^{\mu\nu\rho\sigma}\,B_{\mu\nu}\,W_{\rho\sigma}^j\,\left(\Phi^{\dagger}\,\tau_j\,\Phi\right)\\
\cS_{2}(h)&\to g'\,\epsilon^{\mu\nu\rho\sigma}\,B_{\mu\nu}\,\left[\left(\Phi^{\dagger}\,\overleftrightarrow{D_\sigma \,D_{\rho}}\Phi\right) + 2\,\left(D_\rho \Phi\right)^{\dagger}D_{\sigma}\Phi\right]\\
\cS_{3}(h)&\to\,g\,\epsilon^{\mu\nu\rho\sigma}\,W_{\mu\nu}^i\,\left[\left(\Phi^{\dagger}\,\tau_i\,\overleftrightarrow{D_\sigma \,D_{\rho}}\Phi\right) + 2\,\left(D_\rho \Phi\right)^{\dagger}\,\tau_i\,D_{\sigma}\Phi\right]\,
\end{aligned}
\label{Siblingsd6}
\eeq

\boldmath
\item[For chiral operators connected to $d>6$ linear operators:]
\unboldmath
\beq
\begin{aligned}
\cS_{4}(h)&\to\, g\,W^{\mu\nu}_i\,\left(\Phi^{\dagger}\,\tau^i\,\overleftrightarrow{D_{\mu}}\Phi\right)\,\left(\Phi^{\dagger}\,\overleftrightarrow{D_{\nu}}\Phi\right)
&&(d=8)\\
\cS_{5}(h),\cS_{10}(h)&\to\left(D^{\mu}\Phi\right)^\dagger\left(D^{\nu}\Phi\right)\left(\Phi^{\dagger}\,\overleftrightarrow{D_{\mu}\,D_{\nu}}\Phi\right)
&&(d=8)\\
\cS_{6}(h),\cS_{11}(h)&\to \left(D^{\mu}\Phi\right)^\dagger\left(D_{\mu}\Phi\right)\left(\Phi^{\dagger}\,\overleftrightarrow{D^{\nu}\,D_{\nu}}\Phi\right)
&&(d=8)\\
\cS_{7}(h)&\to \,\epsilon_{ijk}\,W_{\mu\nu}^i\left(\Phi^{\dagger}\,\tau^j\,\overleftrightarrow{D^{\mu}\,D^{\nu}}\Phi\right)\,\left(\Phi^{\dagger}\,\tau^k\,\Phi\right)
&&(d=8)\\
\cS_{8}(h)&\to \,g^2\,\epsilon^{\mu\nu\rho\sigma}\,W_{\mu\nu}^i\,W_{\rho\sigma}^j\,\left(\Phi^{\dagger}\,\tau_i\,\Phi\right)\,\left(\Phi^{\dagger}\,\tau_j\,\Phi\right)
&&(d=8)\\
\cS_{9}(h)&\to \,g\,\epsilon^{\mu\nu\rho\sigma}W_{\mu\nu}^i\,\left(\Phi^{\dagger}\,\tau^i\,\Phi\right)\,\left(\Phi^{\dagger}\,\overleftrightarrow{D_\rho\,D_\sigma}\Phi\right)
&&(d=8)\\
\cS_{12}(h),\cS_{13}(h),\cS_{14}(h)&\to\left(\Phi^{\dagger}\,\overleftrightarrow{D^\mu\,D_{\mu}}\Phi\right)\,D^{\nu}D_{\nu}\left(\Phi^{\dagger}\,\Phi\right)
&&(d=8)\\
\cS_{15}(h),\cS_{16}(h)&\to\left(\Phi^{\dagger}\,\overleftrightarrow{D^{\mu}}\Phi\right)\,\left(\Phi^{\dagger}\,\overleftrightarrow{D_{\mu}}\Phi\right)\,\left(\Phi^{\dagger}\,\overleftrightarrow{D^{\nu}\,D_{\nu}}\Phi\right)
&&(d=10)
\end{aligned}
\label{Siblingsdh6}
\eeq
where in the brackets the dimension of the specific linear operator is explicitly reported.
\end{description}

%
%

\unboldmath
\section{Feynman rules}
\label{AppFR}

\noindent This Appendix provides a complete list of all  Feynman rules
resulting from the CP-odd operators in the Lagrangian
$\Delta\LL_{\text{\cancel CP}}$ in Eqs.~(\ref{LCP4}) and (\ref{CPbasis}). 
Greek indexes denote the flavour of the fermionic legs and are assumed to
be summed over when repeated; whenever they do not appear, it should be
understood that the vertex is flavour diagonal. Moreover, $\mathbf{y}_{f}\,\,(f=U,D,E)$  denotes the eigenvalue of the corresponding Yukawa coupling matrix defined in Eq.~(\ref{Yukawas}). Chirality operators $P_{L,R}$ are defined as
\beq
P_L=\frac{1}{2}\left(1-\gamma^5\right)\,, \qquad\qquad 
P_R=\frac{1}{2}\left(1+\gamma^5\right)\,.
\eeq
Flow in momentum convention is assumed in all diagrams. Only diagrams
with up to four legs are shown and the expansion for $\cF_i(h)$ in Eq.~(\ref{F})
has been adopted, together with the definitions of the $\hat{a}_i$ coefficients in Eq.~(\ref{hata}) and $\hat{b}_i= c_i b_i$. Vertices cubic in $h$ have been omitted below, but for Eq.~(FR.26) which results from the product of two $F_i(h)$ functions, see Footnote 5.  Finally, the
SM and BSM   Lorentz structures are
reported in two distinct columns, on the left and on the right,
respectively. Notice that all the pure gauge and gauge-$h$
interactions have no SM contribution. All quantities entering in the
Feynman rules below have resulted after the 
$Z$-renormalization scheme has been implemented.

\fancypagestyle{mylandscape}{%
  \fancyhf{}
  \fancyfoot{
    \makebox[\textwidth][r]{
      \rlap{\hspace{\footskip}
        \smash{
          \raisebox{\dimexpr.5\baselineskip+\footskip+.8\textheight}{
            \rotatebox{90}{\thepage}}}}}}
  \renewcommand{\headrulewidth}{0pt}
  \renewcommand{\footrulewidth}{0pt}
}

\begin{landscape}
\footnotesize

\addtolength{\hoffset}{-1.5cm}
\addtolength{\voffset}{2cm}

\newcommand{\nr}{\stepcounter{diagram}(FR.\arabic{diagram})}
\newcommand{\Anr}{\stepcounter{diagramDV}(A.\arabic{diagramDV})}
\addtolength{\linewidth}{3cm}
\newcounter{diagram}

\pagestyle{mylandscape}

\begin{center}
\vspace*{0.8cm}
\begin{tabular}{c@{\hspace*{5mm}}>{\centering}p{8cm}l@{\hspace{5mm}}l}

& & \bf SM & \bf Non-SM\\\hline
\\[0.5cm]
    
\nr & \parbox{6cm}{\includegraphics[scale=.4]{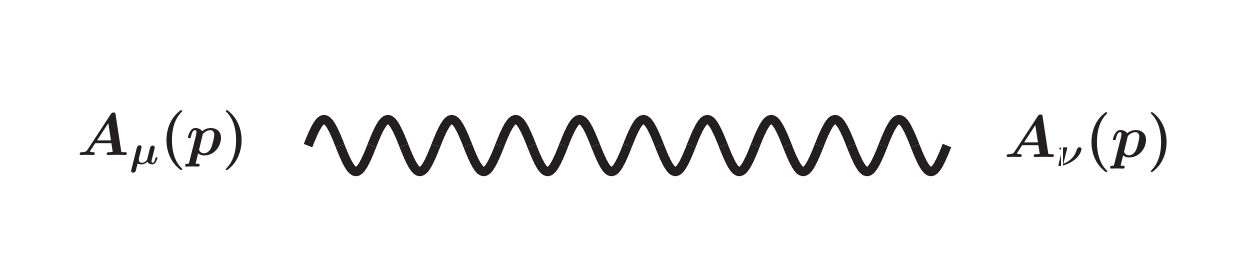}}   	& $\dfrac{-i}{ p^2}\left[g^{\mu\nu}-(1-\eta)\dfrac{p^\mu p^\nu}{p^2}\right]\quad(\eta \text{ is the gauge fixing parameter})$&\\[1.5cm]

\nr&  \parbox{6cm}{\includegraphics[scale=.4]{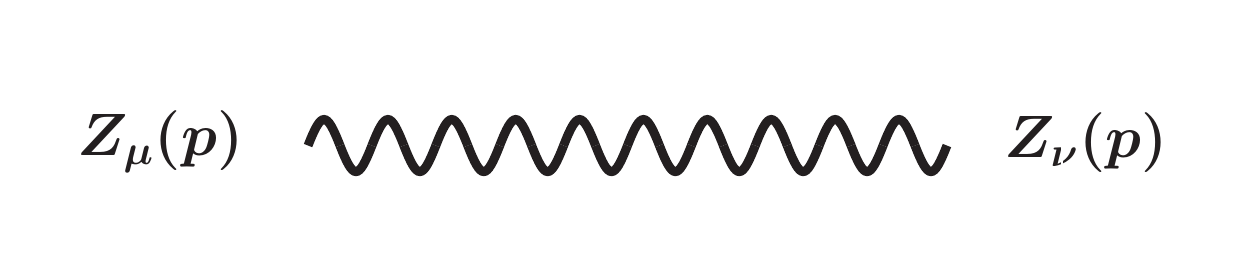}}	& 
$\dfrac{-i}{ p^2-m_Z^2}\left[g^{\mu\nu}-(1-\eta)\dfrac{p^\mu p^\nu}{p^2-\eta\,m_Z^2}\right]$&\\[1.5cm]

\nr&  \parbox{6cm}{\includegraphics[scale=.4]{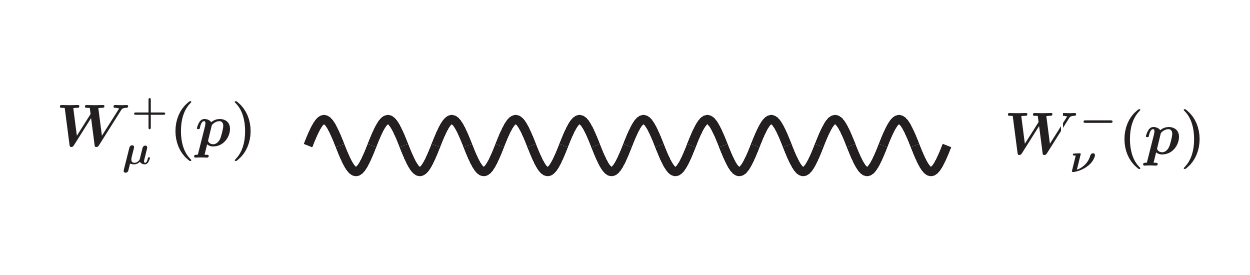}}	& 
$\dfrac{-i}{ p^2-m_W^2}\left[g^{\mu\nu}-(1-\eta)\dfrac{p^\mu p^\nu}{p^2-\eta\,m_W^2}\right]$ &\\[1.5cm]

\nr& \parbox{6cm}{\includegraphics[scale=.4]{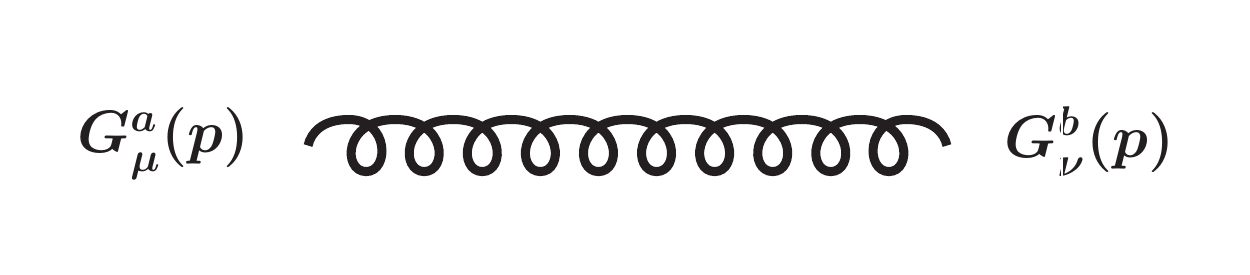}}	& $\dfrac{-ig^{\mu\nu}}{ p^2}\left[g^{\mu\nu}-(1-\eta)\dfrac{p^\mu p^\nu}{p^2}\right]$&\\[1.5cm]
 
\nr& \parbox{6cm}{\includegraphics[scale=.4]{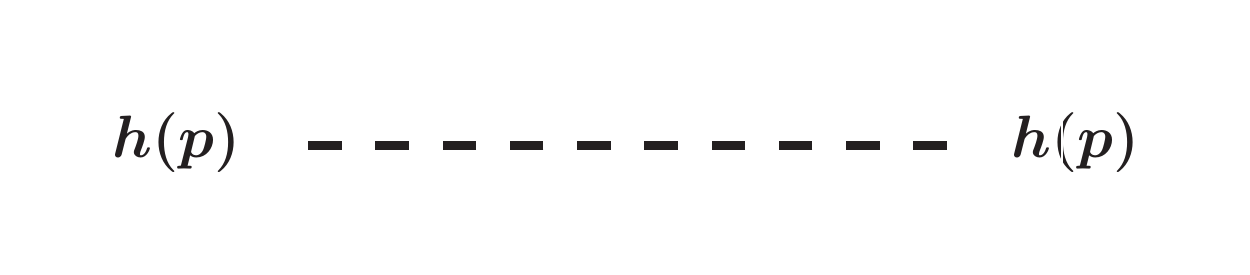}}	& $\dfrac{-i}{ p^2-m_h^2}$&\\[1.5cm]
  
\nr& \parbox{6cm}{\includegraphics[scale=.4]{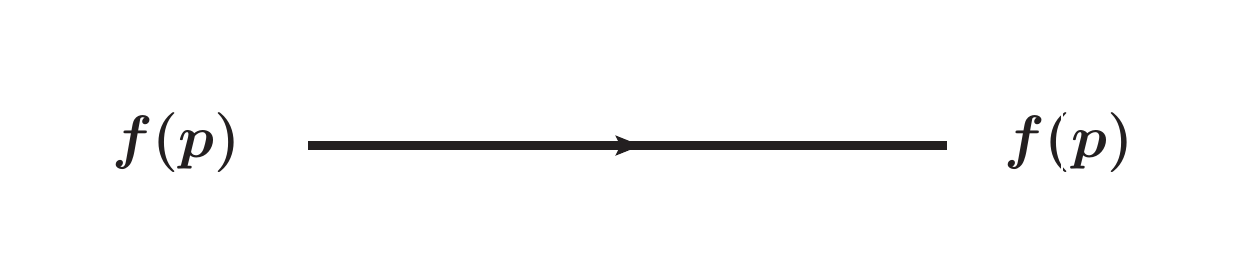}}	& $\dfrac{i(\slashed{p}+m_{f})}{p^2-m_{f}^2};\quad m_{f} = -\dfrac{v\,\mathbf{y}_{f}}{\sqrt2},\qquad f=U,D,E$&
\end{tabular}

\renewcommand{\arraystretch}{5}

\begin{tabular}{c@{\hspace*{3mm}}>{\centering}p{3cm}l@{\hspace*{1cm}}l}
& & \bf SM & \bf Non-SM \\\hline
  

\nr &  \parbox{6cm}{\includegraphics[scale=.35]{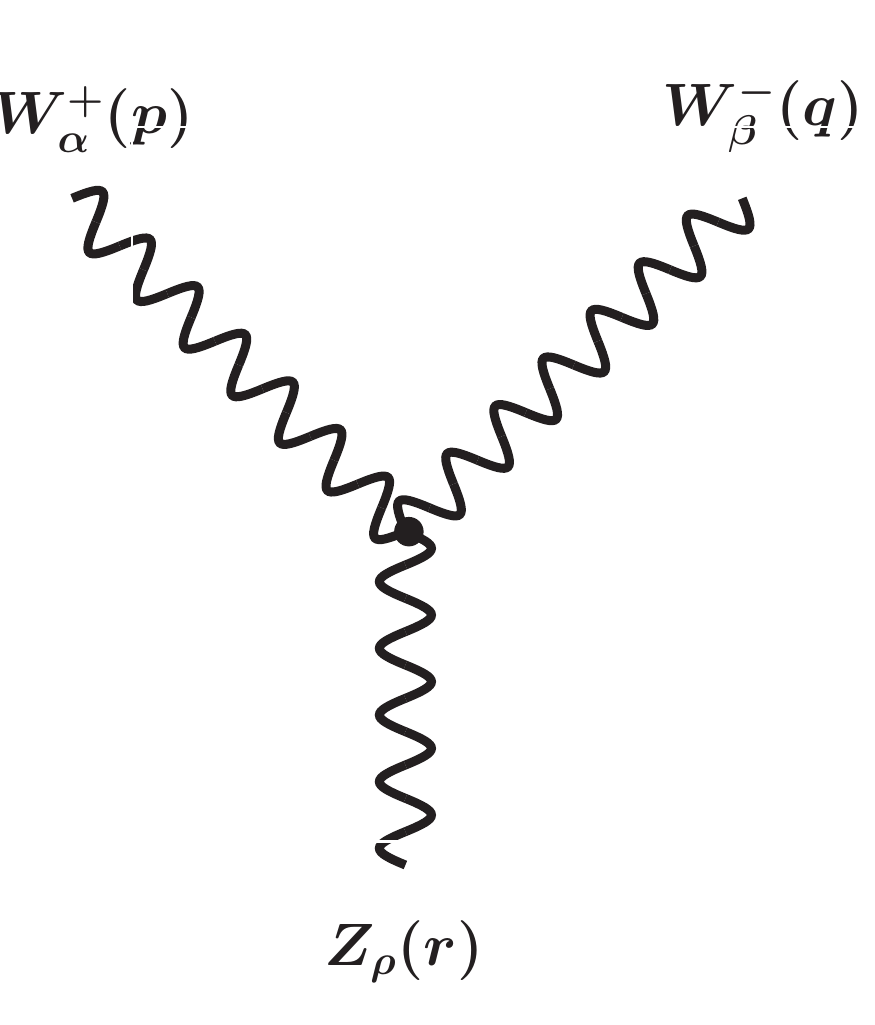}} &   &
$\begin{array}{l}
e^3   \csc ^2\theta _W \csc \left(2 \theta _W\right) \left\{-2 c_{11} g^{\alpha \beta } \left(p^{\rho }+q^{\rho }\right)+c_{10} \left(g^{\beta \rho } p^{\alpha }+g^{\alpha \rho } q^{\beta }\right)+c_4 \left[g^{\beta \rho } \left(p^{\alpha }+r^{\alpha }\right)+g^{\alpha \rho } \left(q^{\beta }+r^{\beta }\right)+g^{\alpha \beta } \left(p^{\rho }+q^{\rho }\right)\right]\right\} +\\[-1.3cm]
+ 8 i e^3   \left[c_8    \cot \theta _W \csc ^2\theta _W-c_1 \csc \left(2 \theta _W\right)\right]\left(p_{\sigma }+q_{\sigma }\right) \epsilon ^{\alpha  \beta  \rho  \sigma }
\end{array}
  $\\

\nr &  \parbox{6cm}{\includegraphics[scale=.35]{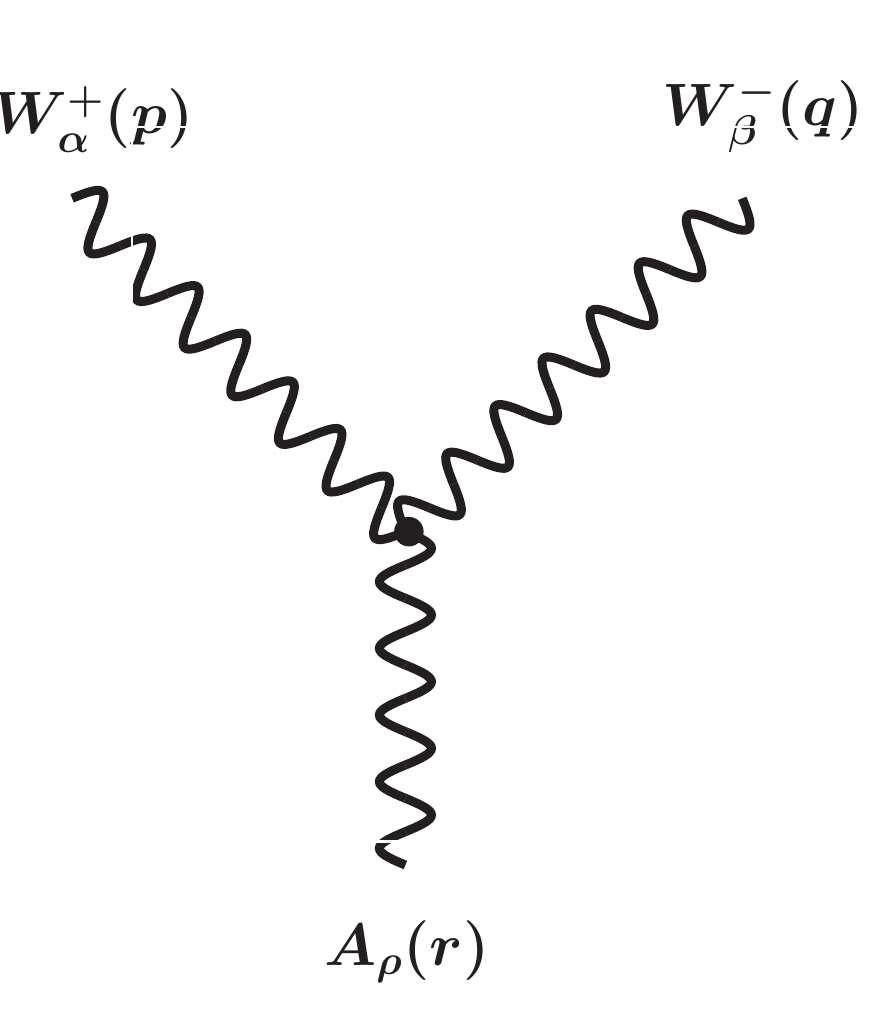}} &   &
$4 i e^3     \csc ^2\theta _W\left(2 c_8   +c_1\right)  \left(p_{\sigma }+q_{\sigma }\right)\epsilon ^{\alpha  \beta  \rho  \sigma }$  \\

\nr &  \parbox{6cm}{\includegraphics[scale=.35]{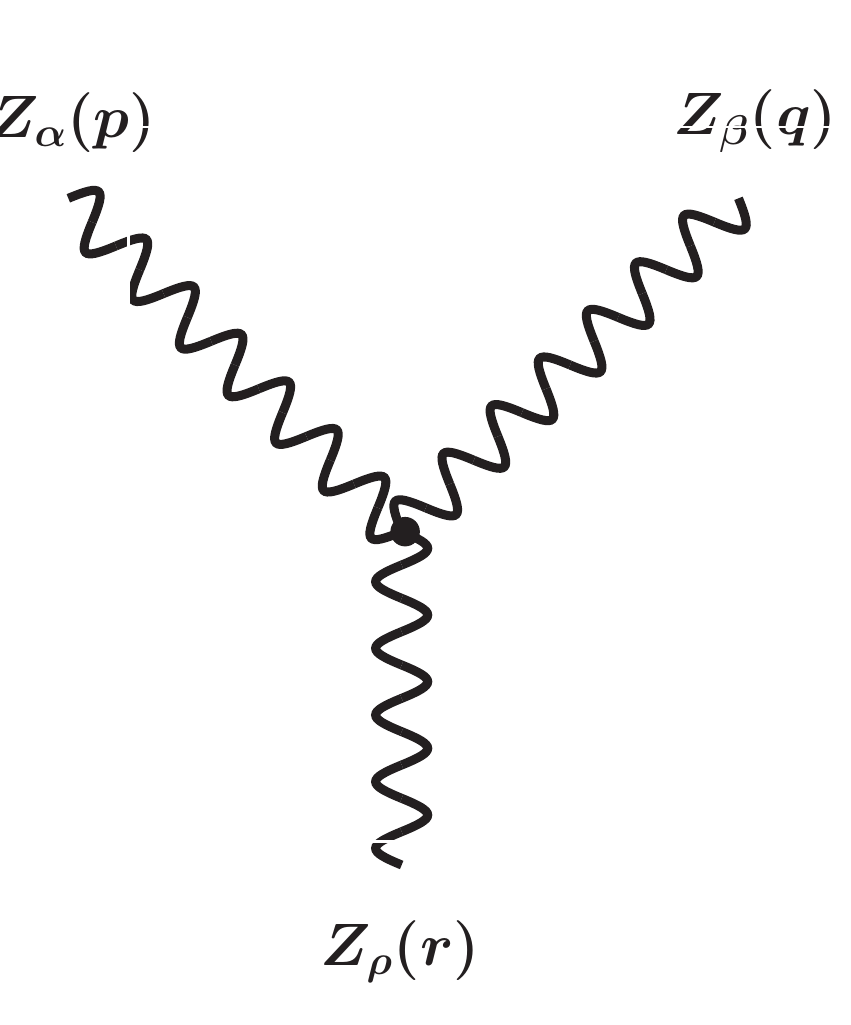}} &   &
$8 e^3   \left(2 c_{16}   +c_{10}+c_{11}\right) \csc ^3\left(2 \theta _W\right) \left[g^{\beta \rho } p^{\alpha }+g^{\alpha \rho } q^{\beta }-g^{\alpha \beta } \left(p^{\rho }+q^{\rho }\right)\right]$
  \tabularnewline


\nr &  \parbox{6cm}{\includegraphics[scale=.35]{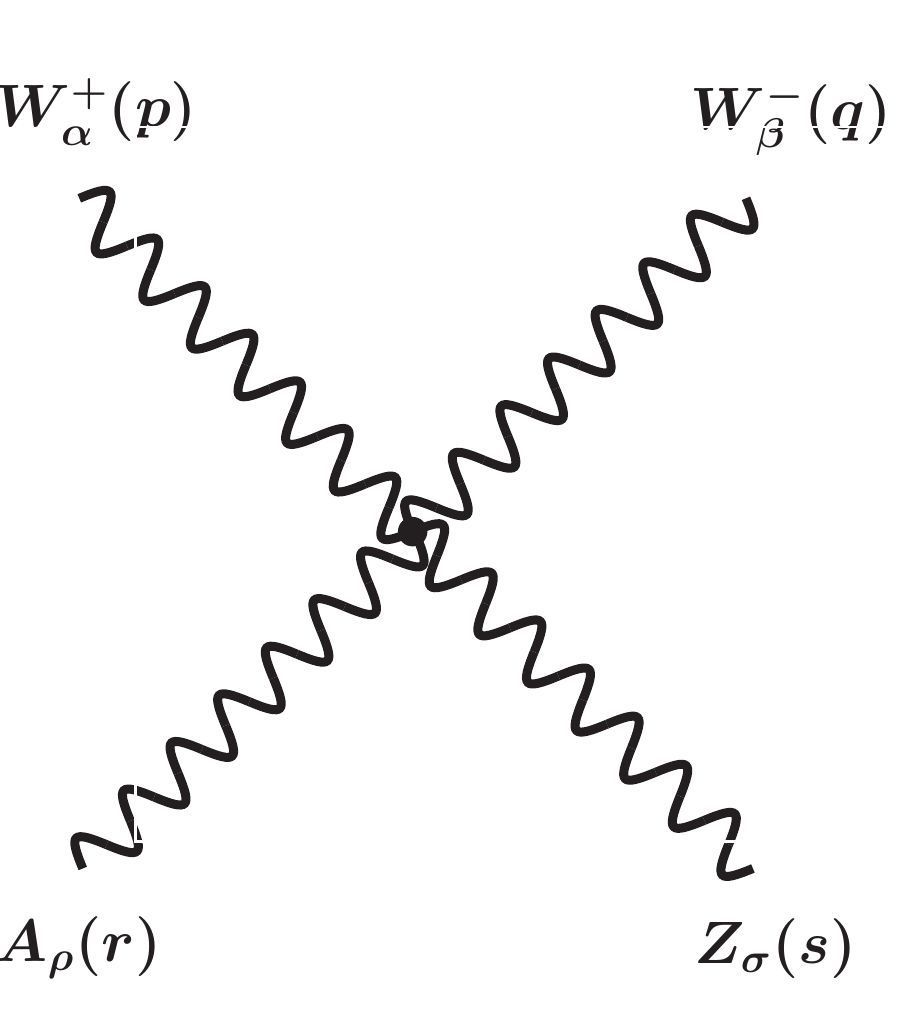}}&   & 
$\left(c_4+c_{10}\right) e^4   \csc ^2\theta _W \csc \left(2 \theta _W\right) \left(g^{\alpha \sigma } g^{\beta \rho }-g^{\alpha \rho } g^{\beta \sigma }\right)$
\tabularnewline

\end{tabular}

\newpage
\renewcommand{\arraystretch}{5}
\begin{tabular}{c@{\hspace*{5mm}}>{\centering}p{3cm}l@{\hspace*{1cm}}l}
& & \bf SM & \bf Non-SM \\\hline
  

\nr &  \parbox{5cm}{\includegraphics[scale=.35]{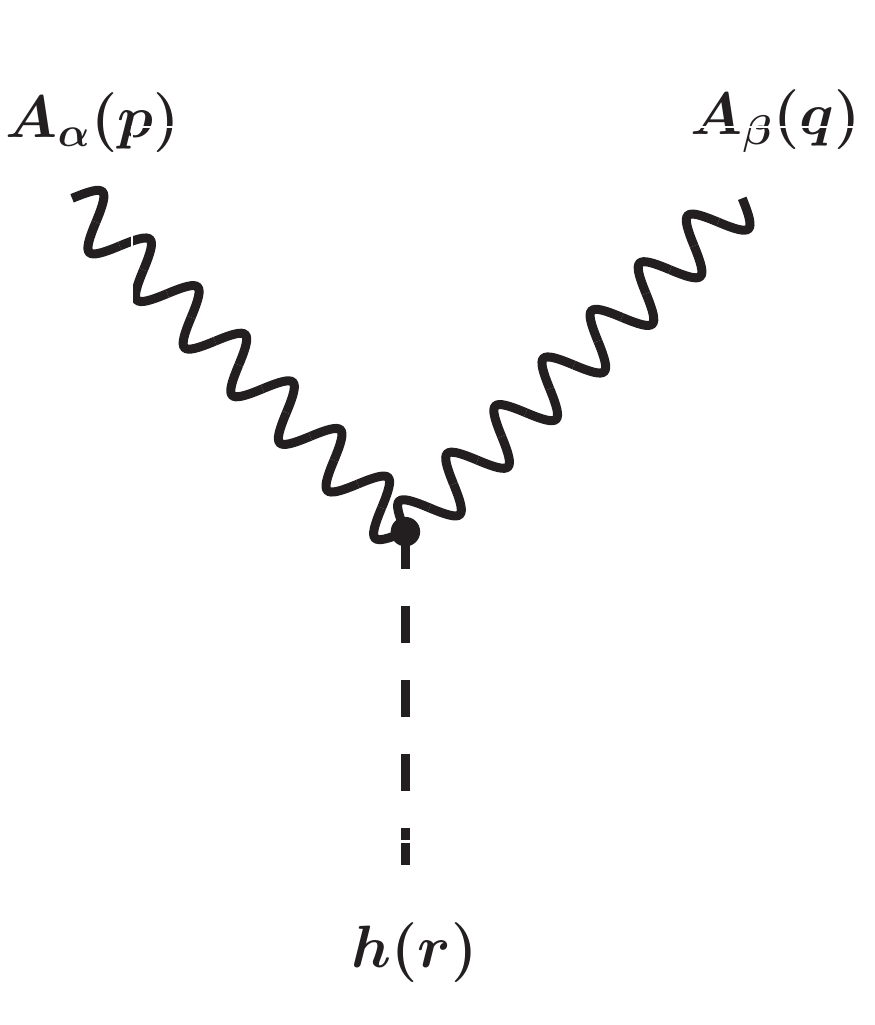}} &   &
$\frac{8 i}{v}e^2     \left[2 \left(-\frac{1}{4}\,\hat{a}_{\tilde{B}}+ \hat{a}_8   + \hat{a}_1\right) -\frac{1}{4}\,\hat{a}_{\tilde{W}}\right] p_{\mu} q_{\nu}\epsilon ^{\alpha  \beta  \mu  \nu }$\\

\nr &  \parbox{5cm}{\includegraphics[scale=.35]{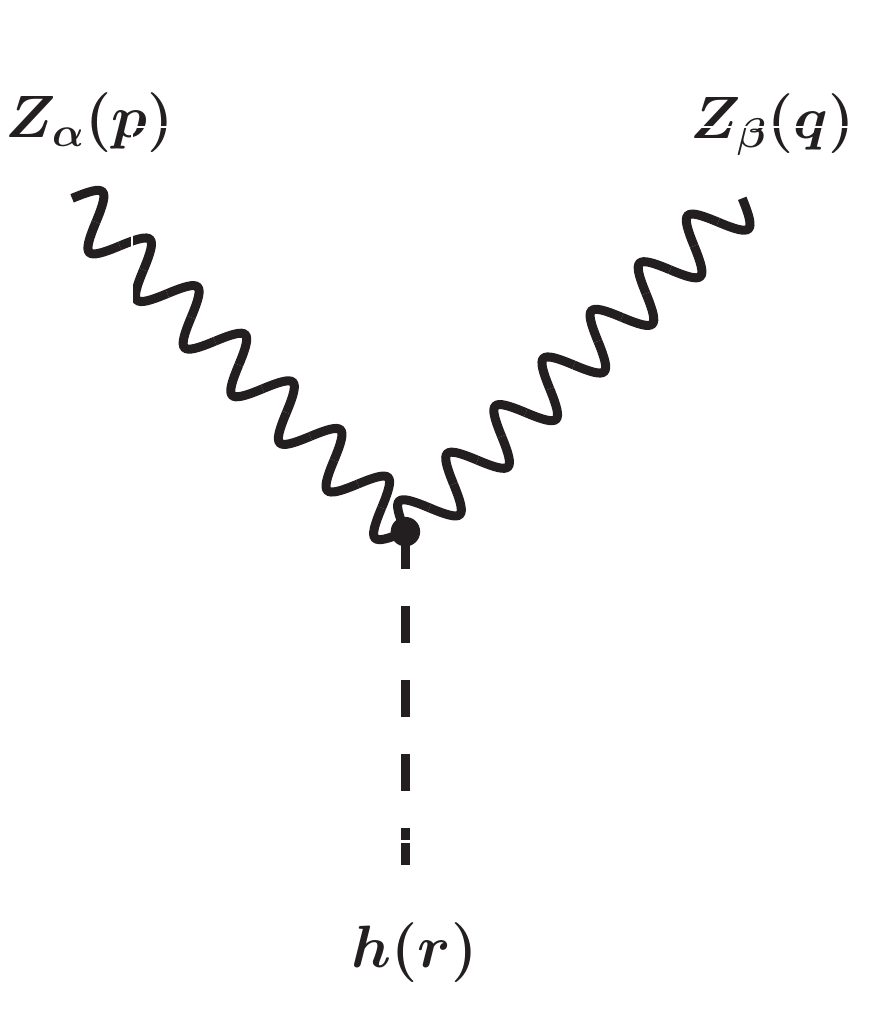}} &   &
$-\frac{4 i }{v}e^2     \left[4 \left(-\frac{1}{4}\,\hat{a}_{\tilde{B}} + \hat{a}_8  +  \hat{a}_1\right)-2 \left(-\frac{1}{2}\,\hat{a}_{\tilde{B}} + \hat{a}_2\right) \sec ^2\theta _W +\csc ^2\theta _W \left(\frac{1}{2}\,\hat{a}_{\tilde{W}}-4\,\hat{a}_8  + 2\,\hat{a}_9   + \hat{a}_3\right) -\frac{1}{2}\,\hat{a}_{\tilde{W}}\right] \,p_{\mu} q_{\nu} \epsilon ^{\alpha  \beta  \mu  \nu }$  \\

\nr &  \parbox{5cm}{\includegraphics[scale=.35]{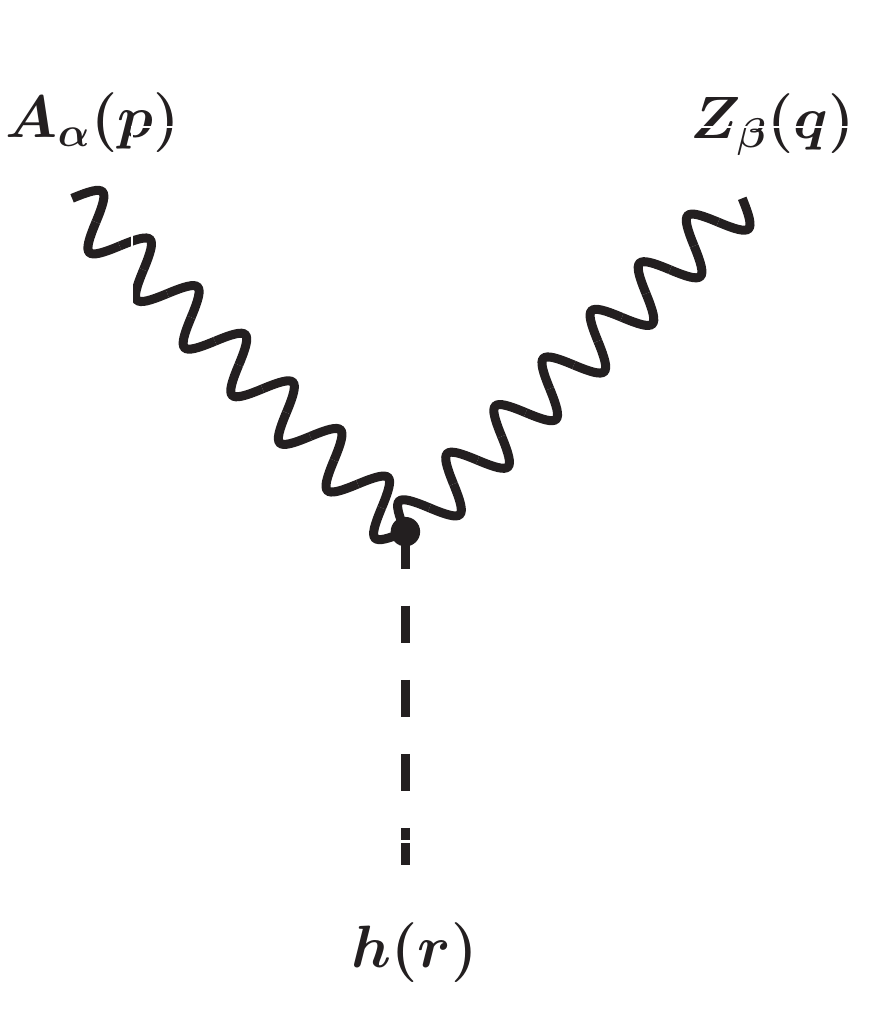}} &   &
$-\frac{4 i }{v}e^2     \left[-\hat{a}_{\tilde{B}} \tan \theta _W-2 \cot \theta _W \left(-\frac{1}{4}\,\hat{a}_{\tilde{W}} + 2\,\hat{a}_8   \right)+\left(2\,\hat{a}_9   +2\,\hat{a}_2 + \hat{a}_3\right) \csc \left(2 \theta _W\right)-4\,\hat{a}_1 \cot \left(2 \theta _W\right)\right]\,p_{\mu} q_{\nu} \epsilon ^{\alpha  \beta  \mu  \nu }$
  \tabularnewline

\end{tabular}

\newpage
\renewcommand{\arraystretch}{5}
\begin{tabular}{c@{\hspace*{5mm}}>{\centering}p{4cm}l@{\hspace*{1cm}}l}
& & \bf SM & \bf Non-SM \\\hline

\nr &  \parbox{6cm}{\includegraphics[scale=.35]{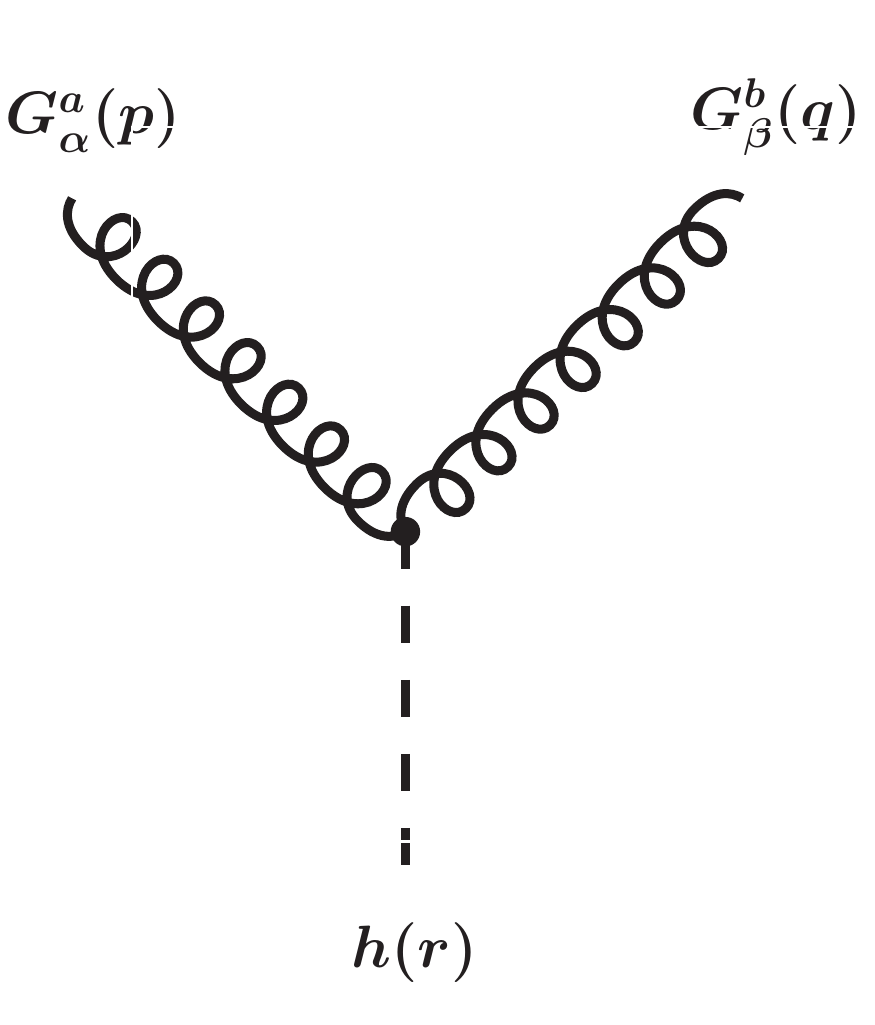}}&   & 
$-\frac{4\,i }{v}\,  \, g_s^2 \,\delta ^{\text{ab}} \, \hat{a}_{\tilde{\mathfrak{G}}}\, p_{\mu} q_{\nu}\epsilon ^{\alpha  \beta  \mu  \nu }$
\tabularnewline

\nr &  \parbox{6cm}{\includegraphics[scale=.35]{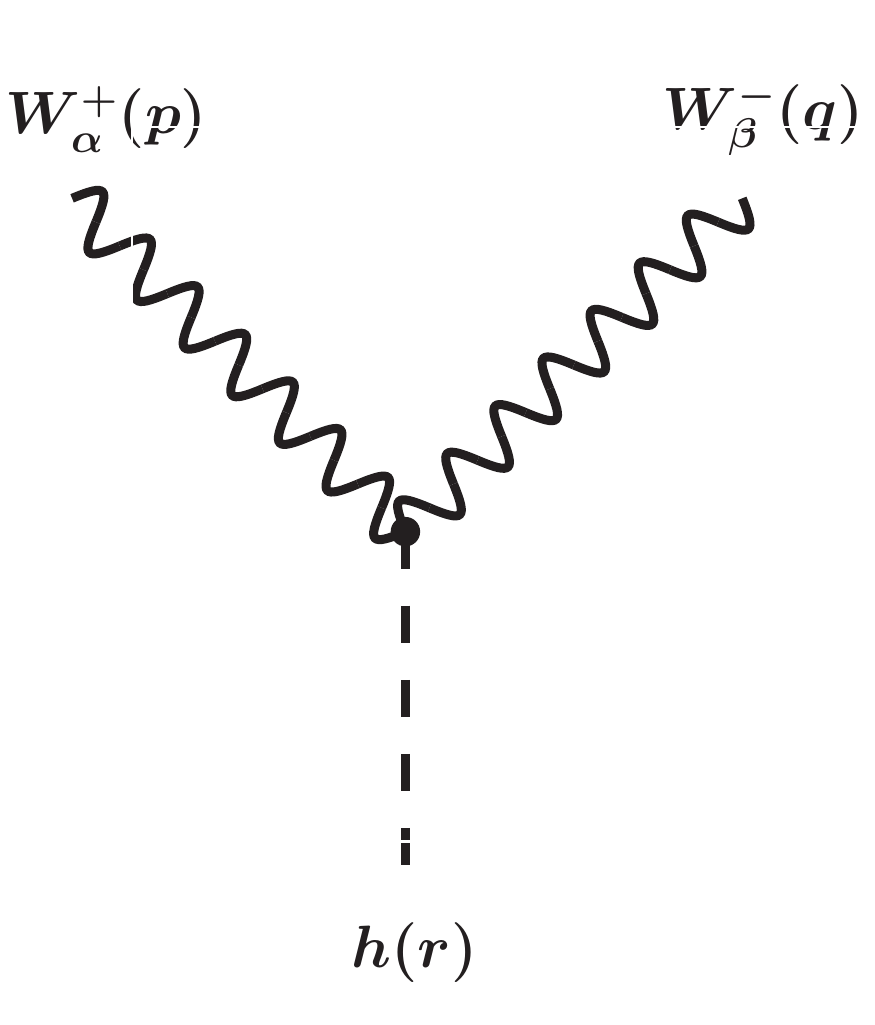}} &   &
$\frac{2}{v}e^2   \csc ^2\theta _W \left[\hat{a}_7\,g^{\alpha \beta } (p\cdot p-q\cdot q)-\left(\hat{a}_7-\hat{a}_{12}\right) \left(p^{\alpha}p^{\beta }-q^{\alpha}q^{\beta }\right)\right] -\frac{4 i}{v}e^2    \csc ^2\theta _W \left(\hat{a}_3+\frac{1}{2}\,\hat{a}_{\tilde{W}}\right)  \,p_{\mu} q_{\nu}\epsilon ^{\alpha  \beta  \mu  \nu }$
  \tabularnewline

\nr &  \parbox{6cm}{\includegraphics[scale=.35]{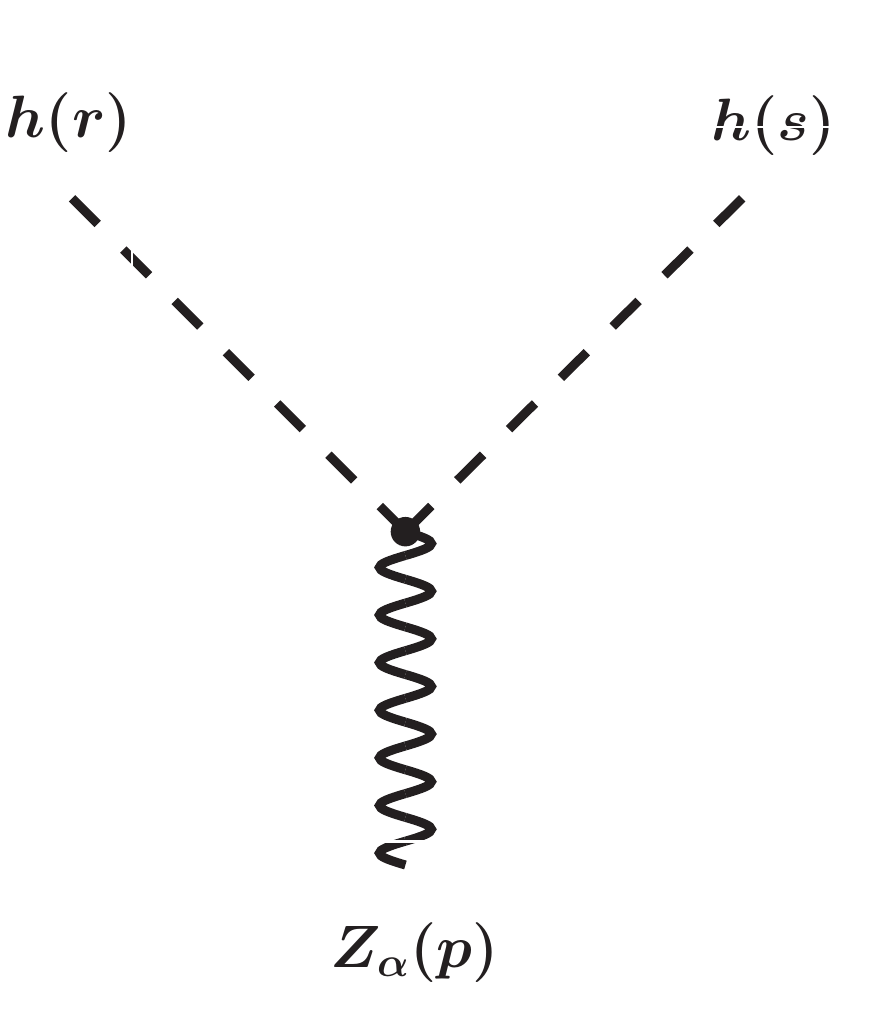}}&   & 
$\frac{2}{v^2}\,e\,\csc \left(2 \theta _W\right)\left\{8\,\hat{a}_{14}\,a'_{14}\,   \left(r\cdot s\right)\, p^{\alpha } + 
v^2\,\left[\left(\hat{a}_{2 D} -\dfrac{\hat{b}_{2D}}{2}\right)\,p^\alpha + \frac{4}{v^2}\,\hat{a}_{13}\,\left(r^2\,r^\alpha + s^2\,s^\alpha\right) + 
\frac{2}{v^2}\,\hat{b}_{13}\,p^2\,p^\alpha\right]\right\} $  

\tabularnewline

\end{tabular}

\newpage
\renewcommand{\arraystretch}{5}
\begin{tabular}{c@{\hspace*{5mm}}>{\centering}p{2cm}l@{\hspace*{1cm}}l}
& & \bf SM & \bf Non-SM \\\hline


\nr &  \parbox{2cm}{\includegraphics[scale=.35]{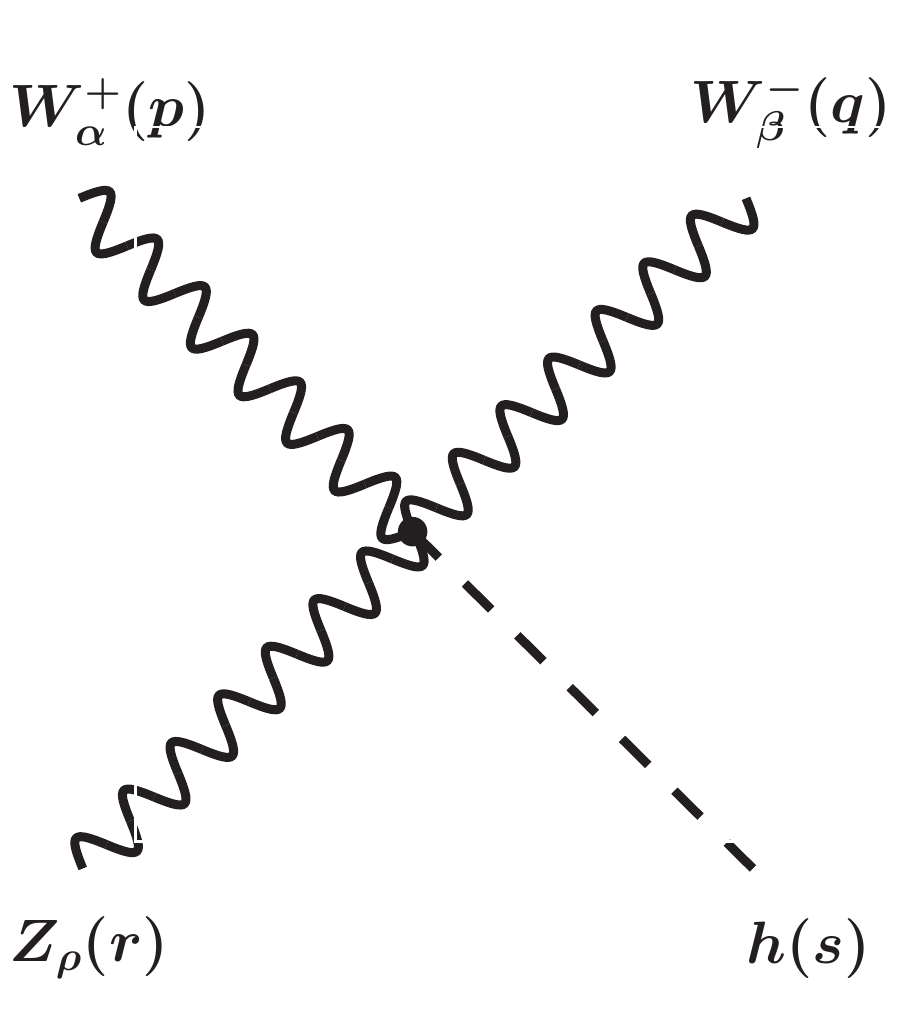}} & & $\begin{array}{l}  
  \frac{1}{v}e^3   \csc ^3\theta _W \sec \theta _W \left(\hat{a}_{10} \,g^{\beta \rho } p^{\alpha } + \hat{a}_7 \left(g^{\beta \rho } \left(p^{\alpha }+q^{\alpha }+r^{\alpha }\right)+g^{\alpha \rho } \left(p^{\beta }+q^{\beta }+r^{\beta }\right)\right)+ \hat{a}_{12} \left(g^{\beta \rho } \left(p^{\alpha }+q^{\alpha }+r^{\alpha }\right)+g^{\alpha \rho } \left(p^{\beta }+q^{\beta }+r^{\beta }\right)\right)+\right. \\[-1cm]
+ \hat{a}_{10} g^{\alpha \rho } q^{\beta }-2\,\hat{a}_6 g^{\alpha \beta } \left(p^{\rho }+q^{\rho }+r^{\rho }\right)-2\,\hat{a}_7 g^{\alpha \beta } \left(p^{\rho }+q^{\rho }+r^{\rho }\right)+2\,\hat{a}_{11}\,g^{\alpha \beta } r^{\rho }+ \\[-1cm]
-\hat{a}_5 \left(g^{\beta \rho } \left(p^{\alpha }+q^{\alpha }+r^{\alpha }\right)+g^{\alpha \rho } \left(p^{\beta }+q^{\beta }+r^{\beta }\right)\right) - \hat{a}_4 \left(g^{\beta \rho } q^{\alpha }+g^{\alpha \rho } p^{\beta }-g^{\alpha \beta } \left(p^{\rho }+q^{\rho }\right)\right) + \\[-1cm]
 \left. + \cos \left(2 \theta _W\right) \left(\left(\hat{a}_7-\hat{a}_{12}\right) \left(g^{\beta \rho } \left(p^{\alpha }+q^{\alpha }+r^{\alpha }\right)+g^{\alpha \rho } \left(p^{\beta }+q^{\beta }+r^{\beta }\right)\right)-2\,\hat{a}_7 g^{\alpha \beta } \left(p^{\rho }+q^{\rho }+r^{\rho }\right)\right)\right) +  \\[-1cm]
+\frac{1}{v}2 i e^3    \csc ^3\theta _W \sec \theta _W  \left\{4 \cos ^2\theta _W \left[\frac{1}{4}\,\hat{a}_{\tilde{W}} \left(p_{\sigma }+q_{\sigma }+r_{\sigma }\right]-2\,\hat{a}_8    r_{\sigma }\right) + 2\,\hat{a}_9    \left(p_{\sigma }+q_{\sigma }+r_{\sigma }\right)+ \right.  \\[-1cm]
\phantom{+\frac{1}{v}2 i e^3    \csc ^3\theta _W \sec \theta _W} \left. + \hat{a}_3 \left[\cos \left(2 \theta _W\right)+2\right] \left(p_{\sigma }+q_{\sigma }+r_{\sigma }\right) + 4\,\hat{a}_1 r_{\sigma } \sin ^2\theta _W\right\}\,\epsilon ^{\alpha  \beta  \rho  \sigma }
  \end{array}$
\tabularnewline

\nr& \parbox{2cm}{\includegraphics[scale=.35]{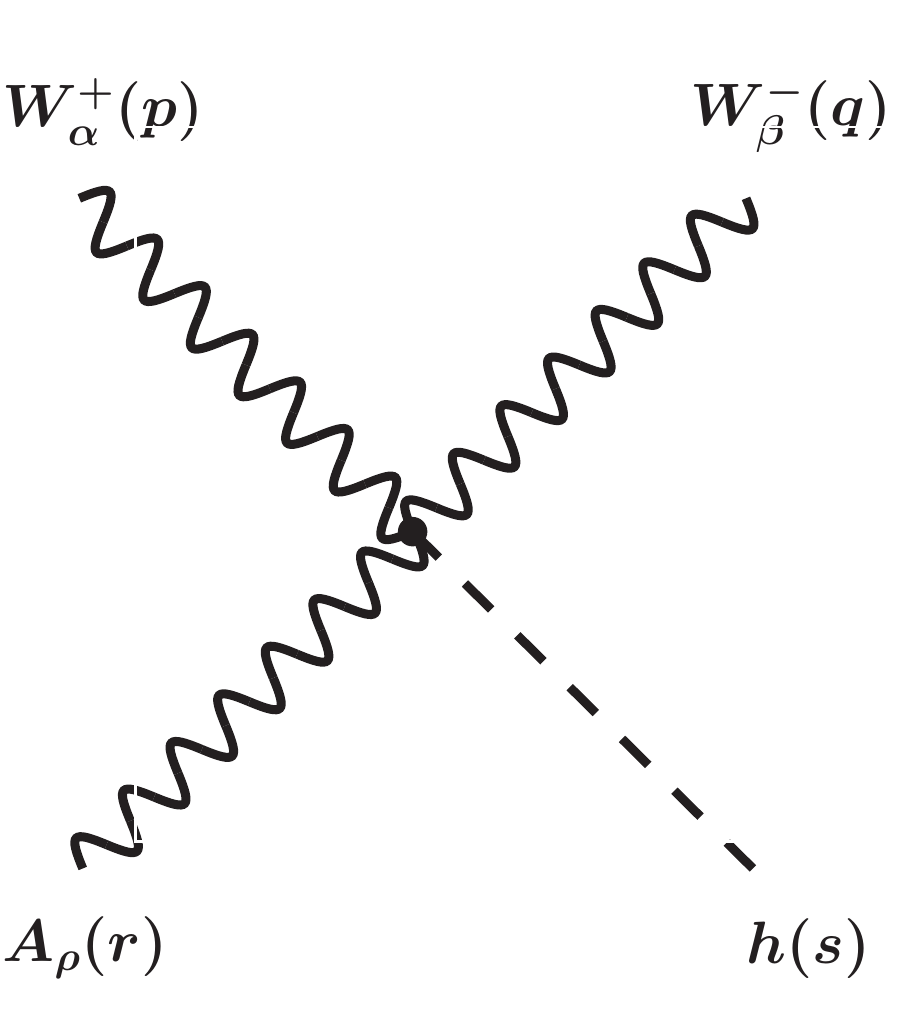}} & & $\begin{array}{l}  
\frac{2}{v}e^3   \csc ^2\theta _W \left\{\left(\hat{a}_7-\hat{a}_{12}\right) \left[g^{\beta \rho } \left(p^{\alpha }+q^{\alpha }+r^{\alpha }\right)+g^{\alpha \rho } \left(p^{\beta }+q^{\beta }+r^{\beta }\right)\right] - 2\,\hat{a}_7 g^{\alpha \beta } \left(p^{\rho }+q^{\rho }+r^{\rho }\right)\right\}
+ \\[-1cm]
 -\frac{4 i }{v}e^3    \csc ^2\theta _W  \left[-\frac{1}{2}\,\hat{a}_{\tilde{W}} \left(p_{\sigma }+q_{\sigma }+r_{\sigma }\right)-\hat{a}_3 \left(p_{\sigma }+q_{\sigma }+r_{\sigma }\right)+2 \left(2\,\hat{a}_8   + \hat{a}_1\right) r_{\sigma }\right]\,\epsilon ^{\alpha  \beta  \rho  \sigma }
  \end{array}$\\

\nr& \parbox{2cm}{\includegraphics[scale=.35]{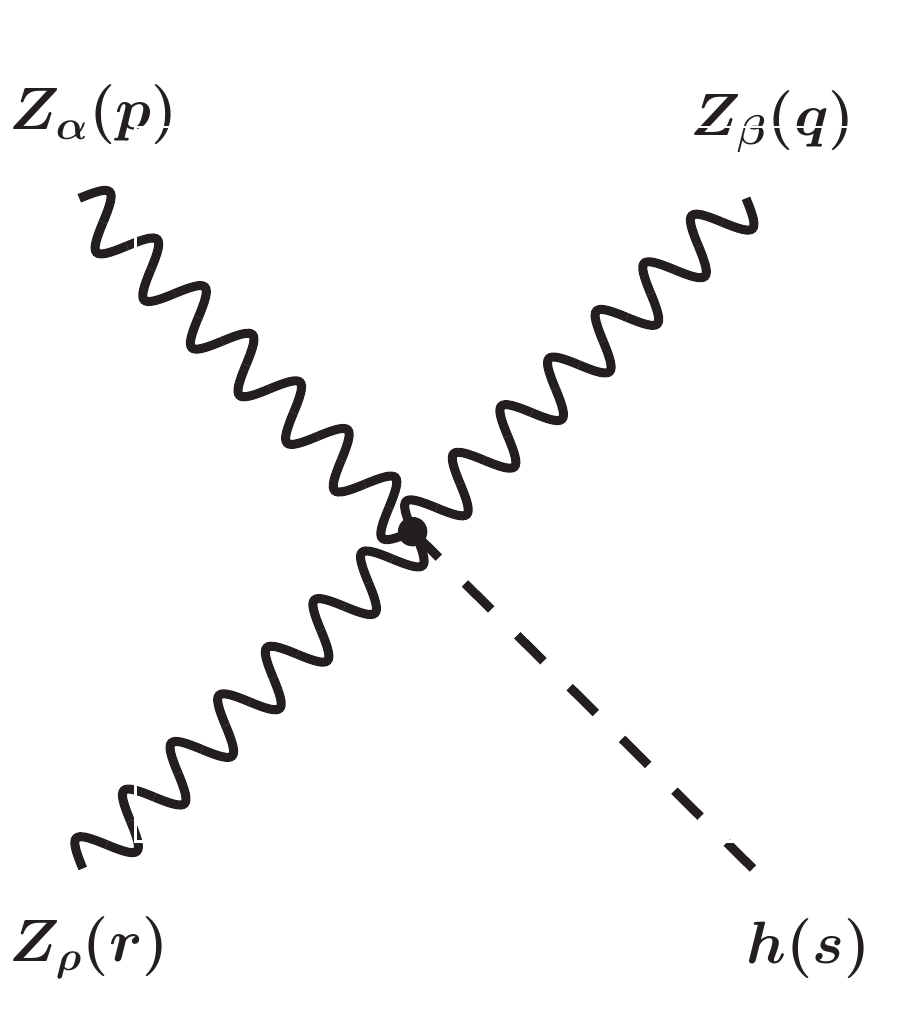}} & & $\begin{array}{l}  
\frac{1}{v}16 e^3   \csc ^3\left(2 \theta _W\right) \left\{\left(2\,\hat{a}_{16}   + \hat{a}_{10} + \hat{a}_{11}\right) \left(g^{\beta \rho } p^{\alpha }+g^{\alpha \rho } q^{\beta }+g^{\alpha \beta } r^{\rho }\right)+\right.
 \\[-1cm]
 \phantom{-\frac{1}{v}8 i e^3   \csc ^3\left(2 \theta _W\right)}
\left. -\left(2\,\hat{a}_{15}   + \hat{a}_5 + \hat{a}_6\right) \left[g^{\beta \rho } \left(p^{\alpha }+q^{\alpha }+r^{\alpha }\right)+g^{\alpha \rho } \left(p^{\beta }+q^{\beta }+r^{\beta }\right)+g^{\alpha \beta } \left(p^{\rho }+q^{\rho }+r^{\rho }\right)\right]\right\}
  \end{array}$\\

\end{tabular}

\newpage
\renewcommand{\arraystretch}{5}
\begin{tabular}{c@{\hspace*{5mm}}>{\centering}p{3cm}l@{\hspace*{1cm}}l}
& & \bf SM & \bf Non-SM \\\hline

   
\nr& \parbox{3cm}{\includegraphics[scale=.35]{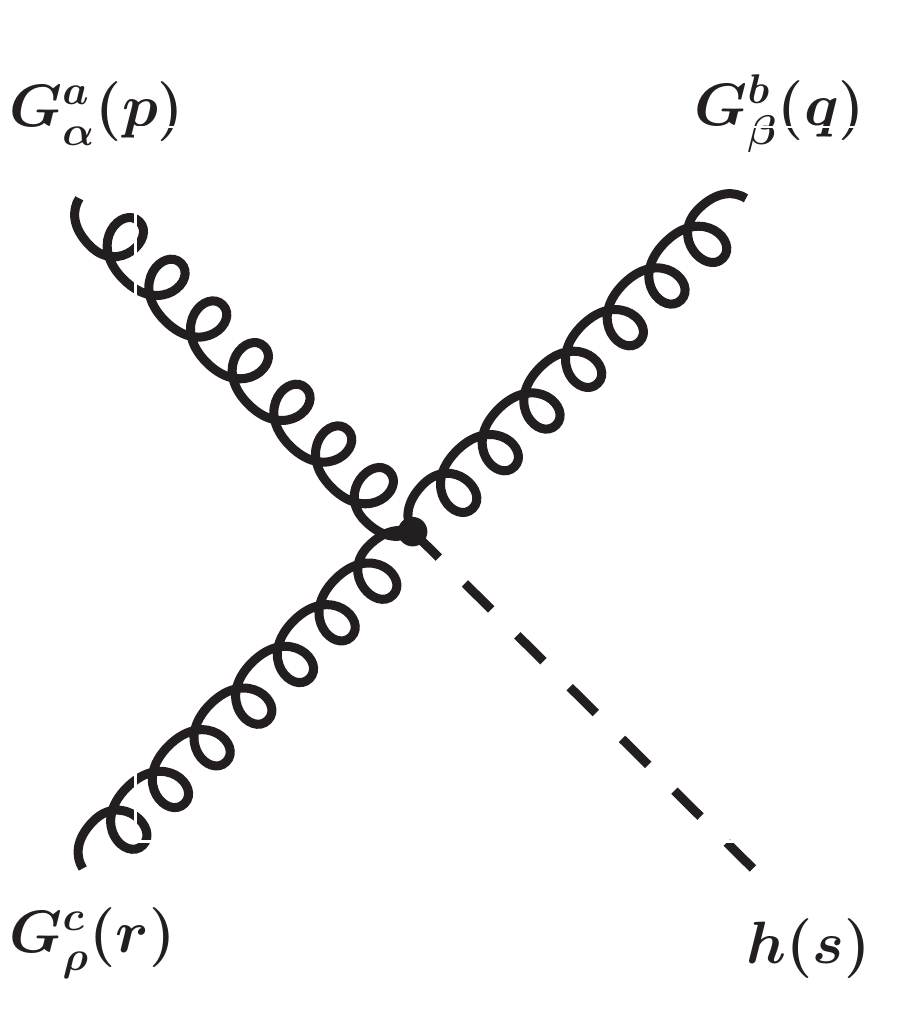}} & &
$\frac{4}{v}    f^{\text{abc}} g_s^3\,\hat{a}_{\tilde{\mathfrak{G}}} \left(p_{\mu }+q_{\mu }+r_{\mu }\right) \epsilon ^{\alpha  \beta  \mu  \rho }
$ \\

\nr& \parbox{3cm}{\includegraphics[scale=.35]{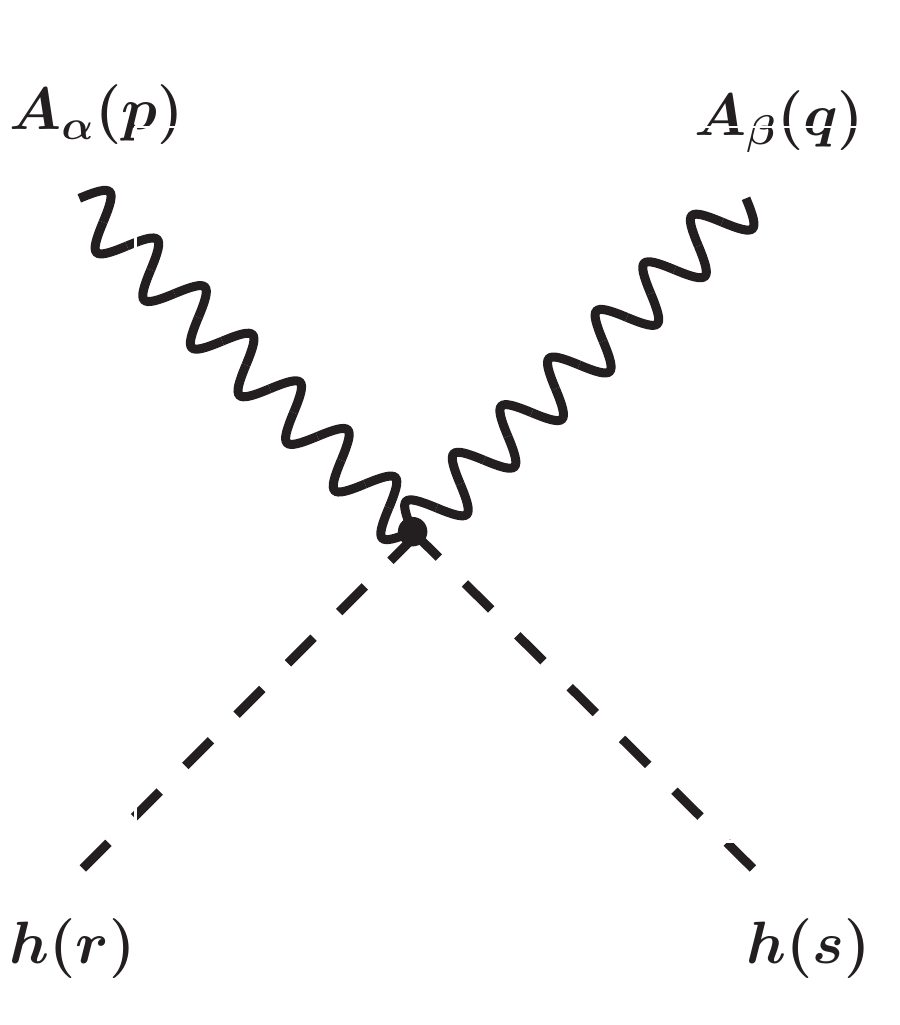}} & & $\frac{8 i}{v^2}e^2     \left[2 \left(-\frac{1}{4}\,\hat{b}_{\tilde{B}} + \hat{b}_8   + \hat{b}_1\right) -\frac{1}{4}\,\hat{b}_{\tilde{W}}\right] p_{\mu} q_{\nu}\epsilon ^{\alpha  \beta  \mu  \nu }$\\

\nr& \parbox{3cm}{\includegraphics[scale=.35]{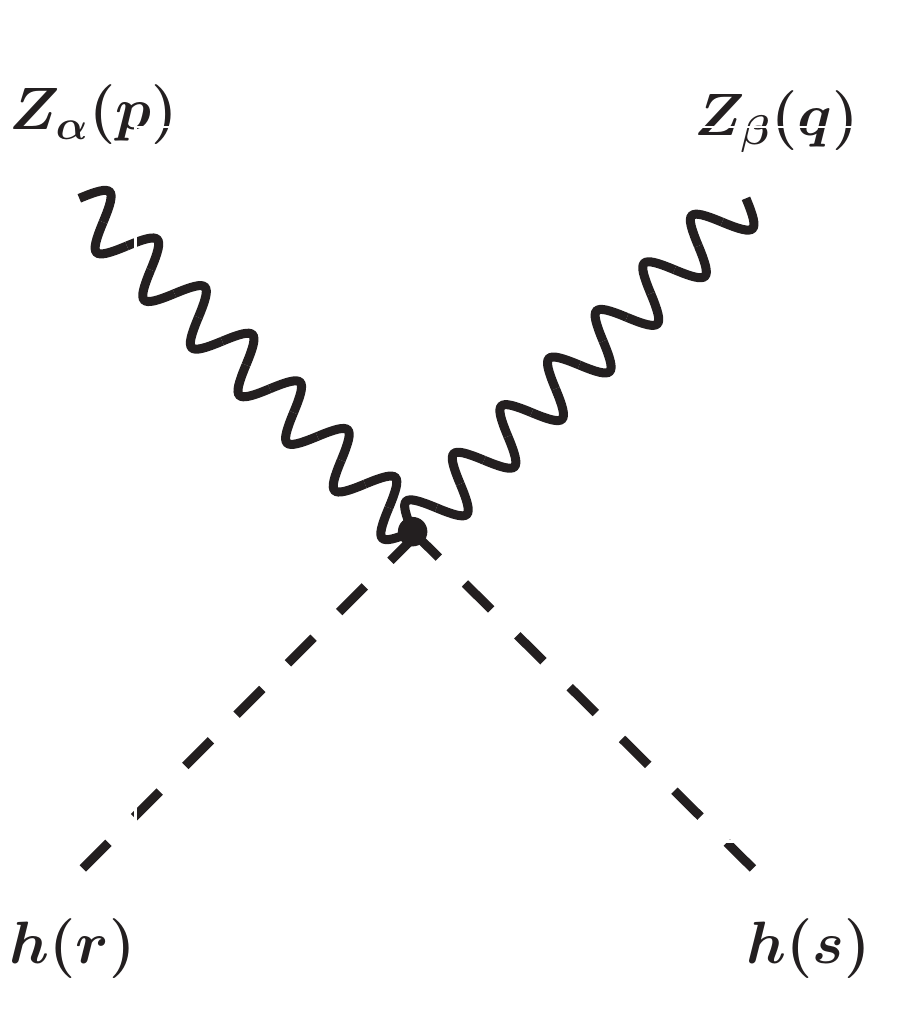}} & & $-\frac{4 i}{v^2}e^2    \left[4 \left(-\frac{1}{4}\,\hat{b}_{\tilde{B}} + \hat{b}_8   + \hat{b}_1\right)-2 \left(-\frac{1}{2}\,\hat{b}_{\tilde{B}} + \hat{b}_2\right) \sec ^2\theta _W + \csc ^2\theta _W \left(\frac{1}{2}\,\hat{b}_{\tilde{W}}-4\,\hat{b}_8   + 2\,\hat{b}_9   + \hat{b}_3\right)-\frac{1}{2}\,\hat{b}_{\tilde{W}}\right]\,p_{\mu} q_{\nu} \epsilon ^{\alpha  \beta  \mu  \nu }$\\

\nr& \parbox{3cm}{\includegraphics[scale=.35]{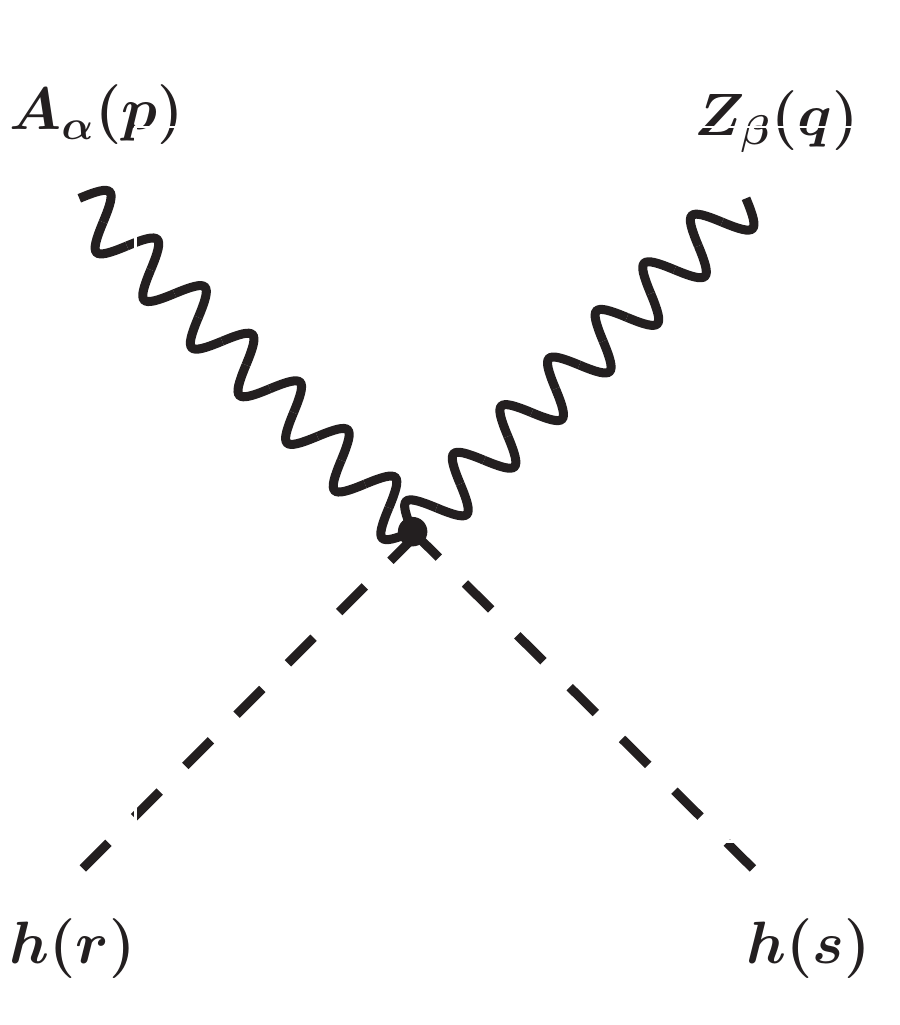}} & & $-\frac{4 i}{v^2}e^2     \left[-\hat{b}_{\tilde{B}} \tan \theta _W-2 \cot \theta _W \left(-\frac{1}{4}\,\hat{b}_{\tilde{W}}+2\,\hat{b}_8   \right)+\left(2\,\hat{b}_9   +2\,\hat{b}_2+ \hat{b}_3\right) \csc \left(2 \theta _W\right)-4\,\hat{b}_1 \cot \left(2 \theta _W\right)\right]\,p_{\mu} q_{\nu} \epsilon ^{\alpha  \beta  \mu  \nu }$\\

\end{tabular}

\newpage
\renewcommand{\arraystretch}{5}
\begin{tabular}{c@{\hspace*{5mm}}>{\centering}p{4cm}l@{\hspace*{1cm}}l}
& & \bf SM & \bf Non-SM \\\hline

\nr& \parbox{3cm}{\includegraphics[scale=.35]{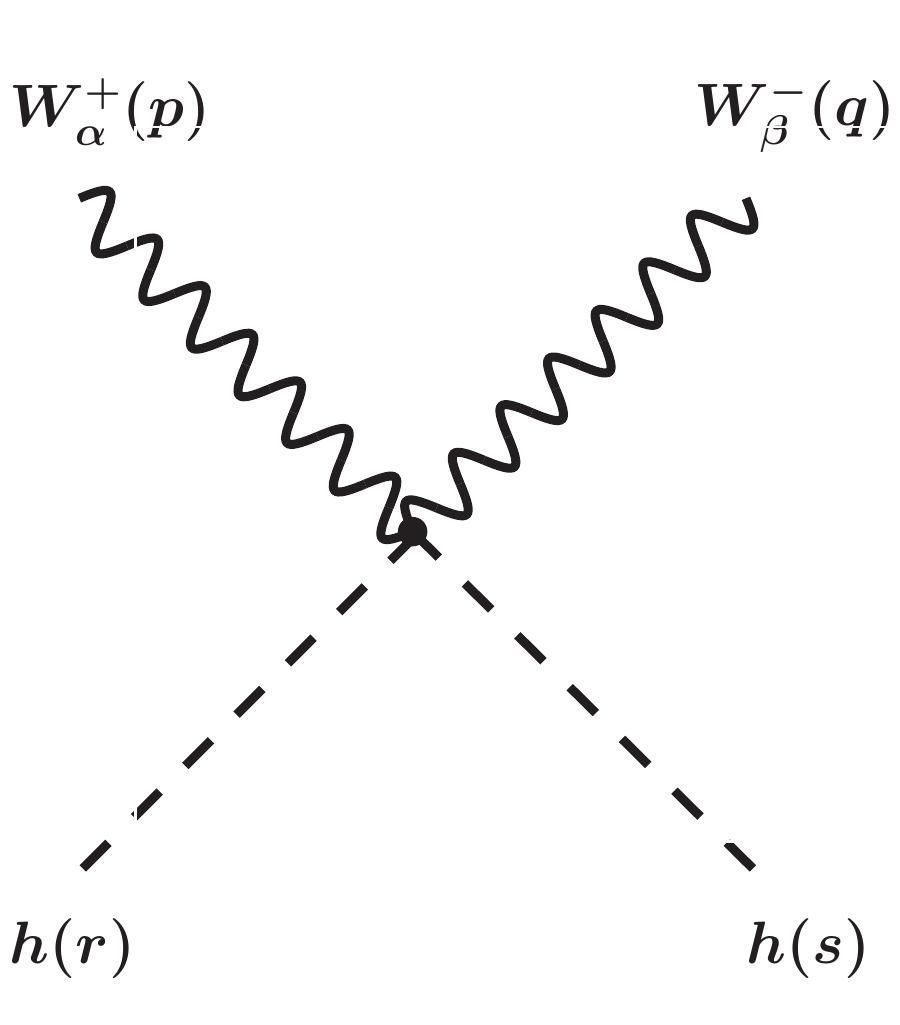}} & &
$\frac{2}{v^2} e^2   \csc ^2\theta _W \left\{\hat{b}_7 \left[g^{\alpha \beta } (q\cdot r-p\cdot r)+p^{\beta } r^{\alpha }-q^{\alpha } r^{\beta }\right]+\hat{b}_{12} \left(q^{\beta } r^{\alpha }-p^{\alpha } r^{\beta }\right)\right\}  - \frac{2 i }{v^2}e^2   \left(\hat{b}_3+\,\hat{b}_{\tilde{W}}\right) \csc ^2\theta _W\,p_{\mu } q_{\nu }\, \epsilon ^{\alpha  \beta  \mu  \nu }$ \\

\nr& \parbox{3cm}{\includegraphics[scale=.35]{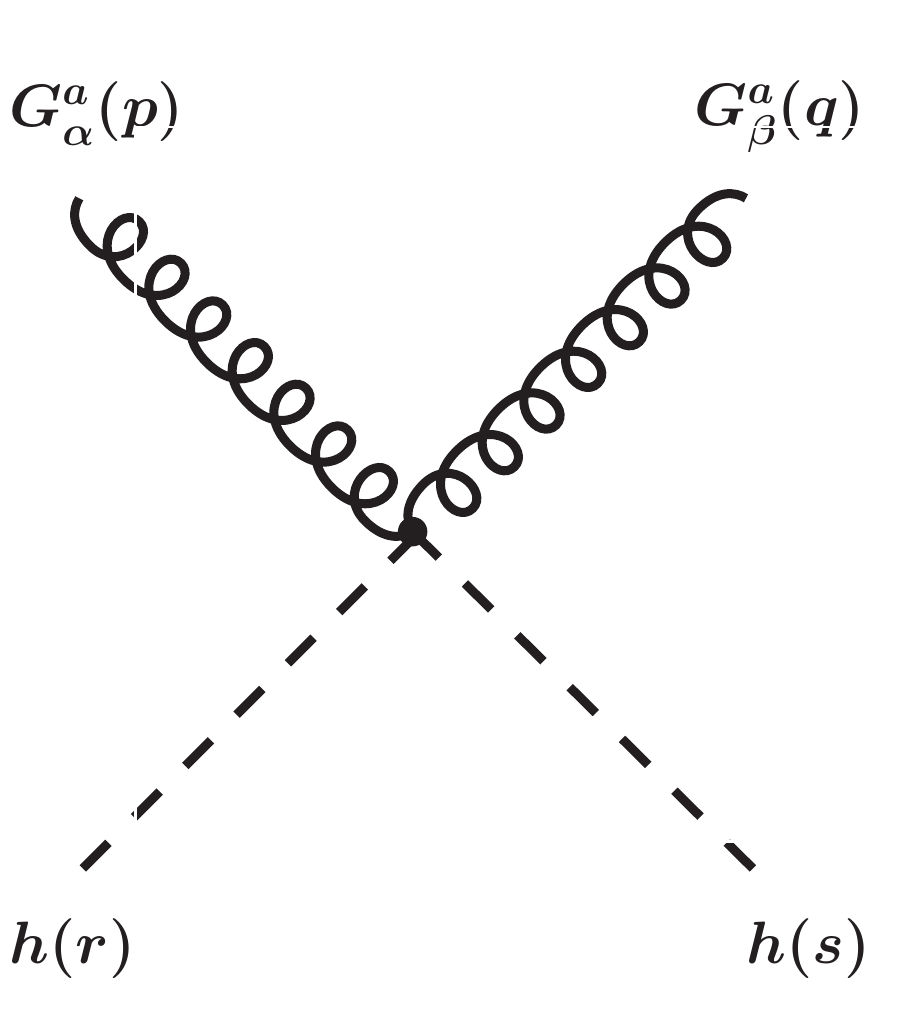}} & & $-\frac{4\,i}{v^2}\,  \, g_s^2 \, \delta ^{\text{ab}}\,\hat{b}_{\tilde{\mathfrak{G}}}\, p_{\mu} q_{\nu}\epsilon ^{\alpha  \beta  \mu  \nu }$\\

\nr& \parbox{3cm}{\includegraphics[scale=.35]{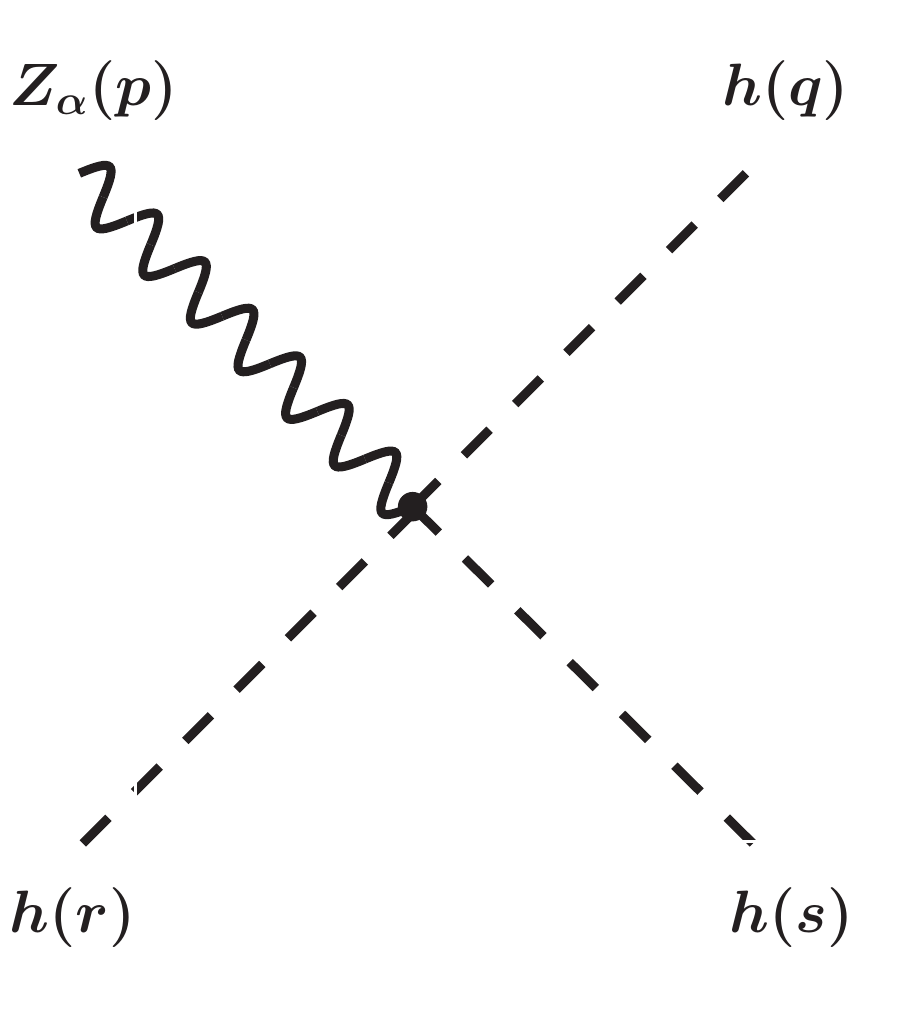}} & & 
$\frac{2}{v^3}\,e\,\csc \left(2 \theta _W\right)\left\{8\,p^{\alpha }\,(r\cdot q+q\cdot s+r\cdot s) \left(\hat{a}_{14}\,b'_{14} + \hat{b}_{14}\, a'_{14}\right) + v^2\,\left[\hat{a}_{2D}\,p^\alpha + \frac{4}{v^2}\,\hat{a}_{13}\,\left(q^2\,q^\alpha  + r^2\,r^\alpha + s^2\,s^\alpha\right)\right]\right\}$

\\

\end{tabular}

\newpage
\renewcommand{\arraystretch}{5}
\begin{tabular}{c@{\hspace*{5mm}}>{\centering}p{5cm}l@{\hspace*{2cm}}l}
& & \bf SM & \bf Non-SM \\\hline


\nr&  \parbox{2.5cm}{\includegraphics[scale=.35]{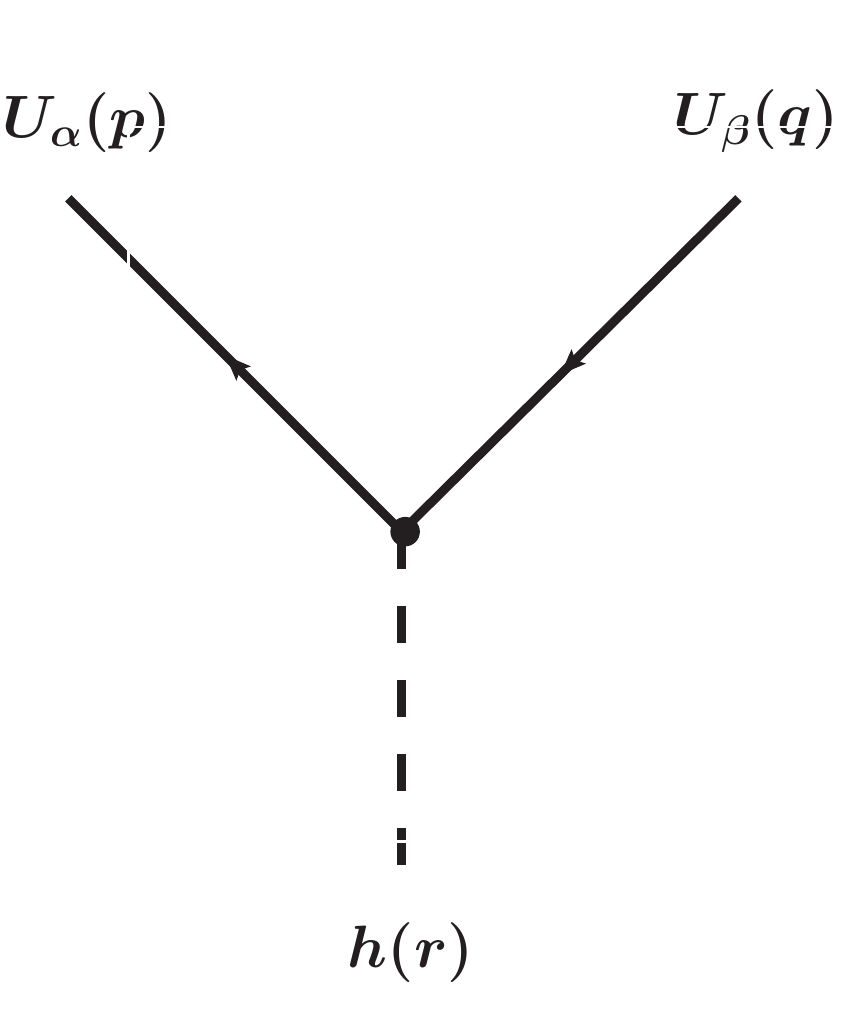}} & $-\frac{i}{\sqrt{2}}\left[P_L \left(Y^\dagger_U\right)_{\alpha\beta} + P_R \left(Y_U\right)_{\alpha\beta}\right]$ & $+\frac{1}{\sqrt{2}\,v^2}\left(\hat{a}_{2 D}\,v^2-4\,\hat{a}_{13}\,r^2 \right)\left[P_L \left(Y^\dagger_U\right)_{\alpha\beta} - P_R \left(Y_U\right)_{\alpha\beta}\right]$ \\
  
\nr&  \parbox{2.5cm}{\includegraphics[scale=.35]{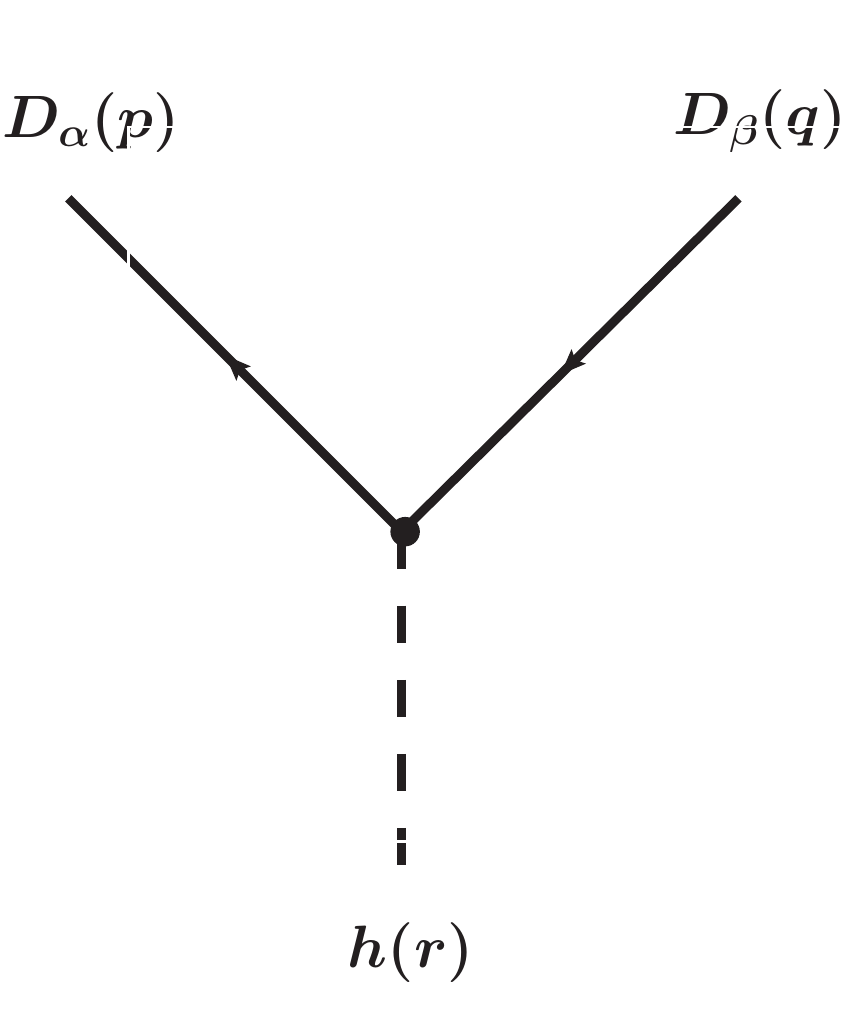}} & $-\frac{i}{\sqrt{2}}\left[P_L \left(Y^\dagger_D\right)_{\alpha\beta} + P_R \left(Y_D\right)_{\alpha\beta}\right]$ &  $-\frac{1}{\sqrt{2}\,v^2}\left(\hat{a}_{2 D}\,v^2-4\,\hat{a}_{13}\,r^2 \right)\left[P_L \left(Y^\dagger_D\right)_{\alpha\beta} - P_R \left(Y_D\right)_{\alpha\beta}\right]$\\
  
\nr&  \parbox{2.5cm}{\includegraphics[scale=.35]{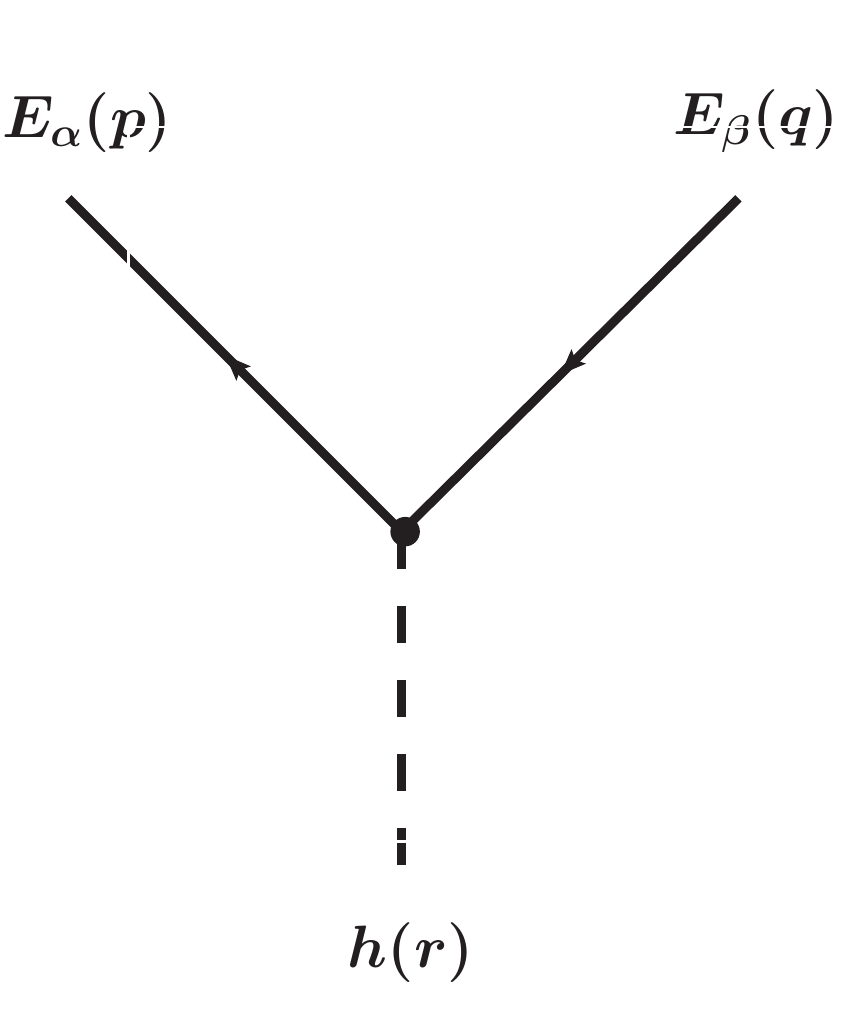}} & $-\frac{i}{\sqrt{2}}\left[P_L \left(Y^\dagger_E\right)_{\alpha\beta} + P_R \left(Y_E\right)_{\alpha\beta}\right]$ & $-\frac{1}{\sqrt{2}\,v^2}\left(\hat{a}_{2 D}\,v^2-4\,\hat{a}_{13}\,r^2 \right)\left[P_L \left(Y^\dagger_E\right)_{\alpha\beta} - P_R \left(Y_E\right)_{\alpha\beta}\right]$\\

\end{tabular}

\newpage
\renewcommand{\arraystretch}{5}
\begin{tabular}{c@{\hspace*{5mm}}>{\centering}p{5cm}l@{\hspace*{2cm}}l}
& & \bf SM & \bf Non-SM \\\hline


\nr&  \parbox{2.5cm}{\includegraphics[scale=.35]{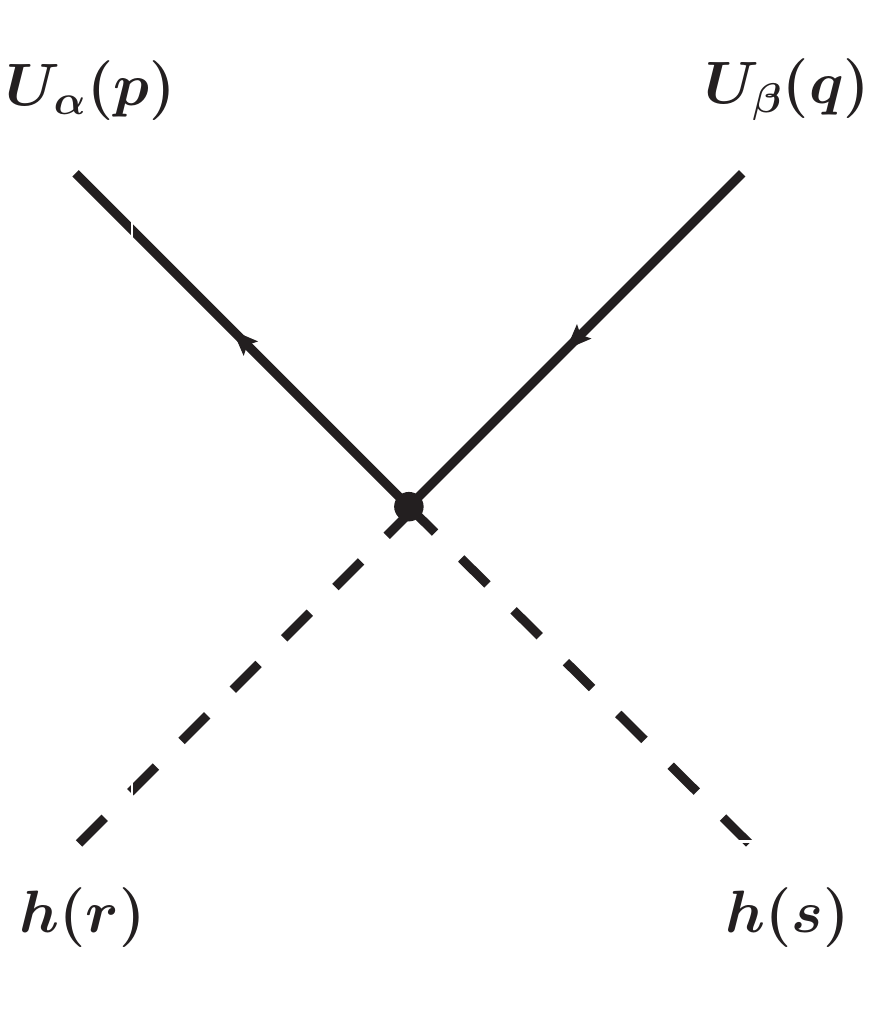}} &  & $+\frac{\sqrt{2}}{v^3}\,\left[\hat{a}_{2 D}\,v^2-2\, \hat{a}_{13}\,\left(r^2+s^2\right)\right]\left[P_L\left(Y^\dagger_U\right)_{\alpha\beta} - P_R \left(Y_U\right)_{\alpha\beta}\right]$ \\
  
\nr&  \parbox{2.5cm}{\includegraphics[scale=.35]{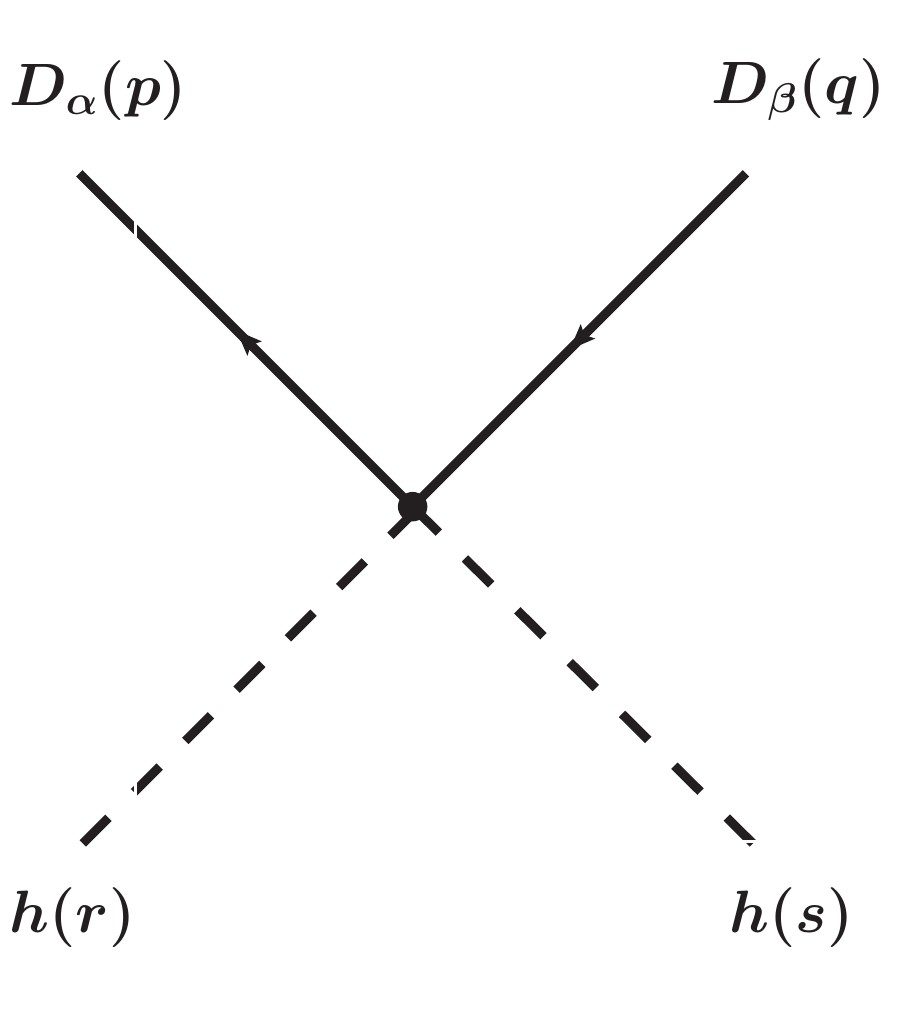}} &  &  $-\frac{\sqrt{2}}{v^3}\, \left[\hat{a}_{2 D}\,v^2-2\,\hat{a}_{13}\,\left(r^2+s^2\right)\right]\left[P_L\left(Y^\dagger_D\right)_{\alpha\beta} - P_R \left(Y_D\right)_{\alpha\beta}\right]$\\
  
\nr&  \parbox{2.5cm}{\includegraphics[scale=.35]{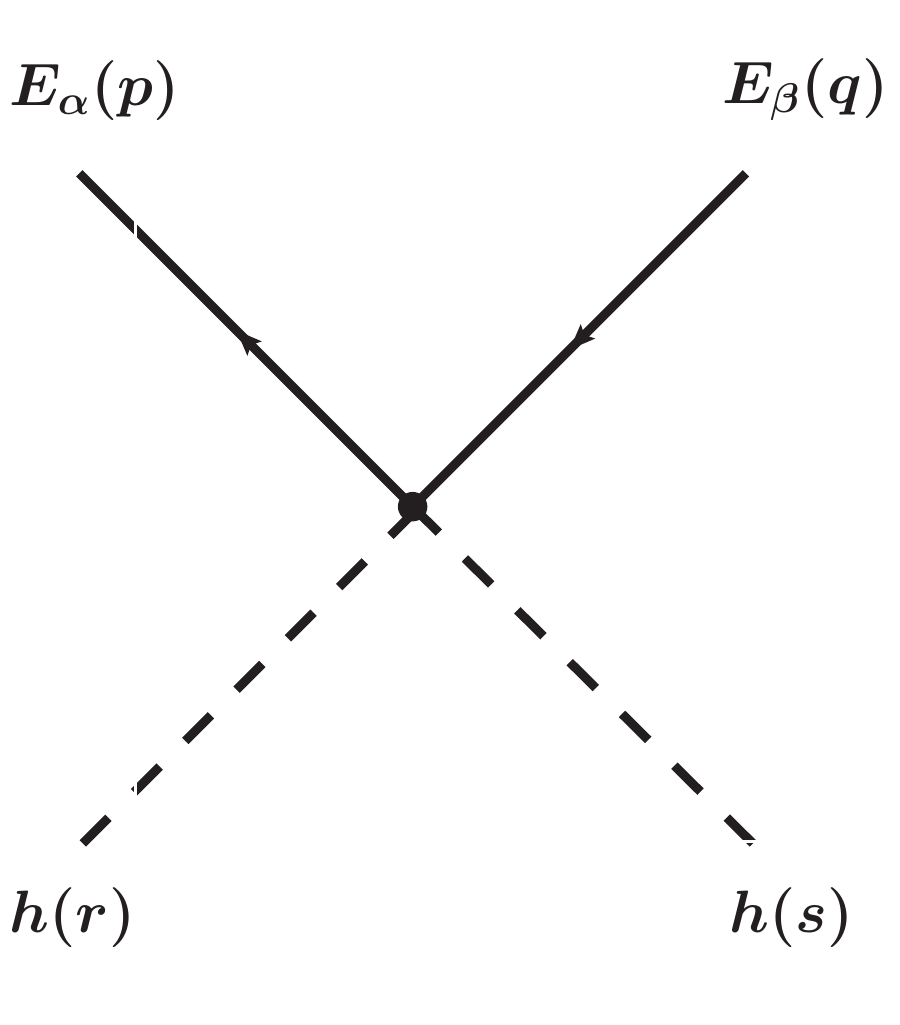}} &  & $-\frac{\sqrt{2}}{v^3}\, \left[\hat{a}_{2 D}\,v^2-2\,\hat{a}_{13}\,\left(r^2+s^2\right)\right]\left[P_L\left(Y^\dagger_E\right)_{\alpha\beta} - P_R \left(Y_E\right)_{\alpha\beta}\right]$\\

\end{tabular}

\end{center}

\end{landscape}

\normalsize

\begin{thebibliography}{100}

\bibitem{CMS:yva}
{\bf CMS} Collaboration, {\it {Combination of standard model Higgs boson
  searches and measurements of the properties of the new boson with a mass near
  125 GeV}},  CMS-PAS-HIG-13-005 (2013).

\bibitem{ATLAScouplings}
{\bf ATLAS} Collaboration, {\it {Updated coupling measurements of the Higgs
  boson with the ATLAS detector using up to 25 fb$^{-1}$ of proton-proton
  collision data}},  ATLAS-CONF-2014-009, ATLAS-COM-CONF-2014-013 (2014).

\bibitem{Aad:2013wqa}
{\bf ATLAS} Collaboration, G.~Aad {\em et.~al.}, {\it {Measurements of Higgs
  Boson Production and Couplings in Diboson Final States with the Atlas
  Detector at the Lhc}},  Phys.Lett. {\bf B726} (2013) 88--119,
  [\href{http://xxx.lanl.gov/abs/1307.1427}{{\tt arXiv:1307.1427}}].

\bibitem{Chatrchyan:2013mxa}
{\bf CMS} Collaboration, S.~Chatrchyan {\em et.~al.}, {\it {Measurement of the
  properties of a Higgs boson in the four-lepton final state}},  Phys.Rev. {\bf
  D89} (2014) 092007, [\href{http://xxx.lanl.gov/abs/1312.5353}{{\tt
  arXiv:1312.5353}}].

\bibitem{Englert:1964et}
F.~Englert and R.~Brout, {\it {Broken Symmetry and the Mass of Gauge Vector
  Mesons}},  Phys.Rev.Lett. {\bf 13} (1964) 321--323.

\bibitem{Higgs:1964ia}
P.~W. Higgs, {\it {Broken Symmetries, Massless Particles and Gauge Fields}},
  Phys.Lett. {\bf 12} (1964) 132--133.

\bibitem{Higgs:1964pj}
P.~W. Higgs, {\it {Broken Symmetries and the Masses of Gauge Bosons}},
  Phys.Rev.Lett. {\bf 13} (1964) 508--509.

\bibitem{Buchmuller:1985jz}
W.~Buchmuller and D.~Wyler, {\it {Effective Lagrangian Analysis of New
  Interactions and Flavor Conservation}},  Nucl.Phys. {\bf B268} (1986) 621.

\bibitem{Grzadkowski:2010es}
B.~Grzadkowski, M.~Iskrzynski, M.~Misiak, and J.~Rosiek, {\it {Dimension-Six
  Terms in the Standard Model Lagrangian}},  JHEP {\bf 1010} (2010) 085,
  [\href{http://xxx.lanl.gov/abs/1008.4884}{{\tt arXiv:1008.4884}}].

\bibitem{Hagiwara:1993ck}
K.~Hagiwara, S.~Ishihara, R.~Szalapski, and D.~Zeppenfeld, {\it {Low-Energy
  Effects of New Interactions in the Electroweak Boson Sector}},  Phys.Rev.
  {\bf D48} (1993) 2182--2203.

\bibitem{Hagiwara:1996kf}
K.~Hagiwara, T.~Hatsukano, and S.~S.~R. Ishihara, {\it {Probing nonstandard
  bosonic interactions via W boson pair production at lepton colliders}},
  Nucl.Phys. {\bf B496} (1997) 66--102,
  [\href{http://xxx.lanl.gov/abs/hep-ph/9612268}{{\tt hep-ph/9612268}}].

\bibitem{Hagiwara:1993qt}
K.~Hagiwara, R.~Szalapski, and D.~Zeppenfeld, {\it {Anomalous Higgs boson
  production and decay}},  Phys.Lett. {\bf B318} (1993) 155--162,
  [\href{http://xxx.lanl.gov/abs/hep-ph/9308347}{{\tt hep-ph/9308347}}].

\bibitem{GonzalezGarcia:1999fq}
M.~Gonzalez-Garcia, {\it {Anomalous Higgs couplings}},  Int.J.Mod.Phys. {\bf
  A14} (1999) 3121--3156, [\href{http://xxx.lanl.gov/abs/hep-ph/9902321}{{\tt
  hep-ph/9902321}}].

\bibitem{Low:2012rj}
I.~Low, J.~Lykken, and G.~Shaughnessy, {\it {Have We Observed the Higgs
  (Imposter)?}},  Phys.Rev. {\bf D86} (2012) 093012,
  [\href{http://xxx.lanl.gov/abs/1207.1093}{{\tt arXiv:1207.1093}}].

\bibitem{Corbett:2012dm}
T.~Corbett, O.~Eboli, J.~Gonzalez-Fraile, and M.~Gonzalez-Garcia, {\it
  {Constraining anomalous Higgs interactions}},  Phys.Rev. {\bf D86} (2012)
  075013, [\href{http://xxx.lanl.gov/abs/1207.1344}{{\tt arXiv:1207.1344}}].

\bibitem{Ellis:2012hz}
J.~Ellis and T.~You, {\it {Global Analysis of the Higgs Candidate with Mass ~
  125 GeV}},  JHEP {\bf 1209} (2012) 123,
  [\href{http://xxx.lanl.gov/abs/1207.1693}{{\tt arXiv:1207.1693}}].

\bibitem{Giardino:2012dp}
P.~P. Giardino, K.~Kannike, M.~Raidal, and A.~Strumia, {\it {Is the Resonance
  at 125 GeV the Higgs Boson?}},  Phys.Lett. {\bf B718} (2012) 469--474,
  [\href{http://xxx.lanl.gov/abs/1207.1347}{{\tt arXiv:1207.1347}}].

\bibitem{Montull:2012ik}
M.~Montull and F.~Riva, {\it {Higgs Discovery: the Beginning Or the End of
  Natural Ewsb?}},  JHEP {\bf 1211} (2012) 018,
  [\href{http://xxx.lanl.gov/abs/1207.1716}{{\tt arXiv:1207.1716}}].

\bibitem{Espinosa:2012im}
J.~Espinosa, C.~Grojean, M.~Muhlleitner, and M.~Trott, {\it {First Glimpses at
  Higgs' Face}},  JHEP {\bf 1212} (2012) 045,
  [\href{http://xxx.lanl.gov/abs/1207.1717}{{\tt arXiv:1207.1717}}].

\bibitem{Carmi:2012in}
D.~Carmi, A.~Falkowski, E.~Kuflik, T.~Volansky, and J.~Zupan, {\it {Higgs After
  the Discovery: a Status Report}},  JHEP {\bf 1210} (2012) 196,
  [\href{http://xxx.lanl.gov/abs/1207.1718}{{\tt arXiv:1207.1718}}].

\bibitem{Banerjee:2012xc}
S.~Banerjee, S.~Mukhopadhyay, and B.~Mukhopadhyaya, {\it {New Higgs
  Interactions and Recent Data from the Lhc and the Tevatron}},  JHEP {\bf
  1210} (2012) 062, [\href{http://xxx.lanl.gov/abs/1207.3588}{{\tt
  arXiv:1207.3588}}].

\bibitem{Bonnet:2012nm}
F.~Bonnet, T.~Ota, M.~Rauch, and W.~Winter, {\it {Interpretation of Precision
  Tests in the Higgs Sector in Terms of Physics Beyond the Standard Model}},
  Phys.Rev. {\bf D86} (2012) 093014,
  [\href{http://xxx.lanl.gov/abs/1207.4599}{{\tt arXiv:1207.4599}}].

\bibitem{Plehn:2012iz}
T.~Plehn and M.~Rauch, {\it {Higgs Couplings After the Discovery}},
  Europhys.Lett. {\bf 100} (2012) 11002,
  [\href{http://xxx.lanl.gov/abs/1207.6108}{{\tt arXiv:1207.6108}}].

\bibitem{Djouadi:2012rh}
A.~Djouadi, {\it {Precision Higgs Coupling Measurements at the LHC Through
  Ratios of Production Cross Sections}},
  \href{http://xxx.lanl.gov/abs/1208.3436}{{\tt arXiv:1208.3436}}.

\bibitem{Batell:2012ca}
B.~Batell, S.~Gori, and L.-T. Wang, {\it {Higgs Couplings and Precision
  Electroweak Data}},  JHEP {\bf 1301} (2013) 139,
  [\href{http://xxx.lanl.gov/abs/1209.6382}{{\tt arXiv:1209.6382}}].

\bibitem{Moreau:2012da}
G.~Moreau, {\it {Constraining Extra-Fermion(S) from the Higgs Boson Data}},
  Phys.Rev. {\bf D87} (2013) 015027,
  [\href{http://xxx.lanl.gov/abs/1210.3977}{{\tt arXiv:1210.3977}}].

\bibitem{Cacciapaglia:2012wb}
G.~Cacciapaglia, A.~Deandrea, G.~D. La~Rochelle, and J.-B. Flament, {\it {Higgs
  Couplings Beyond the Standard Model}},  JHEP {\bf 1303} (2013) 029,
  [\href{http://xxx.lanl.gov/abs/1210.8120}{{\tt arXiv:1210.8120}}].

\bibitem{Azatov:2012qz}
A.~Azatov and J.~Galloway, {\it {Electroweak Symmetry Breaking and the Higgs
  Boson: Confronting Theories at Colliders}},  Int.J.Mod.Phys. {\bf A28} (2013)
  1330004, [\href{http://xxx.lanl.gov/abs/1212.1380}{{\tt arXiv:1212.1380}}].

\bibitem{Masso:2012eq}
E.~Masso and V.~Sanz, {\it {Limits on Anomalous Couplings of the Higgs to
  Electroweak Gauge Bosons from LEP and LHC}},  Phys.Rev. {\bf D87} (2013),
  no.~3 033001, [\href{http://xxx.lanl.gov/abs/1211.1320}{{\tt
  arXiv:1211.1320}}].

\bibitem{Passarino:2012cb}
G.~Passarino, {\it {Nlo Inspired Effective Lagrangians for Higgs Physics}},
  Nucl.Phys. {\bf B868} (2013) 416--458,
  [\href{http://xxx.lanl.gov/abs/1209.5538}{{\tt arXiv:1209.5538}}].

\bibitem{Corbett:2012ja}
T.~Corbett, O.~Eboli, J.~Gonzalez-Fraile, and M.~Gonzalez-Garcia, {\it {Robust
  Determination of the Higgs Couplings: Power to the Data}},  Phys.Rev. {\bf
  D87} (2013) 015022, [\href{http://xxx.lanl.gov/abs/1211.4580}{{\tt
  arXiv:1211.4580}}].

\bibitem{Falkowski:2013dza}
A.~Falkowski, F.~Riva, and A.~Urbano, {\it {Higgs at last}},  JHEP {\bf 1311}
  (2013) 111, [\href{http://xxx.lanl.gov/abs/1303.1812}{{\tt
  arXiv:1303.1812}}].

\bibitem{Giardino:2013bma}
P.~P. Giardino, K.~Kannike, I.~Masina, M.~Raidal, and A.~Strumia, {\it {The
  universal Higgs fit}},  JHEP {\bf 1405} (2014) 046,
  [\href{http://xxx.lanl.gov/abs/1303.3570}{{\tt arXiv:1303.3570}}].

\bibitem{Ellis:2013lra}
J.~Ellis and T.~You, {\it {Updated Global Analysis of Higgs Couplings}},  JHEP
  {\bf 1306} (2013) 103, [\href{http://xxx.lanl.gov/abs/1303.3879}{{\tt
  arXiv:1303.3879}}].

\bibitem{Djouadi:2013qya}
A.~Djouadi and G.~Moreau, {\it {The couplings of the Higgs boson and its CP
  properties from fits of the signal strengths and their ratios at the 7+8 TeV
  LHC}},  Eur.Phys.J. {\bf C73} (2013) 2512,
  [\href{http://xxx.lanl.gov/abs/1303.6591}{{\tt arXiv:1303.6591}}].

\bibitem{Contino:2013kra}
R.~Contino, M.~Ghezzi, C.~Grojean, M.~Muhlleitner, and M.~Spira, {\it
  {Effective Lagrangian for a Light Higgs-Like Scalar}},  JHEP {\bf 1307}
  (2013) 035, [\href{http://xxx.lanl.gov/abs/1303.3876}{{\tt
  arXiv:1303.3876}}].

\bibitem{Dumont:2013wma}
B.~Dumont, S.~Fichet, and G.~von Gersdorff, {\it {A Bayesian View of the Higgs
  Sector with Higher Dimensional Operators}},  JHEP {\bf 1307} (2013) 065,
  [\href{http://xxx.lanl.gov/abs/1304.3369}{{\tt arXiv:1304.3369}}].

\bibitem{Elias-Miro:2013mua}
J.~Elias-Miro, J.~Espinosa, E.~Masso, and A.~Pomarol, {\it {Higgs Windows to
  New Physics Through D = 6 Operators: Constraints and One-Loop Anomalous
  Dimensions}},  \href{http://xxx.lanl.gov/abs/1308.1879}{{\tt
  arXiv:1308.1879}}.

\bibitem{Lopez-Val:2013yba}
D.~Lopez-Val, T.~Plehn, and M.~Rauch, {\it {Measuring Extended Higgs Sectors as
  a Consistent Free Couplings Model}},  JHEP {\bf 1310} (2013) 134,
  [\href{http://xxx.lanl.gov/abs/1308.1979}{{\tt arXiv:1308.1979}}].

\bibitem{Jenkins:2013zja}
E.~E. Jenkins, A.~V. Manohar, and M.~Trott, {\it {Renormalization Group
  Evolution of the Standard Model Dimension Six Operators I: Formalism and
  Lambda Dependence}},  JHEP {\bf 1310} (2013) 087,
  [\href{http://xxx.lanl.gov/abs/1308.2627}{{\tt arXiv:1308.2627}}].

\bibitem{Pomarol:2013zra}
A.~Pomarol and F.~Riva, {\it {Towards the Ultimate Sm Fit to Close in on Higgs
  Physics}},  \href{http://xxx.lanl.gov/abs/1308.2803}{{\tt arXiv:1308.2803}}.

\bibitem{Banerjee:2013apa}
S.~Banerjee, S.~Mukhopadhyay, and B.~Mukhopadhyaya, {\it {Higher Dimensional
  Operators and Lhc Higgs Data : the Role of Modified Kinematics}},  Phys.Rev.
  {\bf D89} (2014) 053010, [\href{http://xxx.lanl.gov/abs/1308.4860}{{\tt
  arXiv:1308.4860}}].

\bibitem{Alloul:2013naa}
A.~Alloul, B.~Fuks, and V.~Sanz, {\it {Phenomenology of the Higgs Effective
  Lagrangian via Feynrules}},  JHEP {\bf 1404} (2014) 110,
  [\href{http://xxx.lanl.gov/abs/1310.5150}{{\tt arXiv:1310.5150}}].

\bibitem{Englert:2014uua}
C.~Englert, A.~Freitas, M.~Muhlleitner, T.~Plehn, M.~Rauch, {\em et.~al.}, {\it
  {Precision Measurements of Higgs Couplings: Implications for New Physics
  Scales}},  \href{http://xxx.lanl.gov/abs/1403.7191}{{\tt arXiv:1403.7191}}.

\bibitem{Bagger:1993zf}
J.~Bagger, V.~D. Barger, K.-m. Cheung, J.~F. Gunion, T.~Han, {\em et.~al.},
  {\it {The Strongly Interacting W W System: Gold Plated Modes}},  Phys.Rev.
  {\bf D49} (1994) 1246--1264,
  [\href{http://xxx.lanl.gov/abs/hep-ph/9306256}{{\tt hep-ph/9306256}}].

\bibitem{Koulovassilopoulos:1993pw}
V.~Koulovassilopoulos and R.~S. Chivukula, {\it {The Phenomenology of a
  Nonstandard Higgs Boson in W(L) W(L) Scattering}},  Phys.Rev. {\bf D50}
  (1994) 3218--3234, [\href{http://xxx.lanl.gov/abs/hep-ph/9312317}{{\tt
  hep-ph/9312317}}].

\bibitem{Burgess:1999ha}
C.~Burgess, J.~Matias, and M.~Pospelov, {\it {A Higgs Or Not a Higgs? What to
  Do If You Discover a New Scalar Particle}},  Int.J.Mod.Phys. {\bf A17} (2002)
  1841--1918, [\href{http://xxx.lanl.gov/abs/hep-ph/9912459}{{\tt
  hep-ph/9912459}}].

\bibitem{Grinstein:2007iv}
B.~Grinstein and M.~Trott, {\it {A Higgs-Higgs Bound State Due to New Physics
  at a TeV}},  Phys.Rev. {\bf D76} (2007) 073002,
  [\href{http://xxx.lanl.gov/abs/0704.1505}{{\tt arXiv:0704.1505}}].

\bibitem{Contino:2010mh}
R.~Contino, C.~Grojean, M.~Moretti, F.~Piccinini, and R.~Rattazzi, {\it {Strong
  Double Higgs Production at the Lhc}},  JHEP {\bf 1005} (2010) 089,
  [\href{http://xxx.lanl.gov/abs/1002.1011}{{\tt arXiv:1002.1011}}].

\bibitem{Azatov:2012bz}
A.~Azatov, R.~Contino, and J.~Galloway, {\it {Model-Independent Bounds on a
  Light Higgs}},  JHEP {\bf 1204} (2012) 127,
  [\href{http://xxx.lanl.gov/abs/1202.3415}{{\tt arXiv:1202.3415}}].
  
\bibitem{Buchalla:2012qq}
G.~Buchalla and O.~Cata,
{\it {Effective Theory of a Dynamically Broken Electroweak Standard Model at NLO}},
JHEP {\bf 1207} (2012) 101,  [\href{http://xxx.lanl.gov/abs/1203.6510}{{\tt
  arXiv:1203.6510}}].

\bibitem{Alonso:2012px}
R.~Alonso, M.~Gavela, L.~Merlo, S.~Rigolin, and J.~Yepes, {\it {The Effective
  Chiral Lagrangian for a Light Dynamical `Higgs'}},  Phys.Lett. {\bf B722}
  (2013) 330--335, [\href{http://xxx.lanl.gov/abs/1212.3305}{{\tt
  arXiv:1212.3305}}].

\bibitem{Alonso:2012pz}
R.~Alonso, M.~Gavela, L.~Merlo, S.~Rigolin, and J.~Yepes, {\it {Flavor with a
  Light Dynamical "Higgs Particle"}},  Phys.Rev. {\bf D87} (2013) 055019,
  [\href{http://xxx.lanl.gov/abs/1212.3307}{{\tt arXiv:1212.3307}}].

\bibitem{Buchalla:2013rka}
G.~Buchalla, O.~Cata, and C.~Krause, {\it {Complete Electroweak Chiral
  Lagrangian with a Light Higgs at NLO}},  Nucl.Phys. {\bf B880} (2014)
  552--573, [\href{http://xxx.lanl.gov/abs/1307.5017}{{\tt arXiv:1307.5017}}].

\bibitem{Brivio:2013pma}
I.~Brivio, T.~Corbett, O.~Eboli, M.~Gavela, J.~Gonzalez-Fraile, {\em et.~al.},
  {\it {Disentangling a dynamical Higgs}},  JHEP {\bf 1403} (2014) 024,
  [\href{http://xxx.lanl.gov/abs/1311.1823}{{\tt arXiv:1311.1823}}].

\bibitem{Brivio:2014pfa}
I.~Brivio, O.~Eboli, M.~Gavela, M.~Gonzalez-Garcia, L.~Merlo, {\em et.~al.},
  {\it {Higgs Ultraviolet Softening}},
  \href{http://xxx.lanl.gov/abs/1405.5412}{{\tt arXiv:1405.5412}}.

\bibitem{Dell'Aquila:1985ve}
J.~R. Dell'Aquila and C.~A. Nelson, {\it {$P$ or {CP} Determination by
  Sequential Decays: V1 V2 Modes With Decays Into $\bar{\ell}(A)$ $\ell (B)$
  And/or $\bar{q}(A)$ $q (B)$}},  Phys.Rev. {\bf D33} (1986) 80.

\bibitem{Dell'Aquila:1985vc}
J.~R. Dell'Aquila and C.~A. Nelson, {\it {Distinguishing a Spin 0 Technipion
  and an Elementary Higgs Boson: V1 V2 Modes With Decays Into $\bar{\ell}(A)$
  $\ell (B)$ And/or $\bar{q}(A)$ $q (B)$}},  Phys.Rev. {\bf D33} (1986) 93.

\bibitem{Dell'Aquila:1985vb}
J.~R. Dell'Aquila and C.~A. Nelson, {\it {Simple Tests for {CP} or $P$
  Violation by Sequential Decays: V1 V2 Modes With Decays Into $\bar{\ell}(A)$
  $\ell (B)$ And/or $\bar{q}(A)$ $q (B)$}},  Phys.Rev. {\bf D33} (1986) 101.

\bibitem{Soni:1993jc}
A.~Soni and R.~Xu, {\it {Probing CP violation via Higgs decays to four
  leptons}},  Phys.Rev. {\bf D48} (1993) 5259--5263,
  [\href{http://xxx.lanl.gov/abs/hep-ph/9301225}{{\tt hep-ph/9301225}}].

\bibitem{Chang:1993jy}
D.~Chang, W.-Y. Keung, and I.~Phillips, {\it {CP odd correlation in the decay
  of neutral Higgs boson into Z Z, W+ W-, or t anti-t}},  Phys.Rev. {\bf D48}
  (1993) 3225--3234, [\href{http://xxx.lanl.gov/abs/hep-ph/9303226}{{\tt
  hep-ph/9303226}}].

\bibitem{Arens:1994wd}
T.~Arens and L.~Sehgal, {\it {Energy spectra and energy correlations in the
  decay H --$>$ Z Z --$>$ mu+ mu- mu+ mu-}},  Z.Phys. {\bf C66} (1995) 89--94,
  [\href{http://xxx.lanl.gov/abs/hep-ph/9409396}{{\tt hep-ph/9409396}}].

\bibitem{Choi:2002jk}
S.~Choi, D.~Miller, M.~Muhlleitner, and P.~Zerwas, {\it {Identifying the Higgs
  spin and parity in decays to Z pairs}},  Phys.Lett. {\bf B553} (2003) 61--71,
  [\href{http://xxx.lanl.gov/abs/hep-ph/0210077}{{\tt hep-ph/0210077}}].

\bibitem{Buszello:2002uu}
C.~Buszello, I.~Fleck, P.~Marquard, and J.~van~der Bij, {\it {Prospective
  analysis of spin- and CP-sensitive variables in H --$>$ Z Z --$>$ l(1)+ l(1)-
  l(2)+ l(2)- at the LHC}},  Eur.Phys.J. {\bf C32} (2004) 209--219,
  [\href{http://xxx.lanl.gov/abs/hep-ph/0212396}{{\tt hep-ph/0212396}}].

\bibitem{Godbole:2007cn}
R.~M. Godbole, D.~Miller, and M.~M. Muhlleitner, {\it {Aspects of CP violation
  in the H ZZ coupling at the LHC}},  JHEP {\bf 0712} (2007) 031,
  [\href{http://xxx.lanl.gov/abs/0708.0458}{{\tt arXiv:0708.0458}}].

\bibitem{Cao:2009ah}
Q.-H. Cao, C.~Jackson, W.-Y. Keung, I.~Low, and J.~Shu, {\it {The Higgs
  Mechanism and Loop-induced Decays of a Scalar into Two Z Bosons}},  Phys.Rev.
  {\bf D81} (2010) 015010, [\href{http://xxx.lanl.gov/abs/0911.3398}{{\tt
  arXiv:0911.3398}}].

\bibitem{Gao:2010qx}
Y.~Gao, A.~V. Gritsan, Z.~Guo, K.~Melnikov, M.~Schulze, {\em et.~al.}, {\it
  {Spin determination of single-produced resonances at hadron colliders}},
  Phys.Rev. {\bf D81} (2010) 075022,
  [\href{http://xxx.lanl.gov/abs/1001.3396}{{\tt arXiv:1001.3396}}].

\bibitem{DeRujula:2010ys}
A.~De~Rujula, J.~Lykken, M.~Pierini, C.~Rogan, and M.~Spiropulu, {\it {Higgs
  look-alikes at the LHC}},  Phys.Rev. {\bf D82} (2010) 013003,
  [\href{http://xxx.lanl.gov/abs/1001.5300}{{\tt arXiv:1001.5300}}].

\bibitem{Plehn:2001nj}
T.~Plehn, D.~L. Rainwater, and D.~Zeppenfeld, {\it {Determining the structure
  of Higgs couplings at the LHC}},  Phys.Rev.Lett. {\bf 88} (2002) 051801,
  [\href{http://xxx.lanl.gov/abs/hep-ph/0105325}{{\tt hep-ph/0105325}}].

\bibitem{Buszello:2006hf}
C.~Buszello and P.~Marquard, {\it {Determination of spin and CP of the Higgs
  boson from WBF}},  \href{http://xxx.lanl.gov/abs/hep-ph/0603209}{{\tt
  hep-ph/0603209}}.

\bibitem{Hankele:2006ma}
V.~Hankele, G.~Klamke, D.~Zeppenfeld, and T.~Figy, {\it {Anomalous Higgs boson
  couplings in vector boson fusion at the CERN LHC}},  Phys.Rev. {\bf D74}
  (2006) 095001, [\href{http://xxx.lanl.gov/abs/hep-ph/0609075}{{\tt
  hep-ph/0609075}}].

\bibitem{Klamke:2007cu}
G.~Klamke and D.~Zeppenfeld, {\it {Higgs plus two jet production via gluon
  fusion as a signal at the CERN LHC}},  JHEP {\bf 0704} (2007) 052,
  [\href{http://xxx.lanl.gov/abs/hep-ph/0703202}{{\tt hep-ph/0703202}}].

\bibitem{Englert:2012ct}
C.~Englert, M.~Spannowsky, and M.~Takeuchi, {\it {Measuring Higgs CP and
  couplings with hadronic event shapes}},  JHEP {\bf 1206} (2012) 108,
  [\href{http://xxx.lanl.gov/abs/1203.5788}{{\tt arXiv:1203.5788}}].

\bibitem{Odagiri:2002nd}
K.~Odagiri, {\it {On azimuthal spin correlations in Higgs plus jet events at
  LHC}},  JHEP {\bf 0303} (2003) 009,
  [\href{http://xxx.lanl.gov/abs/hep-ph/0212215}{{\tt hep-ph/0212215}}].

\bibitem{DelDuca:2006hk}
V.~Del~Duca, G.~Klamke, D.~Zeppenfeld, M.~L. Mangano, M.~Moretti, {\em
  et.~al.}, {\it {Monte Carlo studies of the jet activity in Higgs + 2 jet
  events}},  JHEP {\bf 0610} (2006) 016,
  [\href{http://xxx.lanl.gov/abs/hep-ph/0608158}{{\tt hep-ph/0608158}}].

\bibitem{Andersen:2010zx}
J.~R. Andersen, K.~Arnold, and D.~Zeppenfeld, {\it {Azimuthal Angle
  Correlations for Higgs Boson plus Multi-Jet Events}},  JHEP {\bf 1006} (2010)
  091, [\href{http://xxx.lanl.gov/abs/1001.3822}{{\tt arXiv:1001.3822}}].

\bibitem{Englert:2012xt}
C.~Englert, D.~Goncalves-Netto, K.~Mawatari, and T.~Plehn, {\it {Higgs Quantum
  Numbers in Weak Boson Fusion}},  JHEP {\bf 1301} (2013) 148,
  [\href{http://xxx.lanl.gov/abs/1212.0843}{{\tt arXiv:1212.0843}}].

\bibitem{Djouadi:2013yb}
A.~Djouadi, R.~Godbole, B.~Mellado, and K.~Mohan, {\it {Probing the spin-parity
  of the Higgs boson via jet kinematics in vector boson fusion}},  Phys.Lett.
  {\bf B723} (2013) 307--313, [\href{http://xxx.lanl.gov/abs/1301.4965}{{\tt
  arXiv:1301.4965}}].

\bibitem{Dolan:2014upa}
M.~J. Dolan, P.~Harris, M.~Jankowiak, and M.~Spannowsky, {\it {Constraining
  CP-violating Higgs Sectors at the LHC using gluon fusion}},
  \href{http://xxx.lanl.gov/abs/1406.3322}{{\tt arXiv:1406.3322}}.

\bibitem{Christensen:2010pf}
N.~D. Christensen, T.~Han, and Y.~Li, {\it {Testing CP Violation in ZZH
  Interactions at the LHC}},  Phys.Lett. {\bf B693} (2010) 28--35,
  [\href{http://xxx.lanl.gov/abs/1005.5393}{{\tt arXiv:1005.5393}}].

\bibitem{Desai:2011yj}
N.~Desai, D.~K. Ghosh, and B.~Mukhopadhyaya, {\it {CP-violating HWW couplings
  at the Large Hadron Collider}},  Phys.Rev. {\bf D83} (2011) 113004,
  [\href{http://xxx.lanl.gov/abs/1104.3327}{{\tt arXiv:1104.3327}}].

\bibitem{Ellis:2012xd}
J.~Ellis, D.~S. Hwang, V.~Sanz, and T.~You, {\it {A Fast Track towards the
  `Higgs' Spin and Parity}},  JHEP {\bf 1211} (2012) 134,
  [\href{http://xxx.lanl.gov/abs/1208.6002}{{\tt arXiv:1208.6002}}].

\bibitem{Godbole:2013saa}
R.~Godbole, D.~J. Miller, K.~Mohan, and C.~D. White, {\it {Boosting Higgs CP
  properties via $VH$ Production at the Large Hadron Collider}},  Phys.Lett.
  {\bf B730} (2014) 275--279, [\href{http://xxx.lanl.gov/abs/1306.2573}{{\tt
  arXiv:1306.2573}}].

\bibitem{Delaunay:2013npa}
C.~Delaunay, G.~Perez, H.~de~Sandes, and W.~Skiba, {\it {Higgs Up-Down CP
  Asymmetry at the LHC}},  Phys.Rev. {\bf D89} (2014) 035004,
  [\href{http://xxx.lanl.gov/abs/1308.4930}{{\tt arXiv:1308.4930}}].

\bibitem{Voloshin:2012tv}
M.~Voloshin, {\it {CP Violation in Higgs Diphoton Decay in Models with
  Vectorlike Heavy Fermions}},  Phys.Rev. {\bf D86} (2012) 093016,
  [\href{http://xxx.lanl.gov/abs/1208.4303}{{\tt arXiv:1208.4303}}].

\bibitem{Korchin:2013ifa}
A.~Y. Korchin and V.~A. Kovalchuk, {\it {Polarization effects in the Higgs
  boson decay to $\gamma Z$ and test of $CP$ and $CPT$ symmetries}},  Phys.Rev.
  {\bf D88} (2013), no.~3 036009,
  [\href{http://xxx.lanl.gov/abs/1303.0365}{{\tt arXiv:1303.0365}}].

\bibitem{Bishara:2013vya}
F.~Bishara, Y.~Grossman, R.~Harnik, D.~J. Robinson, J.~Shu, {\em et.~al.}, {\it
  {Probing CP Violation in $h\rightarrow\gamma\gamma$ with Converted Photons}},
   JHEP {\bf 1404} (2014) 084, [\href{http://xxx.lanl.gov/abs/1312.2955}{{\tt
  arXiv:1312.2955}}].

\bibitem{Chen:2014ona}
Y.~Chen, A.~Falkowski, I.~Low, and R.~Vega-Morales, {\it {New Observables for
  CP Violation in Higgs Decays}},
  \href{http://xxx.lanl.gov/abs/1405.6723}{{\tt arXiv:1405.6723}}.

\bibitem{Freitas:2012kw}
A.~Freitas and P.~Schwaller, {\it {Higgs CP Properties From Early LHC Data}},
  Phys.Rev. {\bf D87} (2013), no.~5 055014,
  [\href{http://xxx.lanl.gov/abs/1211.1980}{{\tt arXiv:1211.1980}}].

\bibitem{Belusca-Maito:2014dpa}
H.~Belusca-Maito, {\it {Effective Higgs Lagrangian and Constraints on Higgs
  Couplings}},  \href{http://xxx.lanl.gov/abs/1404.5343}{{\tt
  arXiv:1404.5343}}.
  
\bibitem{Shu:2013uua}
J.~Shu and Y.~Zhang,
{\it {Impact of a CP Violating Higgs Sector: from Lhc to Baryogenesis}},
Phys.\ Rev.\ Lett.\ {\bf 111} (2013) 9, 091801,
\href{http://xxx.lanl.gov/abs/1304.0773}{{\tt arXiv:1304.0773}}.

\bibitem{Cheung:2013kla}
K.~Cheung, J.~S.~Lee and P.~-Y.~Tseng,
{\it{Higgs Precision (Higgcision) Era Begins}},
JHEP {\bf 1305} (2013) 134
\href{http://xxx.lanl.gov/abs/1302.3794}{{\tt  arXiv:1302.3794}}.

\bibitem{Appelquist:1980vg}
T.~Appelquist and C.~W. Bernard, {\it {Strongly Interacting Higgs Bosons}},
  Phys. Rev. {\bf D22} (1980) 200.

\bibitem{Longhitano:1980iz}
A.~C. Longhitano, {\it {Heavy Higgs Bosons in the Weinberg-Salam Model}},
  Phys. Rev. {\bf D22} (1980) 1166.

\bibitem{Longhitano:1980tm}
A.~C. Longhitano, {\it {Low-Energy Impact of a Heavy Higgs Boson Sector}},
  Nucl.Phys. {\bf B188} (1981) 118.

\bibitem{Appelquist:1993ka}
T.~Appelquist and G.-H. Wu, {\it {The Electroweak Chiral Lagrangian and New
  Precision Measurements}},  Phys.Rev. {\bf D48} (1993) 3235--3241,
  [\href{http://xxx.lanl.gov/abs/hep-ph/9304240}{{\tt hep-ph/9304240}}].

\bibitem{Georgi:1986df}
H.~Georgi, D.~B. Kaplan, and L.~Randall, {\it {Manifesting the Invisible Axion
  at Low-Energies}},  Phys.Lett. {\bf B169} (1986) 73.

\bibitem{Hagiwara:1986vm}
K.~Hagiwara, R.~Peccei, D.~Zeppenfeld, and K.~Hikasa, {\it {Probing the Weak
  Boson Sector in E+ E- --> W+ W-}},  Nucl.Phys. {\bf B282} (1987) 253.
  
\bibitem{Appelquist:2004mn}
T.~Appelquist, M.~Piai and R.~Shrock,
{\it {Lepton Dipole Moments in Extended Technicolor Models}},
Phys.\ Lett.\ B {\bf 593} (2004) 175,
[\href{http://xxx.lanl.gov/abs/hep-ph/0401114}{{\tt arXiv:hep-ph/0401114}}].

\bibitem{Marciano:1986eh}
W.~J. Marciano and A.~Queijeiro, {\it {Bound on the W Boson Electric Dipole
  Moment}},  Phys.Rev. {\bf D33} (1986) 3449.

\bibitem{Baron:2013eja}
{\bf ACME} Collaboration, J.~Baron {\em et.~al.}, {\it {Order of Magnitude
  Smaller Limit on the Electric Dipole Moment of the Electron}},  Science {\bf
  343} (2014), no.~6168 269--272,
  [\href{http://xxx.lanl.gov/abs/1310.7534}{{\tt arXiv:1310.7534}}].

\bibitem{Baker:2006ts}
C.~Baker, D.~Doyle, P.~Geltenbort, K.~Green, M.~van~der Grinten, {\em et.~al.},
  {\it {An Improved Experimental Limit on the Electric Dipole Moment of the
  Neutron}},  Phys.Rev.Lett. {\bf 97} (2006) 131801,
  [\href{http://xxx.lanl.gov/abs/hep-ex/0602020}{{\tt hep-ex/0602020}}].

\bibitem{Dawson:2013owa}
S.~Dawson, S.~K. Gupta, and G.~Valencia, {\it {CP violating anomalous couplings
  in $W\gamma$ and $Z\gamma$ production at the LHC}},  Phys.Rev. {\bf D88}
  (2013), no.~3 035008, [\href{http://xxx.lanl.gov/abs/1304.3514}{{\tt
  arXiv:1304.3514}}].

\bibitem{Abbiendi:2000ei}
{\bf OPAL} Collaboration, G.~Abbiendi {\em et.~al.}, {\it {Measurement of $W$
  boson polarizations and CP violating triple gauge couplings from $W^{+}
  W^{-}$ production at LEP}},  Eur.Phys.J. {\bf C19} (2001) 229--240,
  [\href{http://xxx.lanl.gov/abs/hep-ex/0009021}{{\tt hep-ex/0009021}}].

\bibitem{Abdallah:2008sf}
{\bf DELPHI} Collaboration, J.~Abdallah {\em et.~al.}, {\it {Study of W boson
  polarisations and Triple Gauge boson Couplings in the reaction e+e- --$>$W+W-
  at LEP 2}},  Eur.Phys.J. {\bf C54} (2008) 345--364,
  [\href{http://xxx.lanl.gov/abs/0801.1235}{{\tt arXiv:0801.1235}}].

\bibitem{Schael:2004tq}
{\bf ALEPH} Collaboration, S.~Schael {\em et.~al.}, {\it {Improved measurement
  of the triple gauge-boson couplings gamma W W and Z W W in e+ e-
  collisions}},  Phys.Lett. {\bf B614} (2005) 7--26.

\bibitem{PDG}
{\bf Particle Data Group} Collaboration, J.~Beringer {\em et.~al.}, {\it
  {Review of Particle Physics (RPP)}},  Phys.Rev. {\bf D86} (2012) 010001.

\bibitem{Eboli:2010qd}
O.~Eboli, J.~Gonzalez-Fraile, and M.~Gonzalez-Garcia, {\it {Scrutinizing the
  ZW+W- vertex at the Large Hadron Collider at 7 TeV}},  Phys.Lett. {\bf B692}
  (2010) 20--25, [\href{http://xxx.lanl.gov/abs/1006.3562}{{\tt
  arXiv:1006.3562}}].

\bibitem{Aad:2012twa}
{\bf ATLAS} Collaboration, G.~Aad {\em et.~al.}, {\it {Measurement of $WZ$
  production in proton-proton collisions at $\sqrt{s}=7$ TeV with the ATLAS
  detector}},  Eur.Phys.J. {\bf C72} (2012) 2173,
  [\href{http://xxx.lanl.gov/abs/1208.1390}{{\tt arXiv:1208.1390}}].

\bibitem{Christensen:2008py}
N.~D. Christensen and C.~Duhr, {\it {FeynRules - Feynman rules made easy}},
  Comput.Phys.Commun. {\bf 180} (2009) 1614--1641,
  [\href{http://xxx.lanl.gov/abs/0806.4194}{{\tt arXiv:0806.4194}}].

\bibitem{Alwall:2011uj}
J.~Alwall, M.~Herquet, F.~Maltoni, O.~Mattelaer, and T.~Stelzer, {\it {MadGraph
  5 : Going Beyond}},  JHEP {\bf 1106} (2011) 128,
  [\href{http://xxx.lanl.gov/abs/1106.0522}{{\tt arXiv:1106.0522}}].

\bibitem{Sjostrand:2006za}
T.~Sjostrand, S.~Mrenna, and P.~Z. Skands, {\it {PYTHIA 6.4 Physics and
  Manual}},  JHEP {\bf 0605} (2006) 026,
  [\href{http://xxx.lanl.gov/abs/hep-ph/0603175}{{\tt hep-ph/0603175}}].

\bibitem{pgs}
J.~Conway, {\it {PGS4}},
  {\url{http://www.physics.ucdavis.edu/~conway/research/software/pgs/pgs4-general.htm}}.

\bibitem{Dawson:1996ge}
S.~Dawson, X.-G. He, and G.~Valencia, {\it {CP violation in W gamma and Z gamma
  production}},  Phys.Lett. {\bf B390} (1997) 431--436,
  [\href{http://xxx.lanl.gov/abs/hep-ph/9609523}{{\tt hep-ph/9609523}}].

\bibitem{Kumar:2008ng}
J.~Kumar, A.~Rajaraman, and J.~D. Wells, {\it {Probing CP-violation at
  colliders through interference effects in diboson production and decay}},
  Phys.Rev. {\bf D78} (2008) 035014,
  [\href{http://xxx.lanl.gov/abs/0801.2891}{{\tt arXiv:0801.2891}}].

\bibitem{Han:2009ra}
T.~Han and Y.~Li, {\it {Genuine CP-odd Observables at the LHC}},  Phys.Lett.
  {\bf B683} (2010) 278--281, [\href{http://xxx.lanl.gov/abs/0911.2933}{{\tt
  arXiv:0911.2933}}].

\bibitem{Spira:1995rr}
M.~Spira, A.~Djouadi, D.~Graudenz, and P.~Zerwas, {\it {Higgs boson production
  at the LHC}},  Nucl.Phys. {\bf B453} (1995) 17--82,
  [\href{http://xxx.lanl.gov/abs/hep-ph/9504378}{{\tt hep-ph/9504378}}].

\bibitem{McKeen:2012av}
D.~McKeen, M.~Pospelov, and A.~Ritz, {\it {Modified Higgs branching ratios
  versus CP and lepton flavor violation}},  Phys.Rev. {\bf D86} (2012) 113004,
  [\href{http://xxx.lanl.gov/abs/1208.4597}{{\tt arXiv:1208.4597}}].

\bibitem{Chatrchyan:2012jja}
{\bf CMS} Collaboration, S.~Chatrchyan {\em et.~al.}, {\it {Study of the Mass
  and Spin-Parity of the Higgs Boson Candidate Via Its Decays to Z Boson
  Pairs}},  Phys.Rev.Lett. {\bf 110} (2013) 081803,
  [\href{http://xxx.lanl.gov/abs/1212.6639}{{\tt arXiv:1212.6639}}].

\bibitem{Aad:2013xqa}
{\bf ATLAS} Collaboration, G.~Aad {\em et.~al.}, {\it {Evidence for the spin-0
  nature of the Higgs boson using ATLAS data}},  Phys.Lett. {\bf B726} (2013)
  120--144, [\href{http://xxx.lanl.gov/abs/1307.1432}{{\tt arXiv:1307.1432}}].

\bibitem{ATLAS:2013nma}
{\bf ATLAS} Collaboration, {\it {Measurements of the properties of the
  Higgs-like boson in the four lepton decay channel with the ATLAS detector
  using 25 fb−1 of proton-proton collision data}},  ATLAS-CONF-2013-013,
  ATLAS-COM-CONF-2013-018 (2013).

\bibitem{CMS:2013xfa}
{\bf CMS} Collaboration, {\it {Projected Performance of an Upgraded CMS
  Detector at the LHC and HL-LHC: Contribution to the Snowmass Process}},
  \href{http://xxx.lanl.gov/abs/1307.7135}{{\tt arXiv:1307.7135}}.
  

\end{thebibliography}

\providecommand{\href}[2]{#2}\begingroup\raggedright\endgroup

\end{document}